\documentclass[aip,jcp,amsmath,amssymb,floatfix,reprint,citeautoscript,articletitle=false,noeprint]{revtex4-2}

\usepackage{amssymb}
\usepackage{amsmath}
\usepackage{graphicx}
\usepackage{dcolumn}
\usepackage{xcolor}
\usepackage{fancyhdr}
\usepackage[colorlinks,allcolors=black,citecolor=blue,urlcolor=blue]{hyperref}

\newcommand{\response}[1]{\textcolor{black}{#1}}
\newcommand{\tochange}[1]{\textcolor{blue}{#1}}

\begin{document}

\title{Protons accumulate at the graphene-water interface}
\author{Xavier R. Advincula}
\affiliation{Yusuf Hamied Department of Chemistry, University of Cambridge, Lensfield Road, Cambridge, CB2 1EW, UK}
\affiliation{Cavendish Laboratory, Department of Physics, University of Cambridge, Cambridge, CB3 0HE, UK}
\affiliation{Lennard-Jones Centre, University of Cambridge, Trinity Ln, Cambridge, CB2 1TN, UK}
\author{Kara D. Fong}
\email{kdf22@cam.ac.uk}
\affiliation{Yusuf Hamied Department of Chemistry, University of Cambridge, Lensfield Road, Cambridge, CB2 1EW, UK}
\affiliation{Lennard-Jones Centre, University of Cambridge, Trinity Ln, Cambridge, CB2 1TN, UK}
\author{Angelos Michaelides}
\email{am452@cam.ac.uk}
\affiliation{Yusuf Hamied Department of Chemistry, University of Cambridge, Lensfield Road, Cambridge, CB2 1EW, UK}
\affiliation{Lennard-Jones Centre, University of Cambridge, Trinity Ln, Cambridge, CB2 1TN, UK}
\author{Christoph Schran}
\email{cs2121@cam.ac.uk}
\affiliation{Cavendish Laboratory, Department of Physics, University of Cambridge, Cambridge, CB3 0HE, UK}
\affiliation{Lennard-Jones Centre, University of Cambridge, Trinity Ln, Cambridge, CB2 1TN, UK}

\begin{abstract} 
Water's ability to autoionize into hydroxide and hydronium ions profoundly influences surface properties, rendering interfaces either basic or acidic. 
While it is well-established that \response{protons show an affinity to the air-water interface}, a critical knowledge gap exists in technologically relevant surfaces like the graphene-water interface.
Here we use machine learning-based simulations with first-principles accuracy to unravel the behavior of the hydroxide and hydronium ions at the graphene-water interface.
Our findings reveal that \response{protons accumulate at the graphene-water interface}, with the hydronium ion predominantly residing in the first contact layer of water. 
In contrast, the hydroxide ion exhibits a bimodal distribution, found both near the surface and towards the interior layers.
Analysis of the underlying electronic structure reveals \response{local} polarization effects, resulting in counterintuitive charge rearrangement. 
Proton propensity to the graphene-water interface challenges the interpretation of surface experiments and is expected to have far-reaching consequences for ion conductivity, interfacial reactivity, and proton-mediated processes.
\end{abstract}
\maketitle

Water interacts with interfaces in numerous technologies involving areas such as atmospheric chemistry \cite{atm_chem_2000}, water desalination \cite{desalin_2011}, energy production via water splitting \cite{wat_split_2019}, and storage devices \cite{lopez_2020_inter}. %
For these technologies,  understanding the fundamental nature of these interfaces ---whether they 
\response{
accumulate or repel ions
}--- is essential to improve performance and facilitate scientific breakthroughs.
At the molecular level, this is governed by the self-dissociation of water into hydroxide (OH$^-$) and hydronium (H$_3$O$^+$) ions, which ultimately determines the pH of a solution and facilitates proton transfer.
A deeper understanding of how surface interactions influence the propensity for hydronium and hydroxide ions would enable the optimization of these interfacial processes.
Despite significant progress in understanding proton transfer in bulk water and microsolvation \cite{geissler_2001, chandra_2007_solvation, hassanali_2013_pnas, dahms_2017, calio_jacs_2021, netz_spectral_2022, laage_pt_natchem_2024}, describing these processes near interfaces continue to pose both theoretical and experimental challenges.
One of the most enduring and fundamental debates in chemistry has been the 
\response{
nature of the excess proton (hydronium ion) and hydroxide ion at
}
the air-water interface \cite{buch_wat_acidic, mishra_2012, sayka_2013_two, tse_2015, netz_orient_2017}.
The main complexity stems from the dynamic nature of protonic defects, namely hydronium and hydroxide ions, in the aqueous phase \cite{tucker_1995_h3o_oh, tucker_1999_canonical, tucker_oh_canonical_2002}. %
Furthermore, the interplay between directional hydrogen bonds and non-directional van der Waals forces often results in molecular conformations with similar energies \cite{tucker_1995_h3o_oh, tucker_1999_canonical, tucker_oh_canonical_2002}, adding to the challenge.
Experimental analysis is further complicated by varying probing resolutions and interfacial depths \cite{sayka_2013_two}.
Only recently has it been established, both theoretically and experimentally, that the air-water interface
\response{accumulates protons, while hydroxide ions are repelled \cite{das_2019, das_ez_2020, miguel_ph_2024, litman_2024_nature}.
}
However, this understanding may not extend to other technologically relevant interfaces, as the fundamental mechanisms governing the behavior of these ions have been reported to differ across various interfaces \cite{dellago_2003_prl, cao_2010, Gaigeot_2012, sulpizi_silicawater_2012, marx_eproton, marx_phole}. 
Among all interfaces, the graphene-water interface is particularly relevant due to its extensive range of technological applications, from nanofluidics to electronic devices.
In recent years, it has been reported that when water is confined between graphene sheets, it exhibits unique properties, including an anomalously low dielectric constant \cite{fumagalli_dielectric_2018, chris_dufils_2024}, atypical friction behavior \cite{secchi_nat_2016}, and superionic character \cite{vk_chris_2022, jiang_rich_2024}, among other phenomena \cite{algara-siller_square_2015,quasionedimensional}.
Additionally, graphene’s precise synthesis in experimental settings \cite{radha_b_2016} makes it an ideal candidate for providing insights applicable to real-world systems.
However, our current understanding is limited, and previous studies have shown conflicting results.
Grosjean et al. \cite{mlb_oh} reported a physisorbed state of hydroxide in the contact layer at the graphene/water interface, which is rationalized in the context of a macroscopic experimental observation\cite{secchi_2016_oh_prl} and has significant implications for conductance.
\response{
Conversely, de Aquino et al. \cite{aquino_2019} indicated that hydroxide ions are more prevalent in the interior layers, while hydronium ions are more prevalent in the interfacial layers.
More recently, Scalfi et al. \cite{scalfi_2024} revealed that hydronium adsorbs at the graphene-water interface while hydroxide mostly shows only limited adsorption under specific conditions.
While these studies have advanced our understanding, they rely primarily on non-reactive force fields, which may constrain the mechanistic insights they can offer.
As a result, despite valuable efforts in the field, a detailed picture of protonic defects near the graphene-water interface has not been established.
}
In particular, 
\response{detailed mechanistic and thermodynamic insights into hydrogen bonding, the orientational behavior, and interfacial polarization of the hydronium and hydroxide ion at the graphene-water interface have remained unresolved until now.}
The complex interplay between water and the interface \cite{wat_inter_2016, wat_charged_2021}, coupled with the prohibitive computational expense of \textit{ab initio} molecular dynamics (AIMD) simulations needed to adequately sample these reactive systems, has significantly limited progress in this area.
This study aims to address this gap, enhancing our understanding of graphene’s interfacial properties and improving the interpretation of experimental data.

In recent years, machine learning potentials (MLPs) have emerged as an efficient and flexible solution for accurately modeling reactive processes at interfaces. 
These technologies bypass the prohibitive costs of \textit{ab initio} calculations, significantly extending the length and time scales accessible in molecular simulations \cite{behler_2016_pers, butler_2018, montero_2024}.
By accurately representing the potential energy surface of a chosen \textit{ab initio} reference method, such as density functional theory (DFT), MLPs establish a direct structure-energy relationship.
This enables the description of bond-breaking and bond-forming events in complex environments \cite{complex_2021, de_la_puente_acids_2022, car_dissoc_deepmd_2023, kara_pairing_2024}.
This approach is particularly beneficial for our study, as classical force field models either lack the reactivity needed to represent the dynamics of covalent OH bond breaking and formation \cite{mbpol_2013}, or fail to provide the required accuracy, being parameterized mainly for bulk properties \cite{bonthuis_2016_ff}.
MLPs address these shortcomings, offering a reliable and accurate method to investigate the specific characteristics of interfacial phenomena.

This work uses MLP-based simulations to demonstrate that \response{protons accumulate at the graphene-water interface}, while hydroxide ions exhibit a bimodal distribution, being found both close to the surface and in layers farther from the interface.
This surface affinity is due to the hydrogen bond environment of hydronium remaining stable at the interface, while hydroxide's environment is disrupted.
By examining the thermodynamic driving forces, we find that hydronium ions are enthalpically driven to the interface, whereas entropic forces drive hydroxide ions.
When comparing these findings to the air-water interface, we see that graphene significantly influences ionic interactions due to polarization effects.
This response suggests that macroscopic experiments should be interpreted carefully.

\section*{Results and Discussion}

To study protonic defects at the graphene-water interface, we developed an MLP using the MACE architecture \cite{mace_canonical}.
The MLP demonstrates excellent capability in reproducing the potential energy surface of the underlying DFT at the revPBE-D3\cite{revpbed3_1, revpbed3_2} level of theory, known to perform well for water~\cite{angelos_dft_water_2016, tobias_vdw_2016, ondrej_revpbe_2017} and graphene-water\cite{Brandenburg2019} interactions, while also effectively capturing protonic defects \cite{marsalek_mlp_pt}.
\response{We have verified the validity of our results with respect to the functional and dispersion correction and confirmed that the observed trends remain unchanged (see \tochange{Section S2}).}
Our model effectively captures the properties of both types of protonic defects in the water near free-standing graphene surfaces, under conditions ranging from ultra-confined environments to bulk-like settings (see \tochange{Methods} and \tochange{Section S2} for development and validation details).
We used the MLP to perform multi-nanosecond MD simulations, evaluating properties of five systems involving average slit widths of about 6.5, 9.2, 12.2, 14.7, 19.7\;\AA.
These simulations included one layer (1L), two layers (2L), three layers (3L), four layers (4L), and five layers (5L) of water, each containing either a hydronium or a hydroxide ion (see \tochange{Section S1}).

\subsection*{Hydronium resides at the interface, hydroxide does not}
\begin{figure*}
    \centering
    \includegraphics[width=\textwidth]{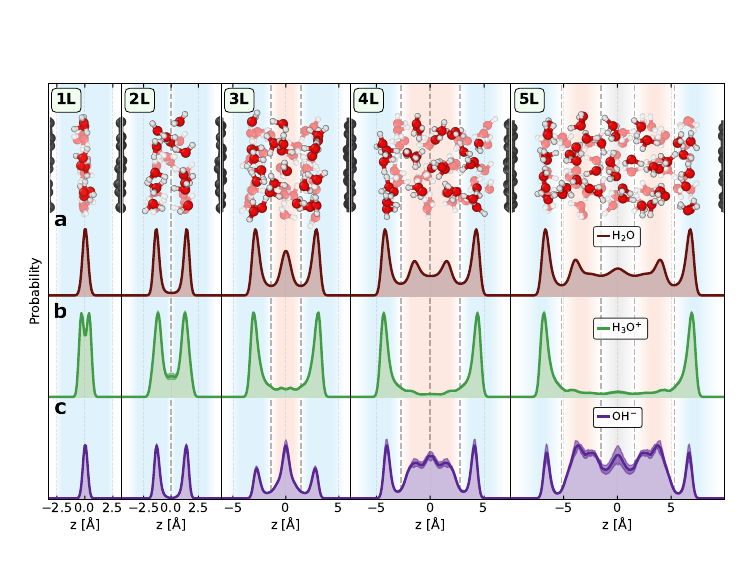}
    \caption{\textbf{Differences in the surface propensity of hydronium and hydroxide in nanoconfined water. }
    Normalized and symmetrized density profiles along the $z$ axis perpendicular to the free-standing graphene sheets obtained from the oxygen atoms in a neutral water system (a), and from the specific oxygen of the protonic defect of acidic (b) and basic systems (c), accompanied by representative snapshots.
    For each of the two species and five slit widths, five independent simulations were conducted, each lasting 4\;ns.
    This approach led to a total of 200\;ns of cumulative simulation time.
    The shaded regions indicate the uncertainty resulting from the standard deviation of the five replicate simulations.
    The background's faded blue, orange, and gray represent the interfacial, intermediate, and bulk-like water layers, respectively.
    The vertical dashed lines indicate the partitioning among these water layers.
    The horizontal axis limits in each plot correspond to the average carbon layer positions.}
    \label{fig:fig1}
\end{figure*}

\begin{figure*}
    \centering
    \includegraphics[width=\textwidth]{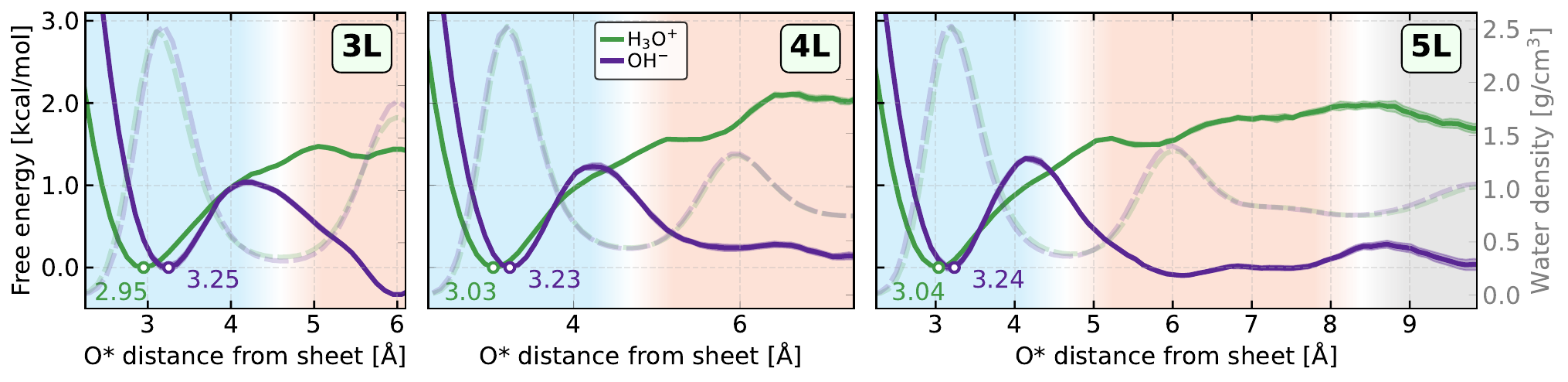}
    \caption{\textbf{Free energy profiles and water structuring of hydronium and hydroxide near graphene sheets.} 
    Free energy profiles for the hydronium and hydroxide ions as a function of their oxygen distance O$^{*}$ to the closest graphene sheet obtained from the symmetrized density profiles. 
    The minima in the interfacial layer, marked with a white dot, serve as the free energy reference point for each ion, with their distances to the interface presented.
    The structuring of the water layers is represented by the water density profiles, which are indicated with corresponding lighter colors and dashed lines.
    The shaded regions indicate the uncertainty resulting from the standard deviation of the five replicate simulations.
    The background's faded blue, orange, and gray represent the interfacial, intermediate, and bulk-like water layers, respectively.
    The horizontal axis is displayed up to half the distance between the two layers.
    }
    \label{fig:fig2}
\end{figure*}

In Fig. \ref{fig:fig1}(a), the water density profiles for the 1L--5L systems reveal distinct layers of water near graphene sheets, consistent with observations of water at solid surfaces \cite{maccarini_density_2007, galli_canonical_2008, tocci_friction, singla_insight_2017, sergi_nanoconf}.
Water forms sharply defined `interfacial' layers in direct contact with the graphene sheets across all systems. 
For the thicker slits, we observe smoother `intermediate' layers, and in the 5L system, these are accompanied by `bulk-like' behavior as the bulk density of water is approached (\response{see \tochange{Section S2}}).
To investigate the hydronium ion, we analyzed the density profiles from the specific oxygen of the protonic defect. %
As depicted in Fig. \ref{fig:fig1}(b), the hydronium ion predominantly resides in the first contact layers of water, the interfacial water layers.
Although the hydronium ion is still observed in the water layers further away from the interface, it appears less frequently there.
For the hydroxide ion, shown in Fig. \ref{fig:fig1}(c), the situation is less clear-cut: it can be present either at the interface or in the interior of the film, generally preferring the layers farther from the interface.
These patterns are consistent across all studied slit widths featuring an intermediate water region (i.e., 3L--5L). 
Furthermore, the flexibility of graphene and its impact on these observations have been examined, confirming consistency even with completely rigid graphene sheets as shown in \tochange{Section S3}.
Using the simulated density profiles, we quantified the surface affinity of hydronium and hydroxide ions by examining their free energy profiles.
To accurately define this affinity, systems must include both interfacial and non-interfacial layers; otherwise, all layers would be considered interfacial.
Therefore, we analyzed systems with an intermediate water region (i.e., 3L--5L).
Our analysis focuses on the distance between the specific oxygen atom of the protonic defect (O$^{*}$) and the closest graphene sheet.
As shown in the free energy profiles in Fig. \ref{fig:fig2}, the hydronium ion is stabilized at the graphene-water interface compared to the bulk, with an energy that is substantially higher than the thermal energy, $k_{\mathrm{B}}T \approx 0.6$\;kcal/mol ($T=300$\;K).
In contrast, the hydroxide ion exhibits a more nuanced stabilization behavior at the interface, generally showing a slight preference for layers farther from the interface. 
The profiles reported herein are validated using umbrella sampling to ensure effective sampling of the phase space and are explicitly compared to previous literature results \cite{mlb_oh} (see \tochange{Section S4}).
The results reported herein highlight the strong preference of the hydronium ion for the interface, in contrast to the hydroxide ion's bimodal distribution, which is found both near the surface and toward the interior layers.

\subsection*{Difference in hydrogen bonding and orientational behavior explain ion stability}
\begin{figure}
    \centering
    \includegraphics[width=0.48\textwidth]{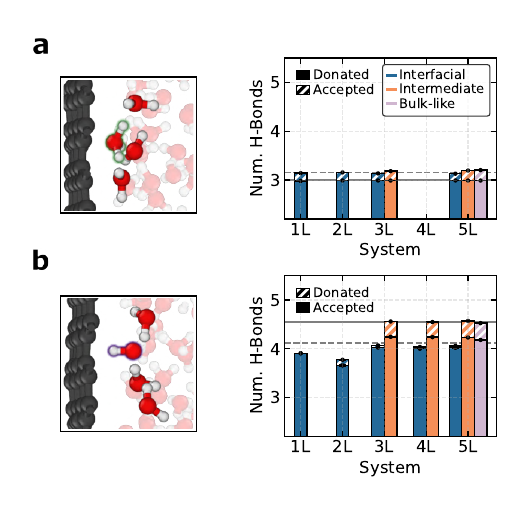}
    \caption{\textbf{Difference in hydrogen bonding behavior of the protonic defects at the graphene-water interface.}
    Average number of hydrogen bonds (accepted or donated, as indicated in each legend) for the hydronium ion (a) and the hydroxide ion (b) across the different water layers \response{with their corresponding bulk reference indicated with horizontal lines}. The representative snapshots show their orientational preferences at the graphene-water interface.
    \response{Hydrogen bonds are counted using the geometric definition provided in Ref.~\citenum{luzar_chandler}}.
    The error bars, smaller than the markers, are obtained from the standard deviation of the five replicate simulations.
    }
    \label{fig:fig3}
\end{figure}

To elucidate the molecular mechanism behind the observed ion behaviors, we investigate the structural characteristics of both defects, focusing on their hydrogen bonding patterns and orientational preferences at the interface.
These factors are crucial for understanding their affinity for specific interfacial locations.
In bulk water, the hydronium ion consistently donates three hydrogen bonds to neighboring water molecules while acting very rarely as an acceptor.
Conversely, the hydroxide ion typically adopts a hypercoordinated square-planar arrangement in bulk water, accepting mostly four hydrogen bonds and transiently donating one. 
This arrangement, common in bulk environments, keeps the local water structure stable and enhances the ion’s stability \cite{tucker_oh_canonical_2002}.
In contrast to these well-established bulk solvation patterns, our analysis at the interface reveals significant differences. 
As shown in Fig. \ref{fig:fig3}(a), the hydronium ion maintains a stable hydrogen bond environment across all layers, consistently donating three hydrogen bonds regardless of its location.
Near the interface, the hydronium ion positions its hydrogen atoms toward the water layers, lying flat as shown in the snapshot.
This orientation maintains the hydrogen bond network of the water molecules in the interior layers (see \tochange{Section S5}) and is influenced by the hydrophobic nature of the hydronium ion’s oxygen, which typically does not accept hydrogen bonds due to its limited availability of lone-pair electrons.
In contrast, as shown in Fig. \ref{fig:fig3}(b), the hydroxide ion at the interface experiences significant hydrogen bond disruption, resulting in it accepting fewer hydrogen bonds than in bulk water and donating almost none due to spatial constraints near the interface.
This disruption changes its typical orientation, with the hydroxide ion's hydrogen atom predominantly facing the interface, impacting its usual fourfold hypercoordinated solvation pattern and significantly reducing its stability.
Overall, this analysis shows that hydronium's hydrogen bonding is not disrupted at the interface whereas the hydroxides' is. 
\subsection*{Hydronium is enthalpically driven to the interface, hydroxide is entropically driven}
\begin{figure}
    \centering
    \includegraphics[width=0.48\textwidth]{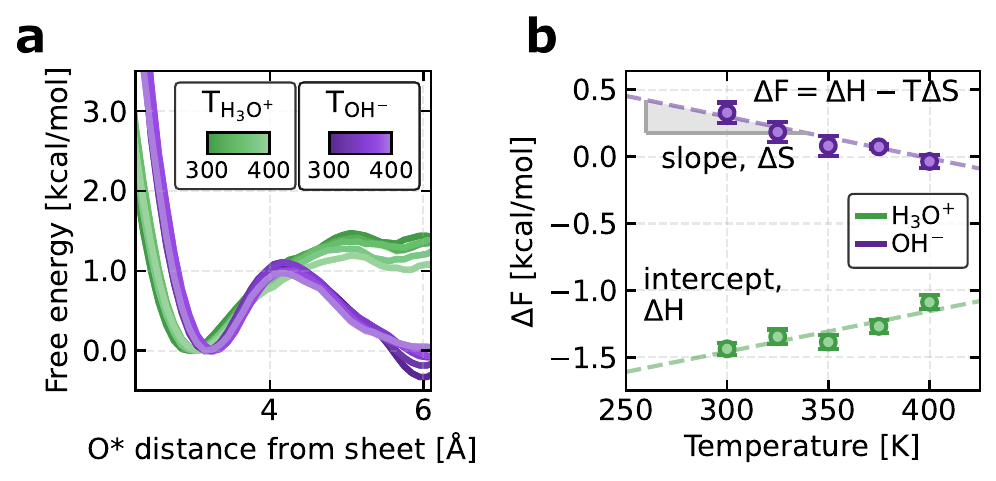}
     \caption{\textbf{Temperature dependence of the free energy for the protonic defects.}
     (a) Free energy profiles for the hydronium and hydroxide ions as a function of their oxygen O$^{*}$ distance to the closest graphene sheet at 300--400\;K for the 3L system.
     The minima in the interfacial layers serve as the reference point for these profiles.
     The shaded regions represent the resulting uncertainty obtained from the standard deviation of the five replicate simulations for each temperature and protonic defect, each propagated for 2.5\;ns.
     (b) Adsorption free energies of the hydronium and hydroxide ions to the interface as a function of temperature, along with their respective linear fits.
     For the hydronium ion, we obtain $\Delta H = -2.4 \pm 0.2$ kcal/mol and $\Delta S = -3.0 \pm 0.6$ cal/mol/K.
     For the hydroxide ion, we obtain $\Delta H = 1.2 \pm 0.3$ kcal/mol and $\Delta S = 3.1 \pm 0.8$ cal/mol/K.
     The error bars, smaller than the markers, are obtained from the standard deviation of the five replicate simulations.
     }
    \label{fig:fig4}
\end{figure}

We now turn our focus to the thermodynamic forces that critically influence ion stability at the interface.
To investigate these forces, we observe how these ions behave within the 3L system featuring an intermediate region, across temperatures from 300 to 400 K, as shown in Fig. \ref{fig:fig4}(a).
First, we compute the adsorption free energy ($\Delta F$) as the difference between the free energy at the interfacial layer and the intermediate layer of the 3L system, measured at the midpoint of the simulation box.
This approach allows us to capture the thermodynamic propensity of ions to either stabilize at the interface or migrate toward more central water layers.
As temperature increases, we see small but significant changes in the adsorption free energies of both protonic defects, as shown in Fig. \ref{fig:fig4}(b).
To explain these differences, we decompose the free energy into enthalpic ($\Delta H$) and entropic ($\Delta S$) contributions.
This is done by performing a linear fit of $\Delta F = \Delta H - T\Delta S$, assuming both changes are temperature-independent within this range and ignoring pressure-volume work contributions under ambient conditions.
This decomposition allows us to understand the driving forces behind ion stability at the interface.
For the hydronium ion, we observe an increase in $\Delta F$ with temperature, thus leading to less stabilization at the interface with temperature.
This highlights that entropy destabilizes the hydronium ion at the interface, while its stability comes from direct interactions with graphene and minimal disruption to the hydrogen bond network of surrounding water molecules, allowing solvent rearrangements that enhance these hydrogen bonds.
This primarily enthalpic interaction, inferred from the intercept of $\Delta F$ at the lower temperature in Fig. \ref{fig:fig4}(b), indicates that enthalpy is the dominant contribution to the proton's surface preference.
Conversely, the behavior of the hydroxide ion at the interface is largely influenced by entropic contributions, which increase its preference for the interface as the temperature rises.
Initially, at 300 K, strong water-ion interactions retain the hydroxide ion predominantly in the bulk due to its well-defined hydration shell.
However, as temperature increases, the system gains entropic stabilization from exploring the additional states of hydroxide physisorbed at the graphene interface.
This entropic drive, reflected in the decreasing slope of $\Delta F$ with temperature in Fig. \ref{fig:fig4}(b), facilitates the exchange of solvent molecules and the hydroxide ions between layers.
These observations demonstrate that hydronium has a preference for the interface due to enthalpic forces, while hydroxide is driven by entropic forces. 
They also reveal the complex balance between these forces at the interface and show how temperature influences ion stability.
\subsection*{Graphene influences the ionic interactions at the interface}
\begin{figure}
    \centering
    \includegraphics[width=0.48\textwidth]{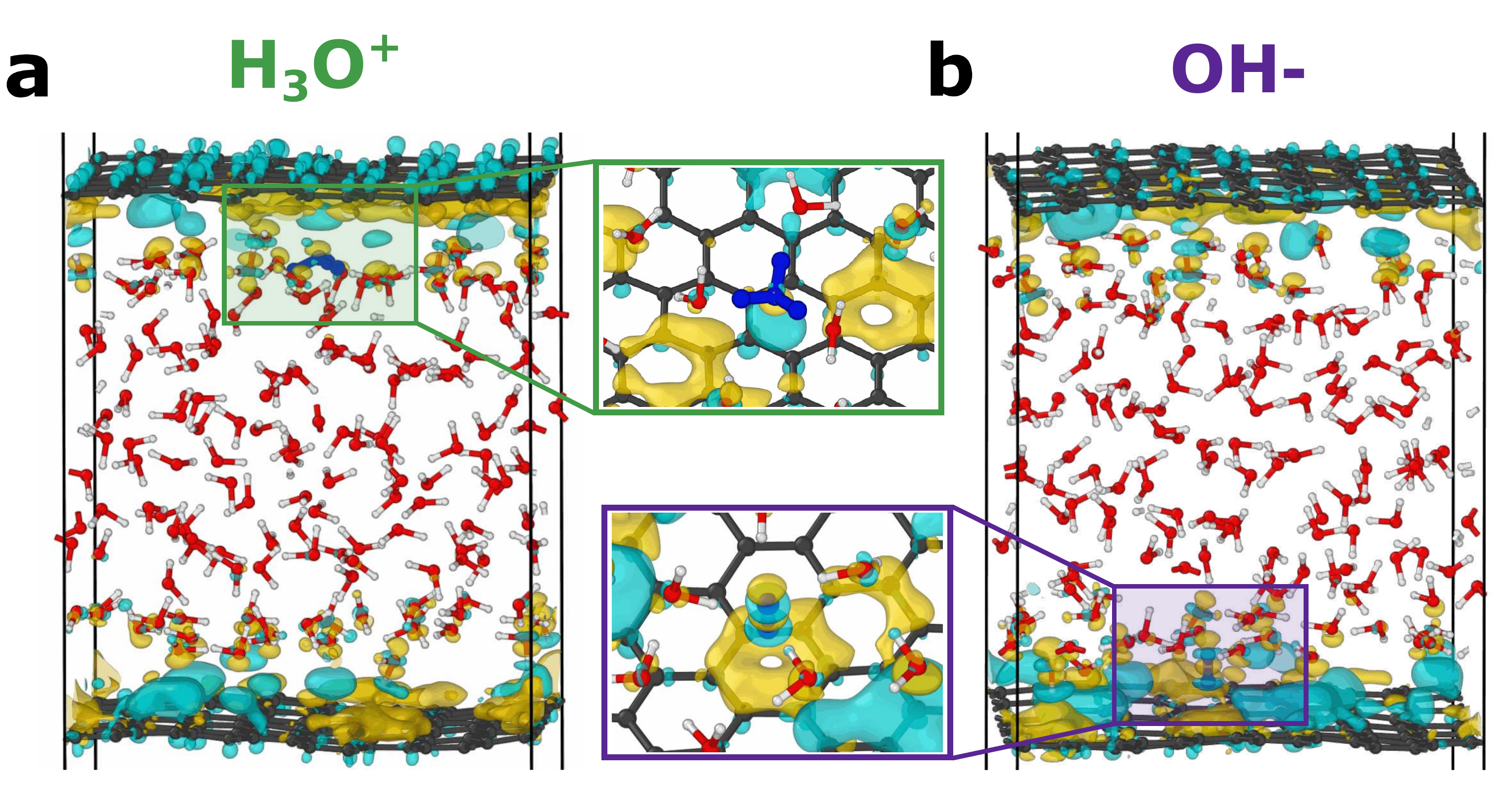}
    \caption{\textbf{The role of graphene and its interaction with the liquid environment.}
    Isosurfaces representing constant electron density differences for representative snapshots in the 5L acidic (a) and basic (b) systems.
    In panel (a), the hydronium ion is colored dark blue, and in panel (b), the hydroxide ion is similarly colored dark blue.
    The solid black lines indicate the edges of the periodic simulation box.
    Blue isosurfaces indicate regions of electron depletion, whereas yellow isosurfaces indicate regions of electron accumulation.
    The units are $0.75\times10^{-5}$ e/\AA$^{3}$. 
    The insets provide a zoomed-in view of the distinct environment experienced by the protonic defects at the interface.
    }
    \label{fig:fig_ed}
\end{figure}

\response{The hydronium and hydroxide surface propensities reported herein bear similarities with those previously observed at the air-water interface\cite{das_ez_2020, chiang_2020}, where the interface is typically described as being enriched with hydronium ions and depleted of hydroxide ions relative to the bulk.}
In our study, similar to what is observed at the air-water interface, hydronium ions are typically closer to the graphene layers than hydroxide ions, indicated by their shorter oxygen distances O$^{*}$ from the sheet (recall Fig. \ref{fig:fig2}). 
However, hydroxide ions at the graphene-water interface have a distinct local (but not global) free energy minimum at the interface, which is absent at the air-water interface.
As discussed above, this difference in affinity arises from direct interactions between the liquid environment (including the protonic defects) and the graphene layers.
Additionally, unlike at the air-water interface where ions can disrupt entropic surface stabilization through capillary wave actions, this effect is diminished or absent at the graphene-water interface, contributing to the differences between graphene-water compared to air-water \cite{otten_2012}.
To assess the significance of the interactions between the liquid environment and the graphene layers, we used DFT to analyze the electron density of the interfaces (see definition in \tochange{Methods}).
In particular, we looked at electron density differences, aimed at capturing the rearrangement of electron density due to the interaction between graphene and the liquid environment.
Key results of this analysis are shown in Fig. \ref{fig:fig_ed} where it can be seen that graphene significantly alters the electron density of water molecules at the interface. 
This charge rearrangement ---which is limited to the contact layers--- therefore creates a distinct hydrogen bonding environment for protonic defects at the graphene-water interface compared to what they experience at the air-water interface.
In addition, \response{interesting local} charge reorganization around the interfacial protonic defects is observed.
As shown in the insets of Fig. \ref{fig:fig_ed}, the oxygen of the hydronium ion \response{exhibits a localized decrease in its negative charge upon interacting with graphene, while the nearby carbon atoms near O$^{*}$ show a slight positive polarization. This localized effect is further quantified through Bader charge analysis (see \tochange{Section S6}).}
\response{When the hydroxide ion points with its hydrogen toward the interface, it can induce electron accumulation above the entire C6 ring. However, Bader charge analysis reveals that this is only a local effect rather than a global charge reorganization (see \tochange{Section S6}).} %
This polarization is more subtle than that caused by the hydronium ion because the dangling hydrogen has more freedom in its orientation of the hydroxide ion leading to variable polarization effects.
\response{Notably, while water molecules may occasionally orient their OH bond toward the graphene and induce similar polarization effects, this alignment is transient. In contrast, hydroxide ions consistently orient their OH group toward the graphene, resulting in a stronger and more persistent impact due to their stable interaction with the surface (see \tochange{Section S6}).}

These findings demonstrate a clear outcome: the response of graphene to ions at the interface is counterintuitive. 
The cation induces charge depletion in nearby carbon atoms, while the anion induces charge accumulation.
To understand this behavior is it crucial to consider their asymmetrical charge arrangements and preferred orientations at the interface.
This intriguing finding emphasizes the need to carefully consider these intricate interactions when interpreting surface experiments at the graphene-water interface, such as zeta potential measurements.

\section*{Conclusion}
Our findings demonstrate a clear preference for the hydronium ion at the graphene-water interface under various confinement regimes, rendering the graphene-water interface
\response{enriched with hydronium ions}.
Enthalpic forces drive the hydronium ion to the interface, maintaining the hydrogen bonding network and enabling energetically favorable interactions with graphene.
Conversely, a subtle balance of entropic and enthalpic contributions generally pushes the hydroxide ion toward the bulk at ambient conditions, despite a clear local free energy minimum in the first contact layer, supporting its preference for optimal water interactions.
\response{In summary, our work emphasizes the complex interplay of hydrogen bonding, orientational preferences, and thermodynamic forces that dictate the stability of hydronium and hydroxide ions at the graphene-water interface.
Coupled with surface polarization effects, leading to counterintuitive charge rearrangements within graphene, these findings provide a comprehensive picture of the two species under confinement and at the graphene-water interface.
}
Importantly, these findings were made possible by leveraging machine learning-based MD simulations of systems encompassing up to 700 atoms and spanning over 200\;ns while retaining first-principles accuracy, far beyond the limits set by AIMD.
This approach is crucial to our study as it allows us to achieve DFT-level accuracy and extensively sample the phase space, providing insights into the microscopic mechanisms driving these interactions.

The mechanistic insights provided in this work are expected to significantly influence and guide future experimental studies. 
Previous work \cite{secchi_2016_oh_prl} has rationalized the macroscopic experimental behavior of nanoconfined electrolytes by postulating the presence of a negative surface charge due to the adsorption of hydroxide ions at the carbon interface. %
However, our findings not only demonstrate a lack of preferential hydroxide adsorption but also a complex relationship between ion adsorption and surface charge, wherein an adsorbed hydroxide can negatively polarize the surface.
The phenomena driving this behavior ---hydrogen bonding and ion orientational effects--- cannot be captured by continuum theories like Poisson-Boltzmann which are conventionally applied to model these systems.
Additionally, the ambiguity in defining interfacial depth can lead to conflicting outcomes, \response{complicating accurate assessments of whether hydronium or hydroxide ions prefer the interface} \cite{sayka_2013_two}.
This underscores the necessity of integrating our simulation findings with experiments that offer an atomistic resolution of interfacial signatures \cite{PETERSEN2008255_2008}.
Surface-sensitive spectroscopy techniques, such as second harmonic generation and sum frequency generation spectroscopies\cite{saykally_hydron_2005, das_ez_2020, litman_2024_nature, mischa_transp}, provide such resolution as they are active only in noncentrosymmetric regions.
Complementary pH measurements within porous carbon could further enhance our understanding of these phenomena. 
When coupled with nuclear magnetic resonance spectroscopy, these techniques could offer new insights into the behavior of these ions at the interface \cite{dongxun_jacs_2024}.
Finally, direct atomic resolution imaging via atomic force microscopy could help elucidate differences in water structure \cite{science_jiang_2024}, providing insights into how interactions affect conductivity in graphene.
This study provides a detailed picture of protonic defects at the graphene-water interface as a foundation for interpreting experimental data and advancing our fundamental understanding of ions at interfaces.
Our prediction of high proton propensity to the graphene-water interface opens up the possibility for technological innovations in nanofluidics, heterogeneous solid-liquid catalysis, and other critical domains that rely on proton-mediated processes.
Finally, \response{given the model system character of our setup, we anticipate that this proton-enriched interface behavior may be} transferable to other systems like water in biological channels, geological formations, and technological nano-devices.

\section*{Methods}
\label{sec:methods}
\textbf{Machine Learning Potentials.} 
In this work, we use the MACE architecture \cite{mace_canonical}, which allows for fast and highly data-efficient training with high-order equivariant message passing and has been proven robust in a wide variety of scenarios \cite{mace_mp_0}.
We developed and validated a MACE MLP model (see \tochange{Section S2}) with two layers, a 6\;\AA\;cutoff distance, 128 equivariant messages, and a maximal message equivariance of $L = 1$.
\response{The MLP captures semi-local interactions through a receptive field that spans the product of the number of layers and the cutoff distance per layer. In this case, the total receptive field is 12 \;\AA, allowing the MLP to account for interactions within this range.
While the model does not explicitly account for long-range effects, the 12\;\AA\;receptive field spans nearly the entire width of the slit in most cases, effectively capturing the relevant electrostatic interactions within the simulation.}
The final energy and force validation root-mean-square error were 0.7\;meV/atom and 17.2\;meV/\AA, respectively.

To accurately represent the potential energy surface of the systems, we train our MLP model using energies and atomic forces obtained from DFT calculations using the CP2K/Quickstep code \cite{cp2k_2020}. 
We specifically used the revPBE-D3 \cite{revpbed3_1, revpbed3_2} functional due to its robust performance in reproducing the structure and dynamics of liquid water  \cite{angelos_dft_water_2016, tobias_vdw_2016, ondrej_revpbe_2017}, while also effectively capturing protonic defects \cite{marsalek_mlp_pt} and the interaction energies between water and graphene \cite{Brandenburg2019}.
Atomic cores are represented using dual-space GTH pseudopotentials \cite{gth_1996}.
The Kohn-Sham orbitals of oxygen and hydrogen atoms are expanded using the TZV2P basis set, while those of carbon atoms are expanded using the DZVP basis set.
An auxiliary plane-wave basis with a cutoff of 1200 Ry was used to represent the density.
\response{We have examined the influence of the functional and dispersion correction on our results and found that the observed trends remain consistent, irrespective of these choices (see \tochange{Section S2}).}
\response{
For interfacial systems, we used a vacuum of 15\,\AA{} to uncouple periodic images in the $z$ direction, leading to negligible interactions between the images as confirmed by the convergence of the energy with respect to the vacuum size.
}
See \tochange{Section S1} for further details\cite{basis_set_2007}.

The MLP model was systematically developed over five generations.
The first generation involved a training set obtained from previous work \cite{quasionedimensional}, enhanced by an active learning procedure \cite{cnnp_2020, complex_2021} to incorporate structures that explicitly account for water-carbon interactions at slit widths of 5 and 6.5\;\AA.
This included conditions ranging from low- to high-density water at various temperatures, including 100, 300, and 600\;K.
The second generation incorporated structures obtained from path integral MD simulations to capture the quantum fluctuations of the nuclei in our model.
For the third generation, we targeted various slit widths, including 6, 10, 15, and 20\;\AA, and included structures corresponding to these dimensions.
In the fourth generation, we conducted an additional round of active learning to refine the model based on the conditions sampled thus far.
This led to the fifth generation, which included configurations of neutral frames containing a protonic defect pair (both a hydronium and a hydroxide ion) under both bulk water conditions and confined conditions at slit widths of 6, 10, 13\;\AA.
To avoid the complications of applying a background charge for charge neutrality, we did not include isolated protonic defects as suggested in Ref.~\citenum{marsalek_mlp_pt}.  
The final model consisted of 3,378 structures, with 1,303 involving graphene interfaces and 2,075 associated with bulk conditions.

\textbf{Molecular Dynamics Simulations.} All MD simulations reported herein, which were based on the MLP, were performed using the ASE software \cite{ase_2017} at a temperature of 300 K, unless explicitly stated otherwise, in the NVT ensemble.
A time step of 0.5 fs was employed, and simulations utilized a Langevin thermostat with a friction coefficient of 2.5\;ps$^{-1}$.
For each of the five slit widths and two species, we conducted five independent simulations.
Each simulation included a 90\;ps equilibration period followed by a 4\;ns production run, resulting in a total of over 200\;ns of simulation time.
Uncertainties in reported values were calculated using the standard deviation from these replicates.
\response{
All systems were simulated in orthorhombic simulation cells employing periodic boundary conditions in all three directions.}
The simulation cells were initially set up by randomly packing several molecules between the graphene sheets to form one to five well-defined layers of water.
\response{To prevent interactions between the periodic images, 15\;\AA\; vacuum (exceeding the model's receptive field) was added in the $z$ direction of these initial configurations.}
To achieve equilibrium density, the graphene sheets were treated in a fully flexible manner, allowing them to adapt without additional constraints.
To validate the findings reported, we conducted additional MLP-based biased simulations using umbrella sampling to compare the free energy profiles reported. See \tochange{Section S1 and S4} for further details.

\textbf{Electronic Structure Analyses.} The electronic properties of the protonic defects at the graphene-water interface were analyzed using the same electronic structure settings used to train our MLP.
However, to reduce the computational cost, a cutoff of 1050 Ry was used. 
To assess the interactions between the liquid environments and the graphene layers, we used DFT to analyze their electron density difference ($\Delta \rho$), defined as $\Delta\rho = \rho_{\text{liq}/\text{gra}} - \rho_{\text{gra}} - \rho_{\text{liq}}$, where $\rho_{\text{liq}/\text{gra}}$, $\rho_{\text{gra}}$, and  $\rho_{\text{liq}}$ are the electron densities of the system under consideration, the isolated graphene surfaces, and the isolated liquid environment, respectively.
Where appropriate, the system's net charge was neutralized using a uniformly charged background.
To determine the response of graphene to the protonic defects at the interface, we performed Bader charge analysis using the same settings reported herein with the reduced cutoff (see \tochange{Section S6}).

\begin{acknowledgements}
    We thank Marie-Laure Bocquet for fruitful discussions and Benjamin X. Shi for technical help with the hybrid functional setup.
    X.R.A. and A.M. acknowledge support from the European Union under the “n-AQUA” European Research Council project (Grant No. 101071937). 
    K.D.F. acknowledges support from Schmidt Science Fellows and Trinity College, Cambridge. 
    C.S acknowledges partial financial support from the Alexander von Humboldt Stiftung and the Deutsche Forschungsgemeinschaft (DFG, German Research Foundation) project number 500244608. 
    Via our membership of the UK's HEC Materials Chemistry Consortium, which is funded by EPSRC (EP/F067496), this work used the ARCHER2 UK National Supercomputing Service (\url{http://www.archer2.ac.uk}).
    This work was also performed using resources provided by the Cambridge Service for Data Driven Discovery (CSD3) operated by the University of Cambridge Research Computing Service (\url{www.csd3.cam.ac.uk}), provided by Dell EMC and Intel using Tier-2 funding from the Engineering and Physical Sciences Research Council (capital grant EP/T022159/1), and DiRAC funding from the Science and Technology Facilities Council (\url{www.dirac.ac.uk}).
    Access to CSD3 was also obtained through a University of Cambridge EPSRC Core Equipment Award (EP/X034712/1).
    We additionally acknowledge computational support and resources from the Cirrus UK National Tier-2 HPC Service at EPCC (\url{http://www.cirrus.ac.uk}) funded by the University of Edinburgh and EPSRC (EP/P020267/1). 
\end{acknowledgements}

\section*{Data Availability}
All data required to reproduce the findings of this work are openly available on GitHub (\url{https://github.com/water-ice-group/graphene-water-protons}).
\section*{Code Availability}
All analysis code supporting the findings of this work will be made openly available on GitHub upon acceptance of this manuscript.
All simulation code used in this study is available open-source, see \href{https://www.cp2k.org/}{CP2K}, \href{https://wiki.fysik.dtu.dk/ase/}{ASE} and \href{https://mace-docs.readthedocs.io/en/latest/index.html}{MACE} for details.

\section*{References}

\begin{thebibliography}{79}%
\makeatletter
\providecommand \@ifxundefined [1]{%
 \@ifx{#1\undefined}
}%
\providecommand \@ifnum [1]{%
 \ifnum #1\expandafter \@firstoftwo
 \else \expandafter \@secondoftwo
 \fi
}%
\providecommand \@ifx [1]{%
 \ifx #1\expandafter \@firstoftwo
 \else \expandafter \@secondoftwo
 \fi
}%
\providecommand \natexlab [1]{#1}%
\providecommand \enquote  [1]{``#1''}%
\providecommand \bibnamefont  [1]{#1}%
\providecommand \bibfnamefont [1]{#1}%
\providecommand \citenamefont [1]{#1}%
\providecommand \href@noop [0]{\@secondoftwo}%
\providecommand \href [0]{\begingroup \@sanitize@url \@href}%
\providecommand \@href[1]{\@@startlink{#1}\@@href}%
\providecommand \@@href[1]{\endgroup#1\@@endlink}%
\providecommand \@sanitize@url [0]{\catcode `\\12\catcode `\$12\catcode
  `\&12\catcode `\#12\catcode `\^12\catcode `\_12\catcode `\%12\relax}%
\providecommand \@@startlink[1]{}%
\providecommand \@@endlink[0]{}%
\providecommand \url  [0]{\begingroup\@sanitize@url \@url }%
\providecommand \@url [1]{\endgroup\@href {#1}{\urlprefix }}%
\providecommand \urlprefix  [0]{URL }%
\providecommand \Eprint [0]{\href }%
\providecommand \doibase [0]{https://doi.org/}%
\providecommand \selectlanguage [0]{\@gobble}%
\providecommand \bibinfo  [0]{\@secondoftwo}%
\providecommand \bibfield  [0]{\@secondoftwo}%
\providecommand \translation [1]{[#1]}%
\providecommand \BibitemOpen [0]{}%
\providecommand \bibitemStop [0]{}%
\providecommand \bibitemNoStop [0]{.\EOS\space}%
\providecommand \EOS [0]{\spacefactor3000\relax}%
\providecommand \BibitemShut  [1]{\csname bibitem#1\endcsname}%
\let\auto@bib@innerbib\@empty
\bibitem [{\citenamefont {Knipping}\ \emph {et~al.}(2000)\citenamefont
  {Knipping}, \citenamefont {Lakin}, \citenamefont {Foster}, \citenamefont
  {Jungwirth}, \citenamefont {Tobias}, \citenamefont {Gerber}, \citenamefont
  {Dabdub},\ and\ \citenamefont {Finlayson-Pitts}}]{atm_chem_2000}%
  \BibitemOpen
  \bibfield  {author} {\bibinfo {author} {\bibfnamefont {E.~M.}\ \bibnamefont
  {Knipping}}, \bibinfo {author} {\bibfnamefont {M.~J.}\ \bibnamefont {Lakin}},
  \bibinfo {author} {\bibfnamefont {K.~L.}\ \bibnamefont {Foster}}, \bibinfo
  {author} {\bibfnamefont {P.}~\bibnamefont {Jungwirth}}, \bibinfo {author}
  {\bibfnamefont {D.~J.}\ \bibnamefont {Tobias}}, \bibinfo {author}
  {\bibfnamefont {R.~B.}\ \bibnamefont {Gerber}}, \bibinfo {author}
  {\bibfnamefont {D.}~\bibnamefont {Dabdub}},\ and\ \bibinfo {author}
  {\bibfnamefont {B.~J.}\ \bibnamefont {Finlayson-Pitts}},\ }\bibfield  {title}
  {\enquote {\bibinfo {title} {Experiments and simulations of ion-enhanced
  interfacial chemistry on aqueous nacl aerosols},}\ }\href
  {https://doi.org/10.1126/science.288.5464.301} {\bibfield  {journal}
  {\bibinfo  {journal} {Science}\ }\textbf {\bibinfo {volume} {288}},\ \bibinfo
  {pages} {301--306} (\bibinfo {year} {2000})}\BibitemShut {NoStop}%
\bibitem [{\citenamefont {Elimelech}\ and\ \citenamefont
  {Phillip}(2011)}]{desalin_2011}%
  \BibitemOpen
  \bibfield  {author} {\bibinfo {author} {\bibfnamefont {M.}~\bibnamefont
  {Elimelech}}\ and\ \bibinfo {author} {\bibfnamefont {W.~A.}\ \bibnamefont
  {Phillip}},\ }\bibfield  {title} {\enquote {\bibinfo {title} {The future of
  seawater desalination: Energy, technology, and the environment},}\ }\href
  {https://doi.org/10.1126/science.1200488} {\bibfield  {journal} {\bibinfo
  {journal} {Science}\ }\textbf {\bibinfo {volume} {333}},\ \bibinfo {pages}
  {712--717} (\bibinfo {year} {2011})}\BibitemShut {NoStop}%
\bibitem [{\citenamefont {Hisatomi}\ and\ \citenamefont
  {Domen}(2019)}]{wat_split_2019}%
  \BibitemOpen
  \bibfield  {author} {\bibinfo {author} {\bibfnamefont {T.}~\bibnamefont
  {Hisatomi}}\ and\ \bibinfo {author} {\bibfnamefont {K.}~\bibnamefont
  {Domen}},\ }\bibfield  {title} {\enquote {\bibinfo {title} {Reaction systems
  for solar hydrogen production via water splitting with particulate
  semiconductor photocatalysts},}\ }\href
  {https://doi.org/10.1038/s41929-019-0242-6} {\bibfield  {journal} {\bibinfo
  {journal} {Nature Catalysis}\ }\textbf {\bibinfo {volume} {2}},\ \bibinfo
  {pages} {387--399} (\bibinfo {year} {2019})}\BibitemShut {NoStop}%
\bibitem [{\citenamefont {Ruiz-Lopez}\ \emph {et~al.}(2020)\citenamefont
  {Ruiz-Lopez}, \citenamefont {Francisco}, \citenamefont {Martins-Costa},\ and\
  \citenamefont {Anglada}}]{lopez_2020_inter}%
  \BibitemOpen
  \bibfield  {author} {\bibinfo {author} {\bibfnamefont {M.~F.}\ \bibnamefont
  {Ruiz-Lopez}}, \bibinfo {author} {\bibfnamefont {J.~S.}\ \bibnamefont
  {Francisco}}, \bibinfo {author} {\bibfnamefont {M.~T.~C.}\ \bibnamefont
  {Martins-Costa}},\ and\ \bibinfo {author} {\bibfnamefont {J.~M.}\
  \bibnamefont {Anglada}},\ }\bibfield  {title} {\enquote {\bibinfo {title}
  {Molecular reactions at aqueous interfaces},}\ }\href
  {https://doi.org/10.1038/s41570-020-0203-2} {\bibfield  {journal} {\bibinfo
  {journal} {Nature Reviews Chemistry}\ }\textbf {\bibinfo {volume} {4}},\
  \bibinfo {pages} {459--475} (\bibinfo {year} {2020})}\BibitemShut {NoStop}%
\bibitem [{\citenamefont {Geissler}\ \emph {et~al.}(2001)\citenamefont
  {Geissler}, \citenamefont {Dellago}, \citenamefont {Chandler}, \citenamefont
  {Hutter},\ and\ \citenamefont {Parrinello}}]{geissler_2001}%
  \BibitemOpen
  \bibfield  {author} {\bibinfo {author} {\bibfnamefont {P.~L.}\ \bibnamefont
  {Geissler}}, \bibinfo {author} {\bibfnamefont {C.}~\bibnamefont {Dellago}},
  \bibinfo {author} {\bibfnamefont {D.}~\bibnamefont {Chandler}}, \bibinfo
  {author} {\bibfnamefont {J.}~\bibnamefont {Hutter}},\ and\ \bibinfo {author}
  {\bibfnamefont {M.}~\bibnamefont {Parrinello}},\ }\bibfield  {title}
  {\enquote {\bibinfo {title} {Autoionization in liquid water},}\ }\href
  {https://doi.org/10.1126/science.1056991} {\bibfield  {journal} {\bibinfo
  {journal} {Science}\ }\textbf {\bibinfo {volume} {291}},\ \bibinfo {pages}
  {2121--2124} (\bibinfo {year} {2001})}\BibitemShut {NoStop}%
\bibitem [{\citenamefont {Chandra}, \citenamefont {Tuckerman},\ and\
  \citenamefont {Marx}(2007)}]{chandra_2007_solvation}%
  \BibitemOpen
  \bibfield  {author} {\bibinfo {author} {\bibfnamefont {A.}~\bibnamefont
  {Chandra}}, \bibinfo {author} {\bibfnamefont {M.~E.}\ \bibnamefont
  {Tuckerman}},\ and\ \bibinfo {author} {\bibfnamefont {D.}~\bibnamefont
  {Marx}},\ }\bibfield  {title} {\enquote {\bibinfo {title} {Connecting
  solvation shell structure to proton transport kinetics in hydrogen--bonded
  networks via population correlation functions},}\ }\href
  {https://doi.org/10.1103/PhysRevLett.99.145901} {\bibfield  {journal}
  {\bibinfo  {journal} {Physical Review Letters}\ }\textbf {\bibinfo {volume}
  {99}},\ \bibinfo {pages} {145901} (\bibinfo {year} {2007})}\BibitemShut
  {NoStop}%
\bibitem [{\citenamefont {Hassanali}\ \emph {et~al.}(2013)\citenamefont
  {Hassanali}, \citenamefont {Giberti}, \citenamefont {Cuny}, \citenamefont
  {Kühne},\ and\ \citenamefont {Parrinello}}]{hassanali_2013_pnas}%
  \BibitemOpen
  \bibfield  {author} {\bibinfo {author} {\bibfnamefont {A.}~\bibnamefont
  {Hassanali}}, \bibinfo {author} {\bibfnamefont {F.}~\bibnamefont {Giberti}},
  \bibinfo {author} {\bibfnamefont {J.}~\bibnamefont {Cuny}}, \bibinfo {author}
  {\bibfnamefont {T.~D.}\ \bibnamefont {Kühne}},\ and\ \bibinfo {author}
  {\bibfnamefont {M.}~\bibnamefont {Parrinello}},\ }\bibfield  {title}
  {\enquote {\bibinfo {title} {Proton transfer through the water gossamer},}\
  }\href {https://doi.org/10.1073/pnas.1306642110} {\bibfield  {journal}
  {\bibinfo  {journal} {Proceedings of the National Academy of Sciences}\
  }\textbf {\bibinfo {volume} {110}},\ \bibinfo {pages} {13723--13728}
  (\bibinfo {year} {2013})}\BibitemShut {NoStop}%
\bibitem [{\citenamefont {Dahms}\ \emph {et~al.}(2017)\citenamefont {Dahms},
  \citenamefont {Fingerhut}, \citenamefont {Nibbering}, \citenamefont {Pines},\
  and\ \citenamefont {Elsaesser}}]{dahms_2017}%
  \BibitemOpen
  \bibfield  {author} {\bibinfo {author} {\bibfnamefont {F.}~\bibnamefont
  {Dahms}}, \bibinfo {author} {\bibfnamefont {B.~P.}\ \bibnamefont
  {Fingerhut}}, \bibinfo {author} {\bibfnamefont {E.~T.~J.}\ \bibnamefont
  {Nibbering}}, \bibinfo {author} {\bibfnamefont {E.}~\bibnamefont {Pines}},\
  and\ \bibinfo {author} {\bibfnamefont {T.}~\bibnamefont {Elsaesser}},\
  }\bibfield  {title} {\enquote {\bibinfo {title} {Large-amplitude transfer
  motion of hydrated excess protons mapped by ultrafast 2d ir spectroscopy},}\
  }\href {https://doi.org/10.1126/science.aan5144} {\bibfield  {journal}
  {\bibinfo  {journal} {Science}\ }\textbf {\bibinfo {volume} {357}},\ \bibinfo
  {pages} {491--495} (\bibinfo {year} {2017})}\BibitemShut {NoStop}%
\bibitem [{\citenamefont {Calio}, \citenamefont {Li},\ and\ \citenamefont
  {Voth}(2021)}]{calio_jacs_2021}%
  \BibitemOpen
  \bibfield  {author} {\bibinfo {author} {\bibfnamefont {P.~B.}\ \bibnamefont
  {Calio}}, \bibinfo {author} {\bibfnamefont {C.}~\bibnamefont {Li}},\ and\
  \bibinfo {author} {\bibfnamefont {G.~A.}\ \bibnamefont {Voth}},\ }\bibfield
  {title} {\enquote {\bibinfo {title} {Resolving the structural debate for the
  hydrated excess proton in water},}\ }\href
  {https://doi.org/10.1021/jacs.1c08552} {\bibfield  {journal} {\bibinfo
  {journal} {Journal of the American Chemical Society}\ }\textbf {\bibinfo
  {volume} {143}},\ \bibinfo {pages} {18672--18683} (\bibinfo {year}
  {2021})}\BibitemShut {NoStop}%
\bibitem [{\citenamefont {Brünig}\ \emph {et~al.}(2022)\citenamefont
  {Brünig}, \citenamefont {Rammler}, \citenamefont {Adams}, \citenamefont
  {Havenith},\ and\ \citenamefont {Netz}}]{netz_spectral_2022}%
  \BibitemOpen
  \bibfield  {author} {\bibinfo {author} {\bibfnamefont {F.~N.}\ \bibnamefont
  {Brünig}}, \bibinfo {author} {\bibfnamefont {M.}~\bibnamefont {Rammler}},
  \bibinfo {author} {\bibfnamefont {E.~M.}\ \bibnamefont {Adams}}, \bibinfo
  {author} {\bibfnamefont {M.}~\bibnamefont {Havenith}},\ and\ \bibinfo
  {author} {\bibfnamefont {R.~R.}\ \bibnamefont {Netz}},\ }\bibfield  {title}
  {\enquote {\bibinfo {title} {Spectral signatures of excess-proton waiting and
  transfer-path dynamics in aqueous hydrochloric acid solutions},}\ }\href
  {https://doi.org/10.1038/s41467-022-31700-x} {\bibfield  {journal} {\bibinfo
  {journal} {Nature Communications}\ }\textbf {\bibinfo {volume} {13}},\
  \bibinfo {pages} {4210} (\bibinfo {year} {2022})}\BibitemShut {NoStop}%
\bibitem [{\citenamefont {Gomez}, \citenamefont {Thompson},\ and\ \citenamefont
  {Laage}(2024)}]{laage_pt_natchem_2024}%
  \BibitemOpen
  \bibfield  {author} {\bibinfo {author} {\bibfnamefont {A.}~\bibnamefont
  {Gomez}}, \bibinfo {author} {\bibfnamefont {W.~H.}\ \bibnamefont
  {Thompson}},\ and\ \bibinfo {author} {\bibfnamefont {D.}~\bibnamefont
  {Laage}},\ }\bibfield  {title} {\enquote {\bibinfo {title}
  {Neural-network-based molecular dynamics simulations reveal that proton
  transport in water is doubly gated by sequential hydrogen-bond exchange},}\
  }\href {https://www.nature.com/articles/s41557-024-01593-y} {\bibfield
  {journal} {\bibinfo  {journal} {Nature Chemistry}\ } (\bibinfo {year}
  {2024})}\BibitemShut {NoStop}%
\bibitem [{\citenamefont {Buch}\ \emph {et~al.}(2007)\citenamefont {Buch},
  \citenamefont {Milet}, \citenamefont {Vácha}, \citenamefont {Jungwirth},\
  and\ \citenamefont {Devlin}}]{buch_wat_acidic}%
  \BibitemOpen
  \bibfield  {author} {\bibinfo {author} {\bibfnamefont {V.}~\bibnamefont
  {Buch}}, \bibinfo {author} {\bibfnamefont {A.}~\bibnamefont {Milet}},
  \bibinfo {author} {\bibfnamefont {R.}~\bibnamefont {Vácha}}, \bibinfo
  {author} {\bibfnamefont {P.}~\bibnamefont {Jungwirth}},\ and\ \bibinfo
  {author} {\bibfnamefont {J.~P.}\ \bibnamefont {Devlin}},\ }\bibfield  {title}
  {\enquote {\bibinfo {title} {Water surface is acidic},}\ }\href
  {https://doi.org/10.1073/pnas.0611285104} {\bibfield  {journal} {\bibinfo
  {journal} {Proceedings of the National Academy of Sciences}\ }\textbf
  {\bibinfo {volume} {104}},\ \bibinfo {pages} {7342--7347} (\bibinfo {year}
  {2007})}\BibitemShut {NoStop}%
\bibitem [{\citenamefont {Mishra}\ \emph {et~al.}(2012)\citenamefont {Mishra},
  \citenamefont {Enami}, \citenamefont {Nielsen}, \citenamefont {Stewart},
  \citenamefont {Hoffmann}, \citenamefont {Goddard},\ and\ \citenamefont
  {Colussi}}]{mishra_2012}%
  \BibitemOpen
  \bibfield  {author} {\bibinfo {author} {\bibfnamefont {H.}~\bibnamefont
  {Mishra}}, \bibinfo {author} {\bibfnamefont {S.}~\bibnamefont {Enami}},
  \bibinfo {author} {\bibfnamefont {R.~J.}\ \bibnamefont {Nielsen}}, \bibinfo
  {author} {\bibfnamefont {L.~A.}\ \bibnamefont {Stewart}}, \bibinfo {author}
  {\bibfnamefont {M.~R.}\ \bibnamefont {Hoffmann}}, \bibinfo {author}
  {\bibfnamefont {W.~A.}\ \bibnamefont {Goddard}},\ and\ \bibinfo {author}
  {\bibfnamefont {A.~J.}\ \bibnamefont {Colussi}},\ }\bibfield  {title}
  {\enquote {\bibinfo {title} {Brønsted basicity of the air–water
  interface},}\ }\href {https://doi.org/10.1073/pnas.1209307109} {\bibfield
  {journal} {\bibinfo  {journal} {Proceedings of the National Academy of
  Sciences}\ }\textbf {\bibinfo {volume} {109}},\ \bibinfo {pages}
  {18679--18683} (\bibinfo {year} {2012})}\BibitemShut {NoStop}%
\bibitem [{\citenamefont {Saykally}(2013)}]{sayka_2013_two}%
  \BibitemOpen
  \bibfield  {author} {\bibinfo {author} {\bibfnamefont {R.~J.}\ \bibnamefont
  {Saykally}},\ }\bibfield  {title} {\enquote {\bibinfo {title} {Two sides of
  the acid–base story},}\ }\href {https://doi.org/10.1038/nchem.1556}
  {\bibfield  {journal} {\bibinfo  {journal} {Nature Chemistry}\ }\textbf
  {\bibinfo {volume} {5}},\ \bibinfo {pages} {82--84} (\bibinfo {year}
  {2013})}\BibitemShut {NoStop}%
\bibitem [{\citenamefont {Tse}\ \emph {et~al.}(2015)\citenamefont {Tse},
  \citenamefont {Chen}, \citenamefont {Lindberg}, \citenamefont {Kumar},\ and\
  \citenamefont {Voth}}]{tse_2015}%
  \BibitemOpen
  \bibfield  {author} {\bibinfo {author} {\bibfnamefont {Y.-L.~S.}\
  \bibnamefont {Tse}}, \bibinfo {author} {\bibfnamefont {C.}~\bibnamefont
  {Chen}}, \bibinfo {author} {\bibfnamefont {G.~E.}\ \bibnamefont {Lindberg}},
  \bibinfo {author} {\bibfnamefont {R.}~\bibnamefont {Kumar}},\ and\ \bibinfo
  {author} {\bibfnamefont {G.~A.}\ \bibnamefont {Voth}},\ }\bibfield  {title}
  {\enquote {\bibinfo {title} {Propensity of hydrated excess protons and
  hydroxide anions for the air–water interface},}\ }\href
  {https://doi.org/10.1021/jacs.5b07232} {\bibfield  {journal} {\bibinfo
  {journal} {Journal of the American Chemical Society}\ }\textbf {\bibinfo
  {volume} {137}},\ \bibinfo {pages} {12610--12616} (\bibinfo {year}
  {2015})}\BibitemShut {NoStop}%
\bibitem [{\citenamefont {Mamatkulov}\ \emph {et~al.}(2017)\citenamefont
  {Mamatkulov}, \citenamefont {Allolio}, \citenamefont {Netz},\ and\
  \citenamefont {Bonthuis}}]{netz_orient_2017}%
  \BibitemOpen
  \bibfield  {author} {\bibinfo {author} {\bibfnamefont {S.~I.}\ \bibnamefont
  {Mamatkulov}}, \bibinfo {author} {\bibfnamefont {C.}~\bibnamefont {Allolio}},
  \bibinfo {author} {\bibfnamefont {R.~R.}\ \bibnamefont {Netz}},\ and\
  \bibinfo {author} {\bibfnamefont {D.~J.}\ \bibnamefont {Bonthuis}},\
  }\bibfield  {title} {\enquote {\bibinfo {title} {Orientation-induced
  adsorption of hydrated protons at the air–water interface},}\ }\href
  {https://doi.org/https://doi.org/10.1002/anie.201707391} {\bibfield
  {journal} {\bibinfo  {journal} {Angewandte Chemie International Edition}\
  }\textbf {\bibinfo {volume} {56}},\ \bibinfo {pages} {15846--15851} (\bibinfo
  {year} {2017})}\BibitemShut {NoStop}%
\bibitem [{\citenamefont {Tuckerman}\ \emph {et~al.}(1995)\citenamefont
  {Tuckerman}, \citenamefont {Laasonen}, \citenamefont {Sprik},\ and\
  \citenamefont {Parrinello}}]{tucker_1995_h3o_oh}%
  \BibitemOpen
  \bibfield  {author} {\bibinfo {author} {\bibfnamefont {M.}~\bibnamefont
  {Tuckerman}}, \bibinfo {author} {\bibfnamefont {K.}~\bibnamefont {Laasonen}},
  \bibinfo {author} {\bibfnamefont {M.}~\bibnamefont {Sprik}},\ and\ \bibinfo
  {author} {\bibfnamefont {M.}~\bibnamefont {Parrinello}},\ }\bibfield  {title}
  {\enquote {\bibinfo {title} {Ab initio molecular dynamics simulation of the
  solvation and transport of hydronium and hydroxyl ions in water},}\ }\href
  {https://doi.org/10.1063/1.469654} {\bibfield  {journal} {\bibinfo  {journal}
  {The Journal of Chemical Physics}\ }\textbf {\bibinfo {volume} {103}},\
  \bibinfo {pages} {150--161} (\bibinfo {year} {1995})}\BibitemShut {NoStop}%
\bibitem [{\citenamefont {Marx}\ \emph {et~al.}(1999)\citenamefont {Marx},
  \citenamefont {Tuckerman}, \citenamefont {Hutter},\ and\ \citenamefont
  {Parrinello}}]{tucker_1999_canonical}%
  \BibitemOpen
  \bibfield  {author} {\bibinfo {author} {\bibfnamefont {D.}~\bibnamefont
  {Marx}}, \bibinfo {author} {\bibfnamefont {M.~E.}\ \bibnamefont {Tuckerman}},
  \bibinfo {author} {\bibfnamefont {J.}~\bibnamefont {Hutter}},\ and\ \bibinfo
  {author} {\bibfnamefont {M.}~\bibnamefont {Parrinello}},\ }\bibfield  {title}
  {\enquote {\bibinfo {title} {The nature of the hydrated excess proton in
  water},}\ }\href {https://doi.org/10.1038/17579} {\bibfield  {journal}
  {\bibinfo  {journal} {Nature}\ }\textbf {\bibinfo {volume} {397}},\ \bibinfo
  {pages} {601--604} (\bibinfo {year} {1999})}\BibitemShut {NoStop}%
\bibitem [{\citenamefont {Tuckerman}, \citenamefont {Marx},\ and\ \citenamefont
  {Parrinello}(2002)}]{tucker_oh_canonical_2002}%
  \BibitemOpen
  \bibfield  {author} {\bibinfo {author} {\bibfnamefont {M.~E.}\ \bibnamefont
  {Tuckerman}}, \bibinfo {author} {\bibfnamefont {D.}~\bibnamefont {Marx}},\
  and\ \bibinfo {author} {\bibfnamefont {M.}~\bibnamefont {Parrinello}},\
  }\bibfield  {title} {\enquote {\bibinfo {title} {The nature and transport
  mechanism of hydrated hydroxide ions in aqueous solution},}\ }\href
  {https://doi.org/10.1038/nature00797} {\bibfield  {journal} {\bibinfo
  {journal} {Nature}\ }\textbf {\bibinfo {volume} {417}},\ \bibinfo {pages}
  {925--929} (\bibinfo {year} {2002})}\BibitemShut {NoStop}%
\bibitem [{\citenamefont {Das}, \citenamefont {Bonn},\ and\ \citenamefont
  {Backus}(2019)}]{das_2019}%
  \BibitemOpen
  \bibfield  {author} {\bibinfo {author} {\bibfnamefont {S.}~\bibnamefont
  {Das}}, \bibinfo {author} {\bibfnamefont {M.}~\bibnamefont {Bonn}},\ and\
  \bibinfo {author} {\bibfnamefont {E.~H.~G.}\ \bibnamefont {Backus}},\
  }\bibfield  {title} {\enquote {\bibinfo {title} {The surface activity of the
  hydrated proton is substantially higher than that of the hydroxide ion},}\
  }\href {https://doi.org/https://doi.org/10.1002/anie.201908420} {\bibfield
  {journal} {\bibinfo  {journal} {Angewandte Chemie International Edition}\
  }\textbf {\bibinfo {volume} {58}},\ \bibinfo {pages} {15636--15639} (\bibinfo
  {year} {2019})}\BibitemShut {NoStop}%
\bibitem [{\citenamefont {Das}\ \emph {et~al.}(2020)\citenamefont {Das},
  \citenamefont {Imoto}, \citenamefont {Sun}, \citenamefont {Nagata},
  \citenamefont {Backus},\ and\ \citenamefont {Bonn}}]{das_ez_2020}%
  \BibitemOpen
  \bibfield  {author} {\bibinfo {author} {\bibfnamefont {S.}~\bibnamefont
  {Das}}, \bibinfo {author} {\bibfnamefont {S.}~\bibnamefont {Imoto}}, \bibinfo
  {author} {\bibfnamefont {S.}~\bibnamefont {Sun}}, \bibinfo {author}
  {\bibfnamefont {Y.}~\bibnamefont {Nagata}}, \bibinfo {author} {\bibfnamefont
  {E.~H.~G.}\ \bibnamefont {Backus}},\ and\ \bibinfo {author} {\bibfnamefont
  {M.}~\bibnamefont {Bonn}},\ }\bibfield  {title} {\enquote {\bibinfo {title}
  {Nature of excess hydrated proton at the water–air interface},}\ }\href
  {https://doi.org/10.1021/jacs.9b10807} {\bibfield  {journal} {\bibinfo
  {journal} {Journal of the American Chemical Society}\ }\textbf {\bibinfo
  {volume} {142}},\ \bibinfo {pages} {945--952} (\bibinfo {year}
  {2020})}\BibitemShut {NoStop}%
\bibitem [{\citenamefont {de~la Puente}\ and\ \citenamefont
  {Laage}(2023)}]{miguel_ph_2024}%
  \BibitemOpen
  \bibfield  {author} {\bibinfo {author} {\bibfnamefont {M.}~\bibnamefont
  {de~la Puente}}\ and\ \bibinfo {author} {\bibfnamefont {D.}~\bibnamefont
  {Laage}},\ }\bibfield  {title} {\enquote {\bibinfo {title} {How the acidity
  of water droplets and films is controlled by the air–water interface},}\
  }\href {https://doi.org/10.1021/jacs.3c07506} {\bibfield  {journal} {\bibinfo
   {journal} {Journal of the American Chemical Society}\ }\textbf {\bibinfo
  {volume} {145}},\ \bibinfo {pages} {25186--25194} (\bibinfo {year}
  {2023})}\BibitemShut {NoStop}%
\bibitem [{\citenamefont {Litman}\ \emph {et~al.}(2024)\citenamefont {Litman},
  \citenamefont {Chiang}, \citenamefont {Seki}, \citenamefont {Nagata},\ and\
  \citenamefont {Bonn}}]{litman_2024_nature}%
  \BibitemOpen
  \bibfield  {author} {\bibinfo {author} {\bibfnamefont {Y.}~\bibnamefont
  {Litman}}, \bibinfo {author} {\bibfnamefont {K.-Y.}\ \bibnamefont {Chiang}},
  \bibinfo {author} {\bibfnamefont {T.}~\bibnamefont {Seki}}, \bibinfo {author}
  {\bibfnamefont {Y.}~\bibnamefont {Nagata}},\ and\ \bibinfo {author}
  {\bibfnamefont {M.}~\bibnamefont {Bonn}},\ }\bibfield  {title} {\enquote
  {\bibinfo {title} {Surface stratification determines the interfacial water
  structure of simple electrolyte solutions},}\ }\href
  {https://doi.org/10.1038/s41557-023-01416-6} {\bibfield  {journal} {\bibinfo
  {journal} {Nature Chemistry}\ }\textbf {\bibinfo {volume} {16}},\ \bibinfo
  {pages} {644--650} (\bibinfo {year} {2024})}\BibitemShut {NoStop}%
\bibitem [{\citenamefont {Dellago}, \citenamefont {Naor},\ and\ \citenamefont
  {Hummer}(2003)}]{dellago_2003_prl}%
  \BibitemOpen
  \bibfield  {author} {\bibinfo {author} {\bibfnamefont {C.}~\bibnamefont
  {Dellago}}, \bibinfo {author} {\bibfnamefont {M.~M.}\ \bibnamefont {Naor}},\
  and\ \bibinfo {author} {\bibfnamefont {G.}~\bibnamefont {Hummer}},\
  }\bibfield  {title} {\enquote {\bibinfo {title} {Proton transport through
  water-filled carbon nanotubes},}\ }\href
  {https://doi.org/10.1103/PhysRevLett.90.105902} {\bibfield  {journal}
  {\bibinfo  {journal} {Physical Review Letters}\ }\textbf {\bibinfo {volume}
  {90}},\ \bibinfo {pages} {105902} (\bibinfo {year} {2003})}\BibitemShut
  {NoStop}%
\bibitem [{\citenamefont {Cao}\ \emph {et~al.}(2010)\citenamefont {Cao},
  \citenamefont {Peng}, \citenamefont {Yan}, \citenamefont {Li}, \citenamefont
  {Li},\ and\ \citenamefont {Voth}}]{cao_2010}%
  \BibitemOpen
  \bibfield  {author} {\bibinfo {author} {\bibfnamefont {Z.}~\bibnamefont
  {Cao}}, \bibinfo {author} {\bibfnamefont {Y.}~\bibnamefont {Peng}}, \bibinfo
  {author} {\bibfnamefont {T.}~\bibnamefont {Yan}}, \bibinfo {author}
  {\bibfnamefont {S.}~\bibnamefont {Li}}, \bibinfo {author} {\bibfnamefont
  {A.}~\bibnamefont {Li}},\ and\ \bibinfo {author} {\bibfnamefont {G.~A.}\
  \bibnamefont {Voth}},\ }\bibfield  {title} {\enquote {\bibinfo {title}
  {Mechanism of fast proton transport along one-dimensional water chains
  confined in carbon nanotubes},}\ }\href {https://doi.org/10.1021/ja1046704}
  {\bibfield  {journal} {\bibinfo  {journal} {Journal of the American Chemical
  Society}\ }\textbf {\bibinfo {volume} {132}},\ \bibinfo {pages}
  {11395--11397} (\bibinfo {year} {2010})}\BibitemShut {NoStop}%
\bibitem [{\citenamefont {Gaigeot}, \citenamefont {Sprik},\ and\ \citenamefont
  {Sulpizi}(2012)}]{Gaigeot_2012}%
  \BibitemOpen
  \bibfield  {author} {\bibinfo {author} {\bibfnamefont {M.-P.}\ \bibnamefont
  {Gaigeot}}, \bibinfo {author} {\bibfnamefont {M.}~\bibnamefont {Sprik}},\
  and\ \bibinfo {author} {\bibfnamefont {M.}~\bibnamefont {Sulpizi}},\
  }\bibfield  {title} {\enquote {\bibinfo {title} {Oxide/water interfaces: how
  the surface chemistry modifies interfacial water properties},}\ }\href
  {https://doi.org/10.1088/0953-8984/24/12/124106} {\bibfield  {journal}
  {\bibinfo  {journal} {Journal of Physics: Condensed Matter}\ }\textbf
  {\bibinfo {volume} {24}},\ \bibinfo {pages} {124106} (\bibinfo {year}
  {2012})}\BibitemShut {NoStop}%
\bibitem [{\citenamefont {Sulpizi}, \citenamefont {Gaigeot},\ and\
  \citenamefont {Sprik}(2012)}]{sulpizi_silicawater_2012}%
  \BibitemOpen
  \bibfield  {author} {\bibinfo {author} {\bibfnamefont {M.}~\bibnamefont
  {Sulpizi}}, \bibinfo {author} {\bibfnamefont {M.-P.}\ \bibnamefont
  {Gaigeot}},\ and\ \bibinfo {author} {\bibfnamefont {M.}~\bibnamefont
  {Sprik}},\ }\bibfield  {title} {\enquote {\bibinfo {title} {The
  silica–water interface: How the silanols determine the surface acidity and
  modulate the water properties},}\ }\href {https://doi.org/10.1021/ct2007154}
  {\bibfield  {journal} {\bibinfo  {journal} {Journal of Chemical Theory and
  Computation}\ }\textbf {\bibinfo {volume} {8}},\ \bibinfo {pages}
  {1037--1047} (\bibinfo {year} {2012})}\BibitemShut {NoStop}%
\bibitem [{\citenamefont {Muñoz-Santiburcio}, \citenamefont {Wittekindt},\
  and\ \citenamefont {Marx}(2013)}]{marx_eproton}%
  \BibitemOpen
  \bibfield  {author} {\bibinfo {author} {\bibfnamefont {D.}~\bibnamefont
  {Muñoz-Santiburcio}}, \bibinfo {author} {\bibfnamefont {C.}~\bibnamefont
  {Wittekindt}},\ and\ \bibinfo {author} {\bibfnamefont {D.}~\bibnamefont
  {Marx}},\ }\bibfield  {title} {\enquote {\bibinfo {title} {Nanoconfinement
  effects on hydrated excess protons in layered materials},}\ }\href
  {https://doi.org/10.1038/ncomms3349} {\bibfield  {journal} {\bibinfo
  {journal} {Nature Communications}\ }\textbf {\bibinfo {volume} {4}},\
  \bibinfo {pages} {2349} (\bibinfo {year} {2013})}\BibitemShut {NoStop}%
\bibitem [{\citenamefont {Muñoz-Santiburcio}\ and\ \citenamefont
  {Marx}(2016)}]{marx_phole}%
  \BibitemOpen
  \bibfield  {author} {\bibinfo {author} {\bibfnamefont {D.}~\bibnamefont
  {Muñoz-Santiburcio}}\ and\ \bibinfo {author} {\bibfnamefont
  {D.}~\bibnamefont {Marx}},\ }\bibfield  {title} {\enquote {\bibinfo {title}
  {On the complex structural diffusion of proton holes in nanoconfined alkaline
  solutions within slit pores},}\ }\href {https://doi.org/10.1038/ncomms12625}
  {\bibfield  {journal} {\bibinfo  {journal} {Nature Communications}\ }\textbf
  {\bibinfo {volume} {7}},\ \bibinfo {pages} {12625} (\bibinfo {year}
  {2016})}\BibitemShut {NoStop}%
\bibitem [{\citenamefont {Fumagalli}\ \emph {et~al.}(2018)\citenamefont
  {Fumagalli}, \citenamefont {Esfandiar}, \citenamefont {Fabregas},
  \citenamefont {Hu}, \citenamefont {Ares}, \citenamefont {Janardanan},
  \citenamefont {Yang}, \citenamefont {Radha}, \citenamefont {Taniguchi},
  \citenamefont {Watanabe}, \citenamefont {Gomila}, \citenamefont {Novoselov},\
  and\ \citenamefont {Geim}}]{fumagalli_dielectric_2018}%
  \BibitemOpen
  \bibfield  {author} {\bibinfo {author} {\bibfnamefont {L.}~\bibnamefont
  {Fumagalli}}, \bibinfo {author} {\bibfnamefont {A.}~\bibnamefont
  {Esfandiar}}, \bibinfo {author} {\bibfnamefont {R.}~\bibnamefont {Fabregas}},
  \bibinfo {author} {\bibfnamefont {S.}~\bibnamefont {Hu}}, \bibinfo {author}
  {\bibfnamefont {P.}~\bibnamefont {Ares}}, \bibinfo {author} {\bibfnamefont
  {A.}~\bibnamefont {Janardanan}}, \bibinfo {author} {\bibfnamefont
  {Q.}~\bibnamefont {Yang}}, \bibinfo {author} {\bibfnamefont {B.}~\bibnamefont
  {Radha}}, \bibinfo {author} {\bibfnamefont {T.}~\bibnamefont {Taniguchi}},
  \bibinfo {author} {\bibfnamefont {K.}~\bibnamefont {Watanabe}}, \bibinfo
  {author} {\bibfnamefont {G.}~\bibnamefont {Gomila}}, \bibinfo {author}
  {\bibfnamefont {K.~S.}\ \bibnamefont {Novoselov}},\ and\ \bibinfo {author}
  {\bibfnamefont {A.~K.}\ \bibnamefont {Geim}},\ }\bibfield  {title} {\enquote
  {\bibinfo {title} {Anomalously low dielectric constant of confined water},}\
  }\href {https://doi.org/10.1126/science.aat4191} {\bibfield  {journal}
  {\bibinfo  {journal} {Science}\ }\textbf {\bibinfo {volume} {360}},\ \bibinfo
  {pages} {1339--1342} (\bibinfo {year} {2018})}\BibitemShut {NoStop}%
\bibitem [{\citenamefont {Dufils}\ \emph {et~al.}(2024)\citenamefont {Dufils},
  \citenamefont {Schran}, \citenamefont {Chen}, \citenamefont {Geim},
  \citenamefont {Fumagalli},\ and\ \citenamefont
  {Michaelides}}]{chris_dufils_2024}%
  \BibitemOpen
  \bibfield  {author} {\bibinfo {author} {\bibfnamefont {T.}~\bibnamefont
  {Dufils}}, \bibinfo {author} {\bibfnamefont {C.}~\bibnamefont {Schran}},
  \bibinfo {author} {\bibfnamefont {J.}~\bibnamefont {Chen}}, \bibinfo {author}
  {\bibfnamefont {A.~K.}\ \bibnamefont {Geim}}, \bibinfo {author}
  {\bibfnamefont {L.}~\bibnamefont {Fumagalli}},\ and\ \bibinfo {author}
  {\bibfnamefont {A.}~\bibnamefont {Michaelides}},\ }\bibfield  {title}
  {\enquote {\bibinfo {title} {Origin of dielectric polarization suppression in
  confined water from first principles},}\ }\href
  {https://doi.org/10.1039/D3SC04740G} {\bibfield  {journal} {\bibinfo
  {journal} {Chemical Science}\ }\textbf {\bibinfo {volume} {15}},\ \bibinfo
  {pages} {516--527} (\bibinfo {year} {2024})}\BibitemShut {NoStop}%
\bibitem [{\citenamefont {Secchi}\ \emph
  {et~al.}(2016{\natexlab{a}})\citenamefont {Secchi}, \citenamefont {Marbach},
  \citenamefont {Niguès}, \citenamefont {Stein}, \citenamefont {Siria},\ and\
  \citenamefont {Bocquet}}]{secchi_nat_2016}%
  \BibitemOpen
  \bibfield  {author} {\bibinfo {author} {\bibfnamefont {E.}~\bibnamefont
  {Secchi}}, \bibinfo {author} {\bibfnamefont {S.}~\bibnamefont {Marbach}},
  \bibinfo {author} {\bibfnamefont {A.}~\bibnamefont {Niguès}}, \bibinfo
  {author} {\bibfnamefont {D.}~\bibnamefont {Stein}}, \bibinfo {author}
  {\bibfnamefont {A.}~\bibnamefont {Siria}},\ and\ \bibinfo {author}
  {\bibfnamefont {L.}~\bibnamefont {Bocquet}},\ }\bibfield  {title} {\enquote
  {\bibinfo {title} {Massive radius-dependent flow slippage in carbon
  nanotubes},}\ }\href {https://doi.org/10.1038/nature19315} {\bibfield
  {journal} {\bibinfo  {journal} {Nature}\ }\textbf {\bibinfo {volume} {537}},\
  \bibinfo {pages} {210--213} (\bibinfo {year}
  {2016}{\natexlab{a}})}\BibitemShut {NoStop}%
\bibitem [{\citenamefont {Kapil}\ \emph {et~al.}(2022)\citenamefont {Kapil},
  \citenamefont {Schran}, \citenamefont {Zen}, \citenamefont {Chen},
  \citenamefont {Pickard},\ and\ \citenamefont {Michaelides}}]{vk_chris_2022}%
  \BibitemOpen
  \bibfield  {author} {\bibinfo {author} {\bibfnamefont {V.}~\bibnamefont
  {Kapil}}, \bibinfo {author} {\bibfnamefont {C.}~\bibnamefont {Schran}},
  \bibinfo {author} {\bibfnamefont {A.}~\bibnamefont {Zen}}, \bibinfo {author}
  {\bibfnamefont {J.}~\bibnamefont {Chen}}, \bibinfo {author} {\bibfnamefont
  {C.~J.}\ \bibnamefont {Pickard}},\ and\ \bibinfo {author} {\bibfnamefont
  {A.}~\bibnamefont {Michaelides}},\ }\bibfield  {title} {\enquote {\bibinfo
  {title} {The first-principles phase diagram of monolayer nanoconfined
  water},}\ }\href {https://doi.org/10.1038/s41586-022-05036-x} {\bibfield
  {journal} {\bibinfo  {journal} {Nature}\ }\textbf {\bibinfo {volume} {609}},\
  \bibinfo {pages} {512--516} (\bibinfo {year} {2022})}\BibitemShut {NoStop}%
\bibitem [{\citenamefont {Jiang}\ \emph {et~al.}(2024)\citenamefont {Jiang},
  \citenamefont {Gao}, \citenamefont {Li}, \citenamefont {Liu}, \citenamefont
  {Zhu}, \citenamefont {Zhu}, \citenamefont {Francisco},\ and\ \citenamefont
  {Zeng}}]{jiang_rich_2024}%
  \BibitemOpen
  \bibfield  {author} {\bibinfo {author} {\bibfnamefont {J.}~\bibnamefont
  {Jiang}}, \bibinfo {author} {\bibfnamefont {Y.}~\bibnamefont {Gao}}, \bibinfo
  {author} {\bibfnamefont {L.}~\bibnamefont {Li}}, \bibinfo {author}
  {\bibfnamefont {Y.}~\bibnamefont {Liu}}, \bibinfo {author} {\bibfnamefont
  {W.}~\bibnamefont {Zhu}}, \bibinfo {author} {\bibfnamefont {C.}~\bibnamefont
  {Zhu}}, \bibinfo {author} {\bibfnamefont {J.~S.}\ \bibnamefont {Francisco}},\
  and\ \bibinfo {author} {\bibfnamefont {X.~C.}\ \bibnamefont {Zeng}},\
  }\bibfield  {title} {\enquote {\bibinfo {title} {Rich proton dynamics and
  phase behaviours of nanoconfined ices},}\ }\href
  {https://doi.org/10.1038/s41567-023-02341-8} {\bibfield  {journal} {\bibinfo
  {journal} {Nature Physics}\ }\textbf {\bibinfo {volume} {20}},\ \bibinfo
  {pages} {456--464} (\bibinfo {year} {2024})}\BibitemShut {NoStop}%
\bibitem [{\citenamefont {Algara-Siller}\ \emph {et~al.}(2015)\citenamefont
  {Algara-Siller}, \citenamefont {Lehtinen}, \citenamefont {Wang},
  \citenamefont {Nair}, \citenamefont {Kaiser}, \citenamefont {Wu},
  \citenamefont {Geim},\ and\ \citenamefont
  {Grigorieva}}]{algara-siller_square_2015}%
  \BibitemOpen
  \bibfield  {author} {\bibinfo {author} {\bibfnamefont {G.}~\bibnamefont
  {Algara-Siller}}, \bibinfo {author} {\bibfnamefont {O.}~\bibnamefont
  {Lehtinen}}, \bibinfo {author} {\bibfnamefont {F.~C.}\ \bibnamefont {Wang}},
  \bibinfo {author} {\bibfnamefont {R.~R.}\ \bibnamefont {Nair}}, \bibinfo
  {author} {\bibfnamefont {U.}~\bibnamefont {Kaiser}}, \bibinfo {author}
  {\bibfnamefont {H.~A.}\ \bibnamefont {Wu}}, \bibinfo {author} {\bibfnamefont
  {A.~K.}\ \bibnamefont {Geim}},\ and\ \bibinfo {author} {\bibfnamefont
  {I.~V.}\ \bibnamefont {Grigorieva}},\ }\bibfield  {title} {\enquote {\bibinfo
  {title} {Square ice in graphene nanocapillaries},}\ }\href
  {https://doi.org/10.1038/nature14295} {\bibfield  {journal} {\bibinfo
  {journal} {Nature}\ }\textbf {\bibinfo {volume} {519}},\ \bibinfo {pages}
  {443--445} (\bibinfo {year} {2015})}\BibitemShut {NoStop}%
\bibitem [{\citenamefont {Ravindra}\ \emph {et~al.}(2024)\citenamefont
  {Ravindra}, \citenamefont {Advincula}, \citenamefont {Schran}, \citenamefont
  {Michaelides},\ and\ \citenamefont {Kapil}}]{quasionedimensional}%
  \BibitemOpen
  \bibfield  {author} {\bibinfo {author} {\bibfnamefont {P.}~\bibnamefont
  {Ravindra}}, \bibinfo {author} {\bibfnamefont {X.~R.}\ \bibnamefont
  {Advincula}}, \bibinfo {author} {\bibfnamefont {C.}~\bibnamefont {Schran}},
  \bibinfo {author} {\bibfnamefont {A.}~\bibnamefont {Michaelides}},\ and\
  \bibinfo {author} {\bibfnamefont {V.}~\bibnamefont {Kapil}},\ }\bibfield
  {title} {\enquote {\bibinfo {title} {Quasi-one-dimensional hydrogen bonding
  in nanoconfined ice},}\ }\href
  {https://www.nature.com/articles/s41467-024-51124-z} {\bibfield  {journal}
  {\bibinfo  {journal} {Nature Communications}\ }\textbf {\bibinfo {volume}
  {15}},\ \bibinfo {pages} {7301} (\bibinfo {year} {2024})}\BibitemShut
  {NoStop}%
\bibitem [{\citenamefont {Radha}\ \emph {et~al.}(2016)\citenamefont {Radha},
  \citenamefont {Esfandiar}, \citenamefont {Wang}, \citenamefont {Rooney},
  \citenamefont {Gopinadhan}, \citenamefont {Keerthi}, \citenamefont
  {Mishchenko}, \citenamefont {Janardanan}, \citenamefont {Blake},
  \citenamefont {Fumagalli}, \citenamefont {Lozada-Hidalgo}, \citenamefont
  {Garaj}, \citenamefont {Haigh}, \citenamefont {Grigorieva}, \citenamefont
  {Wu},\ and\ \citenamefont {Geim}}]{radha_b_2016}%
  \BibitemOpen
  \bibfield  {author} {\bibinfo {author} {\bibfnamefont {B.}~\bibnamefont
  {Radha}}, \bibinfo {author} {\bibfnamefont {A.}~\bibnamefont {Esfandiar}},
  \bibinfo {author} {\bibfnamefont {F.~C.}\ \bibnamefont {Wang}}, \bibinfo
  {author} {\bibfnamefont {A.~P.}\ \bibnamefont {Rooney}}, \bibinfo {author}
  {\bibfnamefont {K.}~\bibnamefont {Gopinadhan}}, \bibinfo {author}
  {\bibfnamefont {A.}~\bibnamefont {Keerthi}}, \bibinfo {author} {\bibfnamefont
  {A.}~\bibnamefont {Mishchenko}}, \bibinfo {author} {\bibfnamefont
  {A.}~\bibnamefont {Janardanan}}, \bibinfo {author} {\bibfnamefont
  {P.}~\bibnamefont {Blake}}, \bibinfo {author} {\bibfnamefont
  {L.}~\bibnamefont {Fumagalli}}, \bibinfo {author} {\bibfnamefont
  {M.}~\bibnamefont {Lozada-Hidalgo}}, \bibinfo {author} {\bibfnamefont
  {S.}~\bibnamefont {Garaj}}, \bibinfo {author} {\bibfnamefont {S.~J.}\
  \bibnamefont {Haigh}}, \bibinfo {author} {\bibfnamefont {I.~V.}\ \bibnamefont
  {Grigorieva}}, \bibinfo {author} {\bibfnamefont {H.~A.}\ \bibnamefont {Wu}},\
  and\ \bibinfo {author} {\bibfnamefont {A.~K.}\ \bibnamefont {Geim}},\
  }\bibfield  {title} {\enquote {\bibinfo {title} {Molecular transport through
  capillaries made with atomic-scale precision},}\ }\href
  {https://doi.org/10.1038/nature19363} {\bibfield  {journal} {\bibinfo
  {journal} {Nature}\ }\textbf {\bibinfo {volume} {538}},\ \bibinfo {pages}
  {222--225} (\bibinfo {year} {2016})}\BibitemShut {NoStop}%
\bibitem [{\citenamefont {Grosjean}, \citenamefont {Bocquet},\ and\
  \citenamefont {Vuilleumier}(2019)}]{mlb_oh}%
  \BibitemOpen
  \bibfield  {author} {\bibinfo {author} {\bibfnamefont {B.}~\bibnamefont
  {Grosjean}}, \bibinfo {author} {\bibfnamefont {M.-L.}\ \bibnamefont
  {Bocquet}},\ and\ \bibinfo {author} {\bibfnamefont {R.}~\bibnamefont
  {Vuilleumier}},\ }\bibfield  {title} {\enquote {\bibinfo {title} {Versatile
  electrification of two-dimensional nanomaterials in water},}\ }\href
  {https://doi.org/10.1038/s41467-019-09708-7} {\bibfield  {journal} {\bibinfo
  {journal} {Nature Communications}\ }\textbf {\bibinfo {volume} {10}},\
  \bibinfo {pages} {1656} (\bibinfo {year} {2019})}\BibitemShut {NoStop}%
\bibitem [{\citenamefont {Secchi}\ \emph
  {et~al.}(2016{\natexlab{b}})\citenamefont {Secchi}, \citenamefont {Niguès},
  \citenamefont {Jubin}, \citenamefont {Siria},\ and\ \citenamefont
  {Bocquet}}]{secchi_2016_oh_prl}%
  \BibitemOpen
  \bibfield  {author} {\bibinfo {author} {\bibfnamefont {E.}~\bibnamefont
  {Secchi}}, \bibinfo {author} {\bibfnamefont {A.}~\bibnamefont {Niguès}},
  \bibinfo {author} {\bibfnamefont {L.}~\bibnamefont {Jubin}}, \bibinfo
  {author} {\bibfnamefont {A.}~\bibnamefont {Siria}},\ and\ \bibinfo {author}
  {\bibfnamefont {L.}~\bibnamefont {Bocquet}},\ }\bibfield  {title} {\enquote
  {\bibinfo {title} {Scaling behavior for ionic transport and its fluctuations
  in individual carbon nanotubes},}\ }\href
  {https://doi.org/10.1103/PhysRevLett.116.154501} {\bibfield  {journal}
  {\bibinfo  {journal} {Physical Review Letters}\ }\textbf {\bibinfo {volume}
  {116}},\ \bibinfo {pages} {154501} (\bibinfo {year}
  {2016}{\natexlab{b}})}\BibitemShut {NoStop}%
\bibitem [{\citenamefont {de~Aquino}\ \emph {et~al.}(2019)\citenamefont
  {de~Aquino}, \citenamefont {Ghorbanfekr-Kalashami}, \citenamefont
  {Neek-Amal},\ and\ \citenamefont {Peeters}}]{aquino_2019}%
  \BibitemOpen
  \bibfield  {author} {\bibinfo {author} {\bibfnamefont {B.~R.~H.}\
  \bibnamefont {de~Aquino}}, \bibinfo {author} {\bibfnamefont {H.}~\bibnamefont
  {Ghorbanfekr-Kalashami}}, \bibinfo {author} {\bibfnamefont {M.}~\bibnamefont
  {Neek-Amal}},\ and\ \bibinfo {author} {\bibfnamefont {F.~M.}\ \bibnamefont
  {Peeters}},\ }\bibfield  {title} {\enquote {\bibinfo {title} {Ionized water
  confined in graphene nanochannels},}\ }\href
  {https://doi.org/10.1039/C9CP00075E} {\bibfield  {journal} {\bibinfo
  {journal} {Phys. Chem. Chem. Phys.}\ }\textbf {\bibinfo {volume} {21}},\
  \bibinfo {pages} {9285--9295} (\bibinfo {year} {2019})}\BibitemShut {NoStop}%
\bibitem [{\citenamefont {Scalfi}\ \emph {et~al.}(2024)\citenamefont {Scalfi},
  \citenamefont {Lehmann}, \citenamefont {dos Santos}, \citenamefont {Becker},\
  and\ \citenamefont {Netz}}]{scalfi_2024}%
  \BibitemOpen
  \bibfield  {author} {\bibinfo {author} {\bibfnamefont {L.}~\bibnamefont
  {Scalfi}}, \bibinfo {author} {\bibfnamefont {L.}~\bibnamefont {Lehmann}},
  \bibinfo {author} {\bibfnamefont {A.~P.}\ \bibnamefont {dos Santos}},
  \bibinfo {author} {\bibfnamefont {M.~R.}\ \bibnamefont {Becker}},\ and\
  \bibinfo {author} {\bibfnamefont {R.~R.}\ \bibnamefont {Netz}},\ }\bibfield
  {title} {\enquote {\bibinfo {title} {{Propensity of hydroxide and hydronium
  ions for the air–water and graphene–water interfaces from ab initio and
  force field simulations}},}\ }\href {https://doi.org/10.1063/5.0226966}
  {\bibfield  {journal} {\bibinfo  {journal} {The Journal of Chemical Physics}\
  }\textbf {\bibinfo {volume} {161}},\ \bibinfo {pages} {144701} (\bibinfo
  {year} {2024})}\BibitemShut {NoStop}%
\bibitem [{\citenamefont {Björneholm}\ \emph {et~al.}(2016)\citenamefont
  {Björneholm}, \citenamefont {Hansen}, \citenamefont {Hodgson}, \citenamefont
  {Liu}, \citenamefont {Limmer}, \citenamefont {Michaelides}, \citenamefont
  {Pedevilla}, \citenamefont {Rossmeisl}, \citenamefont {Shen}, \citenamefont
  {Tocci}, \citenamefont {Tyrode}, \citenamefont {Walz}, \citenamefont
  {Werner},\ and\ \citenamefont {Bluhm}}]{wat_inter_2016}%
  \BibitemOpen
  \bibfield  {author} {\bibinfo {author} {\bibfnamefont {O.}~\bibnamefont
  {Björneholm}}, \bibinfo {author} {\bibfnamefont {M.~H.}\ \bibnamefont
  {Hansen}}, \bibinfo {author} {\bibfnamefont {A.}~\bibnamefont {Hodgson}},
  \bibinfo {author} {\bibfnamefont {L.-M.}\ \bibnamefont {Liu}}, \bibinfo
  {author} {\bibfnamefont {D.~T.}\ \bibnamefont {Limmer}}, \bibinfo {author}
  {\bibfnamefont {A.}~\bibnamefont {Michaelides}}, \bibinfo {author}
  {\bibfnamefont {P.}~\bibnamefont {Pedevilla}}, \bibinfo {author}
  {\bibfnamefont {J.}~\bibnamefont {Rossmeisl}}, \bibinfo {author}
  {\bibfnamefont {H.}~\bibnamefont {Shen}}, \bibinfo {author} {\bibfnamefont
  {G.}~\bibnamefont {Tocci}}, \bibinfo {author} {\bibfnamefont
  {E.}~\bibnamefont {Tyrode}}, \bibinfo {author} {\bibfnamefont {M.-M.}\
  \bibnamefont {Walz}}, \bibinfo {author} {\bibfnamefont {J.}~\bibnamefont
  {Werner}},\ and\ \bibinfo {author} {\bibfnamefont {H.}~\bibnamefont
  {Bluhm}},\ }\bibfield  {title} {\enquote {\bibinfo {title} {Water at
  interfaces},}\ }\href {https://doi.org/10.1021/acs.chemrev.6b00045}
  {\bibfield  {journal} {\bibinfo  {journal} {Chemical Reviews}\ }\textbf
  {\bibinfo {volume} {116}},\ \bibinfo {pages} {7698--7726} (\bibinfo {year}
  {2016})}\BibitemShut {NoStop}%
\bibitem [{\citenamefont {Gonella}\ \emph {et~al.}(2021)\citenamefont
  {Gonella}, \citenamefont {Backus}, \citenamefont {Nagata}, \citenamefont
  {Bonthuis}, \citenamefont {Loche}, \citenamefont {Schlaich}, \citenamefont
  {Netz}, \citenamefont {Kühnle}, \citenamefont {McCrum}, \citenamefont
  {Koper}, \citenamefont {Wolf}, \citenamefont {Winter}, \citenamefont
  {Meijer}, \citenamefont {Campen},\ and\ \citenamefont
  {Bonn}}]{wat_charged_2021}%
  \BibitemOpen
  \bibfield  {author} {\bibinfo {author} {\bibfnamefont {G.}~\bibnamefont
  {Gonella}}, \bibinfo {author} {\bibfnamefont {E.~H.~G.}\ \bibnamefont
  {Backus}}, \bibinfo {author} {\bibfnamefont {Y.}~\bibnamefont {Nagata}},
  \bibinfo {author} {\bibfnamefont {D.~J.}\ \bibnamefont {Bonthuis}}, \bibinfo
  {author} {\bibfnamefont {P.}~\bibnamefont {Loche}}, \bibinfo {author}
  {\bibfnamefont {A.}~\bibnamefont {Schlaich}}, \bibinfo {author}
  {\bibfnamefont {R.~R.}\ \bibnamefont {Netz}}, \bibinfo {author}
  {\bibfnamefont {A.}~\bibnamefont {Kühnle}}, \bibinfo {author} {\bibfnamefont
  {I.~T.}\ \bibnamefont {McCrum}}, \bibinfo {author} {\bibfnamefont {M.~T.~M.}\
  \bibnamefont {Koper}}, \bibinfo {author} {\bibfnamefont {M.}~\bibnamefont
  {Wolf}}, \bibinfo {author} {\bibfnamefont {B.}~\bibnamefont {Winter}},
  \bibinfo {author} {\bibfnamefont {G.}~\bibnamefont {Meijer}}, \bibinfo
  {author} {\bibfnamefont {R.~K.}\ \bibnamefont {Campen}},\ and\ \bibinfo
  {author} {\bibfnamefont {M.}~\bibnamefont {Bonn}},\ }\bibfield  {title}
  {\enquote {\bibinfo {title} {Water at charged interfaces},}\ }\href
  {https://doi.org/10.1038/s41570-021-00293-2} {\bibfield  {journal} {\bibinfo
  {journal} {Nature Reviews Chemistry}\ }\textbf {\bibinfo {volume} {5}},\
  \bibinfo {pages} {466--485} (\bibinfo {year} {2021})}\BibitemShut {NoStop}%
\bibitem [{\citenamefont {Behler}(2016)}]{behler_2016_pers}%
  \BibitemOpen
  \bibfield  {author} {\bibinfo {author} {\bibfnamefont {J.}~\bibnamefont
  {Behler}},\ }\bibfield  {title} {\enquote {\bibinfo {title} {Perspective:
  Machine learning potentials for atomistic simulations},}\ }\href
  {https://doi.org/10.1063/1.4966192} {\bibfield  {journal} {\bibinfo
  {journal} {The Journal of Chemical Physics}\ }\textbf {\bibinfo {volume}
  {145}},\ \bibinfo {pages} {170901} (\bibinfo {year} {2016})}\BibitemShut
  {NoStop}%
\bibitem [{\citenamefont {Butler}\ \emph {et~al.}(2018)\citenamefont {Butler},
  \citenamefont {Davies}, \citenamefont {Cartwright}, \citenamefont {Isayev},\
  and\ \citenamefont {Walsh}}]{butler_2018}%
  \BibitemOpen
  \bibfield  {author} {\bibinfo {author} {\bibfnamefont {K.~T.}\ \bibnamefont
  {Butler}}, \bibinfo {author} {\bibfnamefont {D.~W.}\ \bibnamefont {Davies}},
  \bibinfo {author} {\bibfnamefont {H.}~\bibnamefont {Cartwright}}, \bibinfo
  {author} {\bibfnamefont {O.}~\bibnamefont {Isayev}},\ and\ \bibinfo {author}
  {\bibfnamefont {A.}~\bibnamefont {Walsh}},\ }\bibfield  {title} {\enquote
  {\bibinfo {title} {Machine learning for molecular and materials science},}\
  }\href {https://doi.org/10.1038/s41586-018-0337-2} {\bibfield  {journal}
  {\bibinfo  {journal} {Nature}\ }\textbf {\bibinfo {volume} {559}},\ \bibinfo
  {pages} {547--555} (\bibinfo {year} {2018})}\BibitemShut {NoStop}%
\bibitem [{\citenamefont {Omranpour}\ \emph {et~al.}(2024)\citenamefont
  {Omranpour}, \citenamefont {Hijes}, \citenamefont {Behler},\ and\
  \citenamefont {Dellago}}]{montero_2024}%
  \BibitemOpen
  \bibfield  {author} {\bibinfo {author} {\bibfnamefont {A.}~\bibnamefont
  {Omranpour}}, \bibinfo {author} {\bibfnamefont {P.~M.~D.}\ \bibnamefont
  {Hijes}}, \bibinfo {author} {\bibfnamefont {J.}~\bibnamefont {Behler}},\ and\
  \bibinfo {author} {\bibfnamefont {C.}~\bibnamefont {Dellago}},\ }\bibfield
  {title} {\enquote {\bibinfo {title} {Perspective: Atomistic simulations of
  water and aqueous systems with machine learning potentials},}\ }\href
  {https://doi.org/10.1063/5.0201241} {\bibfield  {journal} {\bibinfo
  {journal} {The Journal of Chemical Physics}\ }\textbf {\bibinfo {volume}
  {160}},\ \bibinfo {pages} {170901} (\bibinfo {year} {2024})}\BibitemShut
  {NoStop}%
\bibitem [{\citenamefont {Schran}\ \emph {et~al.}(2021)\citenamefont {Schran},
  \citenamefont {Thiemann}, \citenamefont {Rowe}, \citenamefont {Müller},
  \citenamefont {Marsalek},\ and\ \citenamefont {Michaelides}}]{complex_2021}%
  \BibitemOpen
  \bibfield  {author} {\bibinfo {author} {\bibfnamefont {C.}~\bibnamefont
  {Schran}}, \bibinfo {author} {\bibfnamefont {F.~L.}\ \bibnamefont
  {Thiemann}}, \bibinfo {author} {\bibfnamefont {P.}~\bibnamefont {Rowe}},
  \bibinfo {author} {\bibfnamefont {E.~A.}\ \bibnamefont {Müller}}, \bibinfo
  {author} {\bibfnamefont {O.}~\bibnamefont {Marsalek}},\ and\ \bibinfo
  {author} {\bibfnamefont {A.}~\bibnamefont {Michaelides}},\ }\bibfield
  {title} {\enquote {\bibinfo {title} {Machine learning potentials for complex
  aqueous systems made simple},}\ }\href
  {https://doi.org/10.1073/pnas.2110077118} {\bibfield  {journal} {\bibinfo
  {journal} {Proceedings of the National Academy of Sciences}\ }\textbf
  {\bibinfo {volume} {118}},\ \bibinfo {pages} {e2110077118} (\bibinfo {year}
  {2021})}\BibitemShut {NoStop}%
\bibitem [{\citenamefont {de~la Puente}\ \emph {et~al.}(2022)\citenamefont
  {de~la Puente}, \citenamefont {David}, \citenamefont {Gomez},\ and\
  \citenamefont {Laage}}]{de_la_puente_acids_2022}%
  \BibitemOpen
  \bibfield  {author} {\bibinfo {author} {\bibfnamefont {M.}~\bibnamefont
  {de~la Puente}}, \bibinfo {author} {\bibfnamefont {R.}~\bibnamefont {David}},
  \bibinfo {author} {\bibfnamefont {A.}~\bibnamefont {Gomez}},\ and\ \bibinfo
  {author} {\bibfnamefont {D.}~\bibnamefont {Laage}},\ }\bibfield  {title}
  {\enquote {\bibinfo {title} {Acids at the {Edge}: {Why} {Nitric} and {Formic}
  {Acid} {Dissociations} at {Air}–{Water} {Interfaces} {Depend} on {Depth}
  and on {Interface} {Specific} {Area}},}\ }\href
  {https://doi.org/10.1021/jacs.2c03099} {\bibfield  {journal} {\bibinfo
  {journal} {Journal of the American Chemical Society}\ }\textbf {\bibinfo
  {volume} {144}},\ \bibinfo {pages} {10524--10529} (\bibinfo {year}
  {2022})}\BibitemShut {NoStop}%
\bibitem [{\citenamefont {Andrade}, \citenamefont {Car},\ and\ \citenamefont
  {Selloni}(2023)}]{car_dissoc_deepmd_2023}%
  \BibitemOpen
  \bibfield  {author} {\bibinfo {author} {\bibfnamefont {M.~C.}\ \bibnamefont
  {Andrade}}, \bibinfo {author} {\bibfnamefont {R.}~\bibnamefont {Car}},\ and\
  \bibinfo {author} {\bibfnamefont {A.}~\bibnamefont {Selloni}},\ }\bibfield
  {title} {\enquote {\bibinfo {title} {Probing the self-ionization of liquid
  water with ab initio deep potential molecular dynamics},}\ }\href
  {https://doi.org/10.1073/pnas.2302468120} {\bibfield  {journal} {\bibinfo
  {journal} {Proceedings of the National Academy of Sciences}\ }\textbf
  {\bibinfo {volume} {120}},\ \bibinfo {pages} {e2302468120} (\bibinfo {year}
  {2023})}\BibitemShut {NoStop}%
\bibitem [{\citenamefont {Fong}\ \emph {et~al.}(2024)\citenamefont {Fong},
  \citenamefont {Sumić}, \citenamefont {O’Neill}, \citenamefont {Schran},
  \citenamefont {Grey},\ and\ \citenamefont {Michaelides}}]{kara_pairing_2024}%
  \BibitemOpen
  \bibfield  {author} {\bibinfo {author} {\bibfnamefont {K.~D.}\ \bibnamefont
  {Fong}}, \bibinfo {author} {\bibfnamefont {B.}~\bibnamefont {Sumić}},
  \bibinfo {author} {\bibfnamefont {N.}~\bibnamefont {O’Neill}}, \bibinfo
  {author} {\bibfnamefont {C.}~\bibnamefont {Schran}}, \bibinfo {author}
  {\bibfnamefont {C.~P.}\ \bibnamefont {Grey}},\ and\ \bibinfo {author}
  {\bibfnamefont {A.}~\bibnamefont {Michaelides}},\ }\bibfield  {title}
  {\enquote {\bibinfo {title} {The interplay of solvation and polarization
  effects on ion pairing in nanoconfined electrolytes},}\ }\href
  {https://doi.org/10.1021/acs.nanolett.4c00890} {\bibfield  {journal}
  {\bibinfo  {journal} {Nano Letters}\ }\textbf {\bibinfo {volume} {24}},\
  \bibinfo {pages} {5024--5030} (\bibinfo {year} {2024})}\BibitemShut {NoStop}%
\bibitem [{\citenamefont {Babin}, \citenamefont {Leforestier},\ and\
  \citenamefont {Paesani}(2013)}]{mbpol_2013}%
  \BibitemOpen
  \bibfield  {author} {\bibinfo {author} {\bibfnamefont {V.}~\bibnamefont
  {Babin}}, \bibinfo {author} {\bibfnamefont {C.}~\bibnamefont {Leforestier}},\
  and\ \bibinfo {author} {\bibfnamefont {F.}~\bibnamefont {Paesani}},\
  }\bibfield  {title} {\enquote {\bibinfo {title} {Development of a “first
  principles” water potential with flexible monomers: Dimer potential energy
  surface, vrt spectrum, and second virial coefficient},}\ }\href
  {https://doi.org/10.1021/ct400863t} {\bibfield  {journal} {\bibinfo
  {journal} {Journal of Chemical Theory and Computation}\ }\textbf {\bibinfo
  {volume} {9}},\ \bibinfo {pages} {5395--5403} (\bibinfo {year}
  {2013})}\BibitemShut {NoStop}%
\bibitem [{\citenamefont {Bonthuis}, \citenamefont {Mamatkulov},\ and\
  \citenamefont {Netz}(2016)}]{bonthuis_2016_ff}%
  \BibitemOpen
  \bibfield  {author} {\bibinfo {author} {\bibfnamefont {D.~J.}\ \bibnamefont
  {Bonthuis}}, \bibinfo {author} {\bibfnamefont {S.~I.}\ \bibnamefont
  {Mamatkulov}},\ and\ \bibinfo {author} {\bibfnamefont {R.~R.}\ \bibnamefont
  {Netz}},\ }\bibfield  {title} {\enquote {\bibinfo {title} {Optimization of
  classical nonpolarizable force fields for oh- and h3o+},}\ }\href
  {https://doi.org/10.1063/1.4942771} {\bibfield  {journal} {\bibinfo
  {journal} {The Journal of Chemical Physics}\ }\textbf {\bibinfo {volume}
  {144}},\ \bibinfo {pages} {104503} (\bibinfo {year} {2016})}\BibitemShut
  {NoStop}%
\bibitem [{\citenamefont {Batatia}\ \emph {et~al.}(2022)\citenamefont
  {Batatia}, \citenamefont {Kovacs}, \citenamefont {Simm}, \citenamefont
  {Ortner},\ and\ \citenamefont {Csanyi}}]{mace_canonical}%
  \BibitemOpen
  \bibfield  {author} {\bibinfo {author} {\bibfnamefont {I.}~\bibnamefont
  {Batatia}}, \bibinfo {author} {\bibfnamefont {D.~P.}\ \bibnamefont {Kovacs}},
  \bibinfo {author} {\bibfnamefont {G.~N.~C.}\ \bibnamefont {Simm}}, \bibinfo
  {author} {\bibfnamefont {C.}~\bibnamefont {Ortner}},\ and\ \bibinfo {author}
  {\bibfnamefont {G.}~\bibnamefont {Csanyi}},\ }\bibfield  {title} {\enquote
  {\bibinfo {title} {Mace: Higher order equivariant message passing neural
  networks for fast and accurate force fields},}\ }\href
  {https://openreview.net/forum?id=YPpSngE-ZU} {\bibfield  {journal} {\bibinfo
  {journal} {Advances in Neural Information Processing Systems}\ } (\bibinfo
  {year} {2022})}\BibitemShut {NoStop}%
\bibitem [{\citenamefont {Perdew}, \citenamefont {Burke},\ and\ \citenamefont
  {Ernzerhof}(1996)}]{revpbed3_1}%
  \BibitemOpen
  \bibfield  {author} {\bibinfo {author} {\bibfnamefont {J.~P.}\ \bibnamefont
  {Perdew}}, \bibinfo {author} {\bibfnamefont {K.}~\bibnamefont {Burke}},\ and\
  \bibinfo {author} {\bibfnamefont {M.}~\bibnamefont {Ernzerhof}},\ }\bibfield
  {title} {\enquote {\bibinfo {title} {Generalized gradient approximation made
  simple},}\ }\href {https://doi.org/10.1103/PhysRevLett.77.3865} {\bibfield
  {journal} {\bibinfo  {journal} {Physical Review Letters}\ }\textbf {\bibinfo
  {volume} {77}},\ \bibinfo {pages} {3865--3868} (\bibinfo {year}
  {1996})}\BibitemShut {NoStop}%
\bibitem [{\citenamefont {Grimme}\ \emph {et~al.}(2010)\citenamefont {Grimme},
  \citenamefont {Antony}, \citenamefont {Ehrlich},\ and\ \citenamefont
  {Krieg}}]{revpbed3_2}%
  \BibitemOpen
  \bibfield  {author} {\bibinfo {author} {\bibfnamefont {S.}~\bibnamefont
  {Grimme}}, \bibinfo {author} {\bibfnamefont {J.}~\bibnamefont {Antony}},
  \bibinfo {author} {\bibfnamefont {S.}~\bibnamefont {Ehrlich}},\ and\ \bibinfo
  {author} {\bibfnamefont {H.}~\bibnamefont {Krieg}},\ }\bibfield  {title}
  {\enquote {\bibinfo {title} {A consistent and accurate ab initio
  parametrization of density functional dispersion correction (dft-d) for the
  94 elements h-pu},}\ }\href {https://doi.org/10.1063/1.3382344} {\bibfield
  {journal} {\bibinfo  {journal} {The Journal of Chemical Physics}\ }\textbf
  {\bibinfo {volume} {132}},\ \bibinfo {pages} {154104} (\bibinfo {year}
  {2010})}\BibitemShut {NoStop}%
\bibitem [{\citenamefont {Gillan}, \citenamefont {Alfè},\ and\ \citenamefont
  {Michaelides}(2016)}]{angelos_dft_water_2016}%
  \BibitemOpen
  \bibfield  {author} {\bibinfo {author} {\bibfnamefont {M.~J.}\ \bibnamefont
  {Gillan}}, \bibinfo {author} {\bibfnamefont {D.}~\bibnamefont {Alfè}},\ and\
  \bibinfo {author} {\bibfnamefont {A.}~\bibnamefont {Michaelides}},\
  }\bibfield  {title} {\enquote {\bibinfo {title} {Perspective: How good is dft
  for water?}}\ }\href {https://doi.org/10.1063/1.4944633} {\bibfield
  {journal} {\bibinfo  {journal} {The Journal of Chemical Physics}\ }\textbf
  {\bibinfo {volume} {144}},\ \bibinfo {pages} {130901} (\bibinfo {year}
  {2016})}\BibitemShut {NoStop}%
\bibitem [{\citenamefont {Morawietz}\ \emph {et~al.}(2016)\citenamefont
  {Morawietz}, \citenamefont {Singraber}, \citenamefont {Dellago},\ and\
  \citenamefont {Behler}}]{tobias_vdw_2016}%
  \BibitemOpen
  \bibfield  {author} {\bibinfo {author} {\bibfnamefont {T.}~\bibnamefont
  {Morawietz}}, \bibinfo {author} {\bibfnamefont {A.}~\bibnamefont
  {Singraber}}, \bibinfo {author} {\bibfnamefont {C.}~\bibnamefont {Dellago}},\
  and\ \bibinfo {author} {\bibfnamefont {J.}~\bibnamefont {Behler}},\
  }\bibfield  {title} {\enquote {\bibinfo {title} {How van der waals
  interactions determine the unique properties of water},}\ }\href
  {https://doi.org/10.1073/pnas.1602375113} {\bibfield  {journal} {\bibinfo
  {journal} {Proceedings of the National Academy of Sciences}\ }\textbf
  {\bibinfo {volume} {113}},\ \bibinfo {pages} {8368--8373} (\bibinfo {year}
  {2016})}\BibitemShut {NoStop}%
\bibitem [{\citenamefont {Marsalek}\ and\ \citenamefont
  {Markland}(2017)}]{ondrej_revpbe_2017}%
  \BibitemOpen
  \bibfield  {author} {\bibinfo {author} {\bibfnamefont {O.}~\bibnamefont
  {Marsalek}}\ and\ \bibinfo {author} {\bibfnamefont {T.~E.}\ \bibnamefont
  {Markland}},\ }\bibfield  {title} {\enquote {\bibinfo {title} {Quantum
  dynamics and spectroscopy of ab initio liquid water: The interplay of
  nuclear and electronic quantum effects},}\ }\href
  {https://doi.org/10.1021/acs.jpclett.7b00391} {\bibfield  {journal} {\bibinfo
   {journal} {The Journal of Physical Chemistry Letters}\ }\textbf {\bibinfo
  {volume} {8}},\ \bibinfo {pages} {1545--1551} (\bibinfo {year}
  {2017})}\BibitemShut {NoStop}%
\bibitem [{\citenamefont {Brandenburg}\ \emph {et~al.}(2019)\citenamefont
  {Brandenburg}, \citenamefont {Zen}, \citenamefont {Alfè},\ and\
  \citenamefont {Michaelides}}]{Brandenburg2019}%
  \BibitemOpen
  \bibfield  {author} {\bibinfo {author} {\bibfnamefont {J.~G.}\ \bibnamefont
  {Brandenburg}}, \bibinfo {author} {\bibfnamefont {A.}~\bibnamefont {Zen}},
  \bibinfo {author} {\bibfnamefont {D.}~\bibnamefont {Alfè}},\ and\ \bibinfo
  {author} {\bibfnamefont {A.}~\bibnamefont {Michaelides}},\ }\bibfield
  {title} {\enquote {\bibinfo {title} {Interaction between water and carbon
  nanostructures: How good are current density functional approximations?}}\
  }\href {https://doi.org/10.1063/1.5121370} {\bibfield  {journal} {\bibinfo
  {journal} {The Journal of Chemical Physics}\ }\textbf {\bibinfo {volume}
  {151}},\ \bibinfo {pages} {164702} (\bibinfo {year} {2019})}\BibitemShut
  {NoStop}%
\bibitem [{\citenamefont {Atsango}\ \emph {et~al.}(2023)\citenamefont
  {Atsango}, \citenamefont {Morawietz}, \citenamefont {Marsalek},\ and\
  \citenamefont {Markland}}]{marsalek_mlp_pt}%
  \BibitemOpen
  \bibfield  {author} {\bibinfo {author} {\bibfnamefont {A.~O.}\ \bibnamefont
  {Atsango}}, \bibinfo {author} {\bibfnamefont {T.}~\bibnamefont {Morawietz}},
  \bibinfo {author} {\bibfnamefont {O.}~\bibnamefont {Marsalek}},\ and\
  \bibinfo {author} {\bibfnamefont {T.~E.}\ \bibnamefont {Markland}},\
  }\bibfield  {title} {\enquote {\bibinfo {title} {Developing machine-learned
  potentials to simultaneously capture the dynamics of excess protons and
  hydroxide ions in classical and path integral simulations},}\ }\href
  {https://doi.org/10.1063/5.0162066} {\bibfield  {journal} {\bibinfo
  {journal} {The Journal of Chemical Physics}\ }\textbf {\bibinfo {volume}
  {159}},\ \bibinfo {pages} {074101} (\bibinfo {year} {2023})}\BibitemShut
  {NoStop}%
\bibitem [{\citenamefont {Maccarini}\ \emph {et~al.}(2007)\citenamefont
  {Maccarini}, \citenamefont {Steitz}, \citenamefont {Himmelhaus},
  \citenamefont {Fick}, \citenamefont {Tatur}, \citenamefont {Wolff},
  \citenamefont {Grunze}, \citenamefont {Janeček},\ and\ \citenamefont
  {Netz}}]{maccarini_density_2007}%
  \BibitemOpen
  \bibfield  {author} {\bibinfo {author} {\bibfnamefont {M.}~\bibnamefont
  {Maccarini}}, \bibinfo {author} {\bibfnamefont {R.}~\bibnamefont {Steitz}},
  \bibinfo {author} {\bibfnamefont {M.}~\bibnamefont {Himmelhaus}}, \bibinfo
  {author} {\bibfnamefont {J.}~\bibnamefont {Fick}}, \bibinfo {author}
  {\bibfnamefont {S.}~\bibnamefont {Tatur}}, \bibinfo {author} {\bibfnamefont
  {M.}~\bibnamefont {Wolff}}, \bibinfo {author} {\bibfnamefont
  {M.}~\bibnamefont {Grunze}}, \bibinfo {author} {\bibfnamefont
  {J.}~\bibnamefont {Janeček}},\ and\ \bibinfo {author} {\bibfnamefont
  {R.~R.}\ \bibnamefont {Netz}},\ }\bibfield  {title} {\enquote {\bibinfo
  {title} {Density depletion at solid-liquid interfaces: a neutron reflectivity
  study},}\ }\href {https://doi.org/10.1021/la061943y} {\bibfield  {journal}
  {\bibinfo  {journal} {Langmuir}\ }\textbf {\bibinfo {volume} {23}},\ \bibinfo
  {pages} {598--608} (\bibinfo {year} {2007})},\ \bibinfo {note} {publisher:
  American Chemical Society}\BibitemShut {NoStop}%
\bibitem [{\citenamefont {Cicero}\ \emph {et~al.}(2008)\citenamefont {Cicero},
  \citenamefont {Grossman}, \citenamefont {Schwegler}, \citenamefont {Gygi},\
  and\ \citenamefont {Galli}}]{galli_canonical_2008}%
  \BibitemOpen
  \bibfield  {author} {\bibinfo {author} {\bibfnamefont {G.}~\bibnamefont
  {Cicero}}, \bibinfo {author} {\bibfnamefont {J.~C.}\ \bibnamefont
  {Grossman}}, \bibinfo {author} {\bibfnamefont {E.}~\bibnamefont {Schwegler}},
  \bibinfo {author} {\bibfnamefont {F.}~\bibnamefont {Gygi}},\ and\ \bibinfo
  {author} {\bibfnamefont {G.}~\bibnamefont {Galli}},\ }\bibfield  {title}
  {\enquote {\bibinfo {title} {Water confined in nanotubes and between graphene
  sheets: A first principle study},}\ }\href
  {https://doi.org/10.1021/ja074418+} {\bibfield  {journal} {\bibinfo
  {journal} {Journal of the American Chemical Society}\ }\textbf {\bibinfo
  {volume} {130}},\ \bibinfo {pages} {1871--1878} (\bibinfo {year}
  {2008})}\BibitemShut {NoStop}%
\bibitem [{\citenamefont {Tocci}, \citenamefont {Joly},\ and\ \citenamefont
  {Michaelides}(2014)}]{tocci_friction}%
  \BibitemOpen
  \bibfield  {author} {\bibinfo {author} {\bibfnamefont {G.}~\bibnamefont
  {Tocci}}, \bibinfo {author} {\bibfnamefont {L.}~\bibnamefont {Joly}},\ and\
  \bibinfo {author} {\bibfnamefont {A.}~\bibnamefont {Michaelides}},\
  }\bibfield  {title} {\enquote {\bibinfo {title} {Friction of water on
  graphene and hexagonal boron nitride from ab initio methods: Very different
  slippage despite very similar interface structures},}\ }\href
  {https://doi.org/10.1021/nl502837d} {\bibfield  {journal} {\bibinfo
  {journal} {Nano Letters}\ }\textbf {\bibinfo {volume} {14}},\ \bibinfo
  {pages} {6872--6877} (\bibinfo {year} {2014})}\BibitemShut {NoStop}%
\bibitem [{\citenamefont {Singla}\ \emph {et~al.}(2017)\citenamefont {Singla},
  \citenamefont {Anim-Danso}, \citenamefont {Islam}, \citenamefont {Ngo},
  \citenamefont {Kim}, \citenamefont {Naik},\ and\ \citenamefont
  {Dhinojwala}}]{singla_insight_2017}%
  \BibitemOpen
  \bibfield  {author} {\bibinfo {author} {\bibfnamefont {S.}~\bibnamefont
  {Singla}}, \bibinfo {author} {\bibfnamefont {E.}~\bibnamefont {Anim-Danso}},
  \bibinfo {author} {\bibfnamefont {A.~E.}\ \bibnamefont {Islam}}, \bibinfo
  {author} {\bibfnamefont {Y.}~\bibnamefont {Ngo}}, \bibinfo {author}
  {\bibfnamefont {S.~S.}\ \bibnamefont {Kim}}, \bibinfo {author} {\bibfnamefont
  {R.~R.}\ \bibnamefont {Naik}},\ and\ \bibinfo {author} {\bibfnamefont
  {A.}~\bibnamefont {Dhinojwala}},\ }\bibfield  {title} {\enquote {\bibinfo
  {title} {Insight on structure of water and ice next to graphene using
  surface-sensitive spectroscopy},}\ }\href
  {https://doi.org/10.1021/acsnano.7b01499} {\bibfield  {journal} {\bibinfo
  {journal} {ACS Nano}\ }\textbf {\bibinfo {volume} {11}},\ \bibinfo {pages}
  {4899--4906} (\bibinfo {year} {2017})}\BibitemShut {NoStop}%
\bibitem [{\citenamefont {Ruiz-Barragan}, \citenamefont {Muñoz-Santiburcio},\
  and\ \citenamefont {Marx}(2019)}]{sergi_nanoconf}%
  \BibitemOpen
  \bibfield  {author} {\bibinfo {author} {\bibfnamefont {S.}~\bibnamefont
  {Ruiz-Barragan}}, \bibinfo {author} {\bibfnamefont {D.}~\bibnamefont
  {Muñoz-Santiburcio}},\ and\ \bibinfo {author} {\bibfnamefont
  {D.}~\bibnamefont {Marx}},\ }\bibfield  {title} {\enquote {\bibinfo {title}
  {Nanoconfined water within graphene slit pores adopts distinct
  confinement-dependent regimes},}\ }\href
  {https://doi.org/10.1021/acs.jpclett.8b03530} {\bibfield  {journal} {\bibinfo
   {journal} {The Journal of Physical Chemistry Letters}\ }\textbf {\bibinfo
  {volume} {10}},\ \bibinfo {pages} {329--334} (\bibinfo {year}
  {2019})}\BibitemShut {NoStop}%
\bibitem [{\citenamefont {Luzar}\ and\ \citenamefont
  {Chandler}(1996)}]{luzar_chandler}%
  \BibitemOpen
  \bibfield  {author} {\bibinfo {author} {\bibfnamefont {A.}~\bibnamefont
  {Luzar}}\ and\ \bibinfo {author} {\bibfnamefont {D.}~\bibnamefont
  {Chandler}},\ }\bibfield  {title} {\enquote {\bibinfo {title} {Hydrogen-bond
  kinetics in liquid water},}\ }\href {https://doi.org/10.1038/379055a0}
  {\bibfield  {journal} {\bibinfo  {journal} {Nature}\ }\textbf {\bibinfo
  {volume} {379}},\ \bibinfo {pages} {55--57} (\bibinfo {year}
  {1996})}\BibitemShut {NoStop}%
\bibitem [{\citenamefont {Chiang}, \citenamefont {Dalstein},\ and\
  \citenamefont {Wen}(2020)}]{chiang_2020}%
  \BibitemOpen
  \bibfield  {author} {\bibinfo {author} {\bibfnamefont {K.-Y.}\ \bibnamefont
  {Chiang}}, \bibinfo {author} {\bibfnamefont {L.}~\bibnamefont {Dalstein}},\
  and\ \bibinfo {author} {\bibfnamefont {Y.-C.}\ \bibnamefont {Wen}},\
  }\bibfield  {title} {\enquote {\bibinfo {title} {Affinity of hydrated protons
  at intrinsic water/vapor interface revealed by ion-induced water
  alignment},}\ }\href {https://doi.org/10.1021/acs.jpclett.9b03520} {\bibfield
   {journal} {\bibinfo  {journal} {The Journal of Physical Chemistry Letters}\
  }\textbf {\bibinfo {volume} {11}},\ \bibinfo {pages} {696--701} (\bibinfo
  {year} {2020})}\BibitemShut {NoStop}%
\bibitem [{\citenamefont {Otten}\ \emph {et~al.}(2012)\citenamefont {Otten},
  \citenamefont {Shaffer}, \citenamefont {Geissler},\ and\ \citenamefont
  {Saykally}}]{otten_2012}%
  \BibitemOpen
  \bibfield  {author} {\bibinfo {author} {\bibfnamefont {D.~E.}\ \bibnamefont
  {Otten}}, \bibinfo {author} {\bibfnamefont {P.~R.}\ \bibnamefont {Shaffer}},
  \bibinfo {author} {\bibfnamefont {P.~L.}\ \bibnamefont {Geissler}},\ and\
  \bibinfo {author} {\bibfnamefont {R.~J.}\ \bibnamefont {Saykally}},\
  }\bibfield  {title} {\enquote {\bibinfo {title} {Elucidating the mechanism of
  selective ion adsorption to the liquid water surface},}\ }\href
  {https://doi.org/10.1073/pnas.1116169109} {\bibfield  {journal} {\bibinfo
  {journal} {Proceedings of the National Academy of Sciences}\ }\textbf
  {\bibinfo {volume} {109}},\ \bibinfo {pages} {701--705} (\bibinfo {year}
  {2012})}\BibitemShut {NoStop}%
\bibitem [{\citenamefont {Petersen}\ and\ \citenamefont
  {Saykally}(2008)}]{PETERSEN2008255_2008}%
  \BibitemOpen
  \bibfield  {author} {\bibinfo {author} {\bibfnamefont {P.~B.}\ \bibnamefont
  {Petersen}}\ and\ \bibinfo {author} {\bibfnamefont {R.~J.}\ \bibnamefont
  {Saykally}},\ }\bibfield  {title} {\enquote {\bibinfo {title} {Is the liquid
  water surface basic or acidic? macroscopic vs. molecular-scale
  investigations},}\ }\href
  {https://doi.org/https://doi.org/10.1016/j.cplett.2008.04.010} {\bibfield
  {journal} {\bibinfo  {journal} {Chemical Physics Letters}\ }\textbf {\bibinfo
  {volume} {458}},\ \bibinfo {pages} {255--261} (\bibinfo {year}
  {2008})}\BibitemShut {NoStop}%
\bibitem [{\citenamefont {Petersen}\ and\ \citenamefont
  {Saykally}(2005)}]{saykally_hydron_2005}%
  \BibitemOpen
  \bibfield  {author} {\bibinfo {author} {\bibfnamefont {P.~B.}\ \bibnamefont
  {Petersen}}\ and\ \bibinfo {author} {\bibfnamefont {R.~J.}\ \bibnamefont
  {Saykally}},\ }\bibfield  {title} {\enquote {\bibinfo {title} {Evidence for
  an enhanced hydronium concentration at the liquid water surface},}\ }\href
  {https://doi.org/10.1021/JP044479J} {\bibfield  {journal} {\bibinfo
  {journal} {Journal of Physical Chemistry B}\ }\textbf {\bibinfo {volume}
  {109}},\ \bibinfo {pages} {7976--7980} (\bibinfo {year} {2005})}\BibitemShut
  {NoStop}%
\bibitem [{\citenamefont {Wang}\ \emph {et~al.}(2024)\citenamefont {Wang},
  \citenamefont {Tang}, \citenamefont {Yu}, \citenamefont {Ohto}, \citenamefont
  {Nagata},\ and\ \citenamefont {Bonn}}]{mischa_transp}%
  \BibitemOpen
  \bibfield  {author} {\bibinfo {author} {\bibfnamefont {Y.}~\bibnamefont
  {Wang}}, \bibinfo {author} {\bibfnamefont {F.}~\bibnamefont {Tang}}, \bibinfo
  {author} {\bibfnamefont {X.}~\bibnamefont {Yu}}, \bibinfo {author}
  {\bibfnamefont {T.}~\bibnamefont {Ohto}}, \bibinfo {author} {\bibfnamefont
  {Y.}~\bibnamefont {Nagata}},\ and\ \bibinfo {author} {\bibfnamefont
  {M.}~\bibnamefont {Bonn}},\ }\bibfield  {title} {\enquote {\bibinfo {title}
  {Heterodyne-detected sum-frequency generation vibrational spectroscopy
  reveals aqueous molecular structure at the suspended graphene/water
  interface},}\ }\href {https://doi.org/https://doi.org/10.1002/anie.202319503}
  {\bibfield  {journal} {\bibinfo  {journal} {Angewandte Chemie International
  Edition}\ }\textbf {\bibinfo {volume} {n/a}},\ \bibinfo {pages} {e202319503}
  (\bibinfo {year} {2024})}\BibitemShut {NoStop}%
\bibitem [{\citenamefont {Lyu}\ \emph {et~al.}(2024)\citenamefont {Lyu},
  \citenamefont {Märker}, \citenamefont {Zhou}, \citenamefont {Zhao},
  \citenamefont {Gunnarsdóttir}, \citenamefont {Niblett}, \citenamefont
  {Forse},\ and\ \citenamefont {Grey}}]{dongxun_jacs_2024}%
  \BibitemOpen
  \bibfield  {author} {\bibinfo {author} {\bibfnamefont {D.}~\bibnamefont
  {Lyu}}, \bibinfo {author} {\bibfnamefont {K.}~\bibnamefont {Märker}},
  \bibinfo {author} {\bibfnamefont {Y.}~\bibnamefont {Zhou}}, \bibinfo {author}
  {\bibfnamefont {E.~W.}\ \bibnamefont {Zhao}}, \bibinfo {author}
  {\bibfnamefont {A.~B.}\ \bibnamefont {Gunnarsdóttir}}, \bibinfo {author}
  {\bibfnamefont {S.~P.}\ \bibnamefont {Niblett}}, \bibinfo {author}
  {\bibfnamefont {A.~C.}\ \bibnamefont {Forse}},\ and\ \bibinfo {author}
  {\bibfnamefont {C.~P.}\ \bibnamefont {Grey}},\ }\bibfield  {title} {\enquote
  {\bibinfo {title} {Understanding sorption of aqueous electrolytes in porous
  carbon by nmr spectroscopy},}\ }\href {https://doi.org/10.1021/jacs.3c14807}
  {\bibfield  {journal} {\bibinfo  {journal} {Journal of the American Chemical
  Society}\ }\textbf {\bibinfo {volume} {146}},\ \bibinfo {pages} {9897--9910}
  (\bibinfo {year} {2024})}\BibitemShut {NoStop}%
\bibitem [{\citenamefont {Wu}\ \emph {et~al.}(2024)\citenamefont {Wu},
  \citenamefont {Zhao}, \citenamefont {Lin}, \citenamefont {Song},
  \citenamefont {Qi}, \citenamefont {Jiang}, \citenamefont {Yuan},
  \citenamefont {Cheng}, \citenamefont {Zhao}, \citenamefont {Tian},
  \citenamefont {Wang}, \citenamefont {Wu}, \citenamefont {Bian}, \citenamefont
  {Liu}, \citenamefont {Xu}, \citenamefont {Zeng}, \citenamefont {Wang},\ and\
  \citenamefont {Jiang}}]{science_jiang_2024}%
  \BibitemOpen
  \bibfield  {author} {\bibinfo {author} {\bibfnamefont {D.}~\bibnamefont
  {Wu}}, \bibinfo {author} {\bibfnamefont {Z.}~\bibnamefont {Zhao}}, \bibinfo
  {author} {\bibfnamefont {B.}~\bibnamefont {Lin}}, \bibinfo {author}
  {\bibfnamefont {Y.}~\bibnamefont {Song}}, \bibinfo {author} {\bibfnamefont
  {J.}~\bibnamefont {Qi}}, \bibinfo {author} {\bibfnamefont {J.}~\bibnamefont
  {Jiang}}, \bibinfo {author} {\bibfnamefont {Z.}~\bibnamefont {Yuan}},
  \bibinfo {author} {\bibfnamefont {B.}~\bibnamefont {Cheng}}, \bibinfo
  {author} {\bibfnamefont {M.}~\bibnamefont {Zhao}}, \bibinfo {author}
  {\bibfnamefont {Y.}~\bibnamefont {Tian}}, \bibinfo {author} {\bibfnamefont
  {Z.}~\bibnamefont {Wang}}, \bibinfo {author} {\bibfnamefont {M.}~\bibnamefont
  {Wu}}, \bibinfo {author} {\bibfnamefont {K.}~\bibnamefont {Bian}}, \bibinfo
  {author} {\bibfnamefont {K.-H.}\ \bibnamefont {Liu}}, \bibinfo {author}
  {\bibfnamefont {L.-M.}\ \bibnamefont {Xu}}, \bibinfo {author} {\bibfnamefont
  {X.~C.}\ \bibnamefont {Zeng}}, \bibinfo {author} {\bibfnamefont {E.-G.}\
  \bibnamefont {Wang}},\ and\ \bibinfo {author} {\bibfnamefont
  {Y.}~\bibnamefont {Jiang}},\ }\bibfield  {title} {\enquote {\bibinfo {title}
  {Probing structural superlubricity of two-dimensional water transport with
  atomic resolution},}\ }\href {https://doi.org/10.1126/science.ado1544}
  {\bibfield  {journal} {\bibinfo  {journal} {Science}\ }\textbf {\bibinfo
  {volume} {384}},\ \bibinfo {pages} {1254--1259} (\bibinfo {year}
  {2024})}\BibitemShut {NoStop}%
\bibitem [{\citenamefont {Batatia}\ \emph {et~al.}(2024)\citenamefont
  {Batatia}, \citenamefont {Benner}, \citenamefont {Chiang}, \citenamefont
  {Elena}, \citenamefont {Kovács}, \citenamefont {Riebesell}, \citenamefont
  {Advincula}, \citenamefont {Asta}, \citenamefont {Avaylon}, \citenamefont
  {Baldwin}, \citenamefont {Berger}, \citenamefont {Bernstein}, \citenamefont
  {Bhowmik}, \citenamefont {Blau}, \citenamefont {Cărare}, \citenamefont
  {Darby}, \citenamefont {De}, \citenamefont {Pia}, \citenamefont {Deringer},
  \citenamefont {Elijošius}, \citenamefont {El-Machachi}, \citenamefont
  {Falcioni}, \citenamefont {Fako}, \citenamefont {Ferrari}, \citenamefont
  {Genreith-Schriever}, \citenamefont {George}, \citenamefont {Goodall},
  \citenamefont {Grey}, \citenamefont {Grigorev}, \citenamefont {Han},
  \citenamefont {Handley}, \citenamefont {Heenen}, \citenamefont {Hermansson},
  \citenamefont {Holm}, \citenamefont {Jaafar}, \citenamefont {Hofmann},
  \citenamefont {Jakob}, \citenamefont {Jung}, \citenamefont {Kapil},
  \citenamefont {Kaplan}, \citenamefont {Karimitari}, \citenamefont {Kermode},
  \citenamefont {Kroupa}, \citenamefont {Kullgren}, \citenamefont {Kuner},
  \citenamefont {Kuryla}, \citenamefont {Liepuoniute}, \citenamefont {Margraf},
  \citenamefont {Magdău}, \citenamefont {Michaelides}, \citenamefont {Moore},
  \citenamefont {Naik}, \citenamefont {Niblett}, \citenamefont {Norwood},
  \citenamefont {O'Neill}, \citenamefont {Ortner}, \citenamefont {Persson},
  \citenamefont {Reuter}, \citenamefont {Rosen}, \citenamefont {Schaaf},
  \citenamefont {Schran}, \citenamefont {Shi}, \citenamefont {Sivonxay},
  \citenamefont {Stenczel}, \citenamefont {Svahn}, \citenamefont {Sutton},
  \citenamefont {Swinburne}, \citenamefont {Tilly}, \citenamefont {van~der
  Oord}, \citenamefont {Varga-Umbrich}, \citenamefont {Vegge}, \citenamefont
  {Vondrák}, \citenamefont {Wang}, \citenamefont {Witt}, \citenamefont
  {Zills},\ and\ \citenamefont {Csányi}}]{mace_mp_0}%
  \BibitemOpen
  \bibfield  {author} {\bibinfo {author} {\bibfnamefont {I.}~\bibnamefont
  {Batatia}}, \bibinfo {author} {\bibfnamefont {P.}~\bibnamefont {Benner}},
  \bibinfo {author} {\bibfnamefont {Y.}~\bibnamefont {Chiang}}, \bibinfo
  {author} {\bibfnamefont {A.~M.}\ \bibnamefont {Elena}}, \bibinfo {author}
  {\bibfnamefont {D.~P.}\ \bibnamefont {Kovács}}, \bibinfo {author}
  {\bibfnamefont {J.}~\bibnamefont {Riebesell}}, \bibinfo {author}
  {\bibfnamefont {X.~R.}\ \bibnamefont {Advincula}}, \bibinfo {author}
  {\bibfnamefont {M.}~\bibnamefont {Asta}}, \bibinfo {author} {\bibfnamefont
  {M.}~\bibnamefont {Avaylon}}, \bibinfo {author} {\bibfnamefont {W.~J.}\
  \bibnamefont {Baldwin}}, \bibinfo {author} {\bibfnamefont {F.}~\bibnamefont
  {Berger}}, \bibinfo {author} {\bibfnamefont {N.}~\bibnamefont {Bernstein}},
  \bibinfo {author} {\bibfnamefont {A.}~\bibnamefont {Bhowmik}}, \bibinfo
  {author} {\bibfnamefont {S.~M.}\ \bibnamefont {Blau}}, \bibinfo {author}
  {\bibfnamefont {V.}~\bibnamefont {Cărare}}, \bibinfo {author} {\bibfnamefont
  {J.~P.}\ \bibnamefont {Darby}}, \bibinfo {author} {\bibfnamefont
  {S.}~\bibnamefont {De}}, \bibinfo {author} {\bibfnamefont {F.~D.}\
  \bibnamefont {Pia}}, \bibinfo {author} {\bibfnamefont {V.~L.}\ \bibnamefont
  {Deringer}}, \bibinfo {author} {\bibfnamefont {R.}~\bibnamefont
  {Elijošius}}, \bibinfo {author} {\bibfnamefont {Z.}~\bibnamefont
  {El-Machachi}}, \bibinfo {author} {\bibfnamefont {F.}~\bibnamefont
  {Falcioni}}, \bibinfo {author} {\bibfnamefont {E.}~\bibnamefont {Fako}},
  \bibinfo {author} {\bibfnamefont {A.~C.}\ \bibnamefont {Ferrari}}, \bibinfo
  {author} {\bibfnamefont {A.}~\bibnamefont {Genreith-Schriever}}, \bibinfo
  {author} {\bibfnamefont {J.}~\bibnamefont {George}}, \bibinfo {author}
  {\bibfnamefont {R.~E.~A.}\ \bibnamefont {Goodall}}, \bibinfo {author}
  {\bibfnamefont {C.~P.}\ \bibnamefont {Grey}}, \bibinfo {author}
  {\bibfnamefont {P.}~\bibnamefont {Grigorev}}, \bibinfo {author}
  {\bibfnamefont {S.}~\bibnamefont {Han}}, \bibinfo {author} {\bibfnamefont
  {W.}~\bibnamefont {Handley}}, \bibinfo {author} {\bibfnamefont {H.~H.}\
  \bibnamefont {Heenen}}, \bibinfo {author} {\bibfnamefont {K.}~\bibnamefont
  {Hermansson}}, \bibinfo {author} {\bibfnamefont {C.}~\bibnamefont {Holm}},
  \bibinfo {author} {\bibfnamefont {J.}~\bibnamefont {Jaafar}}, \bibinfo
  {author} {\bibfnamefont {S.}~\bibnamefont {Hofmann}}, \bibinfo {author}
  {\bibfnamefont {K.~S.}\ \bibnamefont {Jakob}}, \bibinfo {author}
  {\bibfnamefont {H.}~\bibnamefont {Jung}}, \bibinfo {author} {\bibfnamefont
  {V.}~\bibnamefont {Kapil}}, \bibinfo {author} {\bibfnamefont {A.~D.}\
  \bibnamefont {Kaplan}}, \bibinfo {author} {\bibfnamefont {N.}~\bibnamefont
  {Karimitari}}, \bibinfo {author} {\bibfnamefont {J.~R.}\ \bibnamefont
  {Kermode}}, \bibinfo {author} {\bibfnamefont {N.}~\bibnamefont {Kroupa}},
  \bibinfo {author} {\bibfnamefont {J.}~\bibnamefont {Kullgren}}, \bibinfo
  {author} {\bibfnamefont {M.~C.}\ \bibnamefont {Kuner}}, \bibinfo {author}
  {\bibfnamefont {D.}~\bibnamefont {Kuryla}}, \bibinfo {author} {\bibfnamefont
  {G.}~\bibnamefont {Liepuoniute}}, \bibinfo {author} {\bibfnamefont {J.~T.}\
  \bibnamefont {Margraf}}, \bibinfo {author} {\bibfnamefont {I.-B.}\
  \bibnamefont {Magdău}}, \bibinfo {author} {\bibfnamefont {A.}~\bibnamefont
  {Michaelides}}, \bibinfo {author} {\bibfnamefont {J.~H.}\ \bibnamefont
  {Moore}}, \bibinfo {author} {\bibfnamefont {A.~A.}\ \bibnamefont {Naik}},
  \bibinfo {author} {\bibfnamefont {S.~P.}\ \bibnamefont {Niblett}}, \bibinfo
  {author} {\bibfnamefont {S.~W.}\ \bibnamefont {Norwood}}, \bibinfo {author}
  {\bibfnamefont {N.}~\bibnamefont {O'Neill}}, \bibinfo {author} {\bibfnamefont
  {C.}~\bibnamefont {Ortner}}, \bibinfo {author} {\bibfnamefont {K.~A.}\
  \bibnamefont {Persson}}, \bibinfo {author} {\bibfnamefont {K.}~\bibnamefont
  {Reuter}}, \bibinfo {author} {\bibfnamefont {A.~S.}\ \bibnamefont {Rosen}},
  \bibinfo {author} {\bibfnamefont {L.~L.}\ \bibnamefont {Schaaf}}, \bibinfo
  {author} {\bibfnamefont {C.}~\bibnamefont {Schran}}, \bibinfo {author}
  {\bibfnamefont {B.~X.}\ \bibnamefont {Shi}}, \bibinfo {author} {\bibfnamefont
  {E.}~\bibnamefont {Sivonxay}}, \bibinfo {author} {\bibfnamefont {T.~K.}\
  \bibnamefont {Stenczel}}, \bibinfo {author} {\bibfnamefont {V.}~\bibnamefont
  {Svahn}}, \bibinfo {author} {\bibfnamefont {C.}~\bibnamefont {Sutton}},
  \bibinfo {author} {\bibfnamefont {T.~D.}\ \bibnamefont {Swinburne}}, \bibinfo
  {author} {\bibfnamefont {J.}~\bibnamefont {Tilly}}, \bibinfo {author}
  {\bibfnamefont {C.}~\bibnamefont {van~der Oord}}, \bibinfo {author}
  {\bibfnamefont {E.}~\bibnamefont {Varga-Umbrich}}, \bibinfo {author}
  {\bibfnamefont {T.}~\bibnamefont {Vegge}}, \bibinfo {author} {\bibfnamefont
  {M.}~\bibnamefont {Vondrák}}, \bibinfo {author} {\bibfnamefont
  {Y.}~\bibnamefont {Wang}}, \bibinfo {author} {\bibfnamefont {W.~C.}\
  \bibnamefont {Witt}}, \bibinfo {author} {\bibfnamefont {F.}~\bibnamefont
  {Zills}},\ and\ \bibinfo {author} {\bibfnamefont {G.}~\bibnamefont
  {Csányi}},\ }\href@noop {} {\enquote {\bibinfo {title} {A foundation model
  for atomistic materials chemistry},}\ } (\bibinfo {year} {2024}),\ \bibinfo
  {note} {{Preprint at \url{ https://arxiv.org/abs/2401.00096}}}\BibitemShut
  {NoStop}%
\bibitem [{\citenamefont {Kühne}\ \emph {et~al.}(2020)\citenamefont {Kühne},
  \citenamefont {Iannuzzi}, \citenamefont {Ben}, \citenamefont {Rybkin},
  \citenamefont {Seewald}, \citenamefont {Stein}, \citenamefont {Laino},
  \citenamefont {Khaliullin}, \citenamefont {Schütt}, \citenamefont
  {Schiffmann}, \citenamefont {Golze}, \citenamefont {Wilhelm}, \citenamefont
  {Chulkov}, \citenamefont {Bani-Hashemian}, \citenamefont {Weber},
  \citenamefont {Borštnik}, \citenamefont {Taillefumier}, \citenamefont
  {Jakobovits}, \citenamefont {Lazzaro}, \citenamefont {Pabst}, \citenamefont
  {Müller}, \citenamefont {Schade}, \citenamefont {Guidon}, \citenamefont
  {Andermatt}, \citenamefont {Holmberg}, \citenamefont {Schenter},
  \citenamefont {Hehn}, \citenamefont {Bussy}, \citenamefont {Belleflamme},
  \citenamefont {Tabacchi}, \citenamefont {Glöß}, \citenamefont {Lass},
  \citenamefont {Bethune}, \citenamefont {Mundy}, \citenamefont {Plessl},
  \citenamefont {Watkins}, \citenamefont {VandeVondele}, \citenamefont
  {Krack},\ and\ \citenamefont {Hutter}}]{cp2k_2020}%
  \BibitemOpen
  \bibfield  {author} {\bibinfo {author} {\bibfnamefont {T.~D.}\ \bibnamefont
  {Kühne}}, \bibinfo {author} {\bibfnamefont {M.}~\bibnamefont {Iannuzzi}},
  \bibinfo {author} {\bibfnamefont {M.~D.}\ \bibnamefont {Ben}}, \bibinfo
  {author} {\bibfnamefont {V.~V.}\ \bibnamefont {Rybkin}}, \bibinfo {author}
  {\bibfnamefont {P.}~\bibnamefont {Seewald}}, \bibinfo {author} {\bibfnamefont
  {F.}~\bibnamefont {Stein}}, \bibinfo {author} {\bibfnamefont
  {T.}~\bibnamefont {Laino}}, \bibinfo {author} {\bibfnamefont {R.~Z.}\
  \bibnamefont {Khaliullin}}, \bibinfo {author} {\bibfnamefont
  {O.}~\bibnamefont {Schütt}}, \bibinfo {author} {\bibfnamefont
  {F.}~\bibnamefont {Schiffmann}}, \bibinfo {author} {\bibfnamefont
  {D.}~\bibnamefont {Golze}}, \bibinfo {author} {\bibfnamefont
  {J.}~\bibnamefont {Wilhelm}}, \bibinfo {author} {\bibfnamefont
  {S.}~\bibnamefont {Chulkov}}, \bibinfo {author} {\bibfnamefont {M.~H.}\
  \bibnamefont {Bani-Hashemian}}, \bibinfo {author} {\bibfnamefont
  {V.}~\bibnamefont {Weber}}, \bibinfo {author} {\bibfnamefont
  {U.}~\bibnamefont {Borštnik}}, \bibinfo {author} {\bibfnamefont
  {M.}~\bibnamefont {Taillefumier}}, \bibinfo {author} {\bibfnamefont {A.~S.}\
  \bibnamefont {Jakobovits}}, \bibinfo {author} {\bibfnamefont
  {A.}~\bibnamefont {Lazzaro}}, \bibinfo {author} {\bibfnamefont
  {H.}~\bibnamefont {Pabst}}, \bibinfo {author} {\bibfnamefont
  {T.}~\bibnamefont {Müller}}, \bibinfo {author} {\bibfnamefont
  {R.}~\bibnamefont {Schade}}, \bibinfo {author} {\bibfnamefont
  {M.}~\bibnamefont {Guidon}}, \bibinfo {author} {\bibfnamefont
  {S.}~\bibnamefont {Andermatt}}, \bibinfo {author} {\bibfnamefont
  {N.}~\bibnamefont {Holmberg}}, \bibinfo {author} {\bibfnamefont {G.~K.}\
  \bibnamefont {Schenter}}, \bibinfo {author} {\bibfnamefont {A.}~\bibnamefont
  {Hehn}}, \bibinfo {author} {\bibfnamefont {A.}~\bibnamefont {Bussy}},
  \bibinfo {author} {\bibfnamefont {F.}~\bibnamefont {Belleflamme}}, \bibinfo
  {author} {\bibfnamefont {G.}~\bibnamefont {Tabacchi}}, \bibinfo {author}
  {\bibfnamefont {A.}~\bibnamefont {Glöß}}, \bibinfo {author} {\bibfnamefont
  {M.}~\bibnamefont {Lass}}, \bibinfo {author} {\bibfnamefont {I.}~\bibnamefont
  {Bethune}}, \bibinfo {author} {\bibfnamefont {C.~J.}\ \bibnamefont {Mundy}},
  \bibinfo {author} {\bibfnamefont {C.}~\bibnamefont {Plessl}}, \bibinfo
  {author} {\bibfnamefont {M.}~\bibnamefont {Watkins}}, \bibinfo {author}
  {\bibfnamefont {J.}~\bibnamefont {VandeVondele}}, \bibinfo {author}
  {\bibfnamefont {M.}~\bibnamefont {Krack}},\ and\ \bibinfo {author}
  {\bibfnamefont {J.}~\bibnamefont {Hutter}},\ }\bibfield  {title} {\enquote
  {\bibinfo {title} {Cp2k: An electronic structure and molecular dynamics
  software package - quickstep: Efficient and accurate electronic structure
  calculations},}\ }\href {https://doi.org/10.1063/5.0007045} {\bibfield
  {journal} {\bibinfo  {journal} {The Journal of Chemical Physics}\ }\textbf
  {\bibinfo {volume} {152}},\ \bibinfo {pages} {194103} (\bibinfo {year}
  {2020})}\BibitemShut {NoStop}%
\bibitem [{\citenamefont {Goedecker}, \citenamefont {Teter},\ and\
  \citenamefont {Hutter}(1996)}]{gth_1996}%
  \BibitemOpen
  \bibfield  {author} {\bibinfo {author} {\bibfnamefont {S.}~\bibnamefont
  {Goedecker}}, \bibinfo {author} {\bibfnamefont {M.}~\bibnamefont {Teter}},\
  and\ \bibinfo {author} {\bibfnamefont {J.}~\bibnamefont {Hutter}},\
  }\bibfield  {title} {\enquote {\bibinfo {title} {Separable dual-space
  gaussian pseudopotentials},}\ }\href
  {https://doi.org/10.1103/PhysRevB.54.1703} {\bibfield  {journal} {\bibinfo
  {journal} {Physical Review B}\ }\textbf {\bibinfo {volume} {54}},\ \bibinfo
  {pages} {1703--1710} (\bibinfo {year} {1996})}\BibitemShut {NoStop}%
\bibitem [{\citenamefont {VandeVondele}\ and\ \citenamefont
  {Hutter}(2007)}]{basis_set_2007}%
  \BibitemOpen
  \bibfield  {author} {\bibinfo {author} {\bibfnamefont {J.}~\bibnamefont
  {VandeVondele}}\ and\ \bibinfo {author} {\bibfnamefont {J.}~\bibnamefont
  {Hutter}},\ }\bibfield  {title} {\enquote {\bibinfo {title} {Gaussian basis
  sets for accurate calculations on molecular systems in gas and condensed
  phases},}\ }\href {https://doi.org/10.1063/1.2770708} {\bibfield  {journal}
  {\bibinfo  {journal} {The Journal of Chemical Physics}\ }\textbf {\bibinfo
  {volume} {127}},\ \bibinfo {pages} {114105} (\bibinfo {year}
  {2007})}\BibitemShut {NoStop}%
\bibitem [{\citenamefont {Schran}, \citenamefont {Brezina},\ and\ \citenamefont
  {Marsalek}(2020)}]{cnnp_2020}%
  \BibitemOpen
  \bibfield  {author} {\bibinfo {author} {\bibfnamefont {C.}~\bibnamefont
  {Schran}}, \bibinfo {author} {\bibfnamefont {K.}~\bibnamefont {Brezina}},\
  and\ \bibinfo {author} {\bibfnamefont {O.}~\bibnamefont {Marsalek}},\
  }\bibfield  {title} {\enquote {\bibinfo {title} {Committee neural network
  potentials control generalization errors and enable active learning},}\
  }\href {https://doi.org/10.1063/5.0016004} {\bibfield  {journal} {\bibinfo
  {journal} {The Journal of Chemical Physics}\ }\textbf {\bibinfo {volume}
  {153}},\ \bibinfo {pages} {104105} (\bibinfo {year} {2020})}\BibitemShut
  {NoStop}%
\bibitem [{\citenamefont {Larsen}\ \emph {et~al.}(2017)\citenamefont {Larsen},
  \citenamefont {Mortensen}, \citenamefont {Blomqvist}, \citenamefont
  {Castelli}, \citenamefont {Christensen}, \citenamefont {Dułak},
  \citenamefont {Friis}, \citenamefont {Groves}, \citenamefont {Hammer},
  \citenamefont {Hargus}, \citenamefont {Hermes}, \citenamefont {Jennings},
  \citenamefont {Jensen}, \citenamefont {Kermode}, \citenamefont {Kitchin},
  \citenamefont {Kolsbjerg}, \citenamefont {Kubal}, \citenamefont {Kaasbjerg},
  \citenamefont {Lysgaard}, \citenamefont {Maronsson}, \citenamefont {Maxson},
  \citenamefont {Olsen}, \citenamefont {Pastewka}, \citenamefont {Peterson},
  \citenamefont {Rostgaard}, \citenamefont {Schiøtz}, \citenamefont {Schütt},
  \citenamefont {Strange}, \citenamefont {Thygesen}, \citenamefont {Vegge},
  \citenamefont {Vilhelmsen}, \citenamefont {Walter}, \citenamefont {Zeng},\
  and\ \citenamefont {Jacobsen}}]{ase_2017}%
  \BibitemOpen
  \bibfield  {author} {\bibinfo {author} {\bibfnamefont {A.~H.}\ \bibnamefont
  {Larsen}}, \bibinfo {author} {\bibfnamefont {J.~J.}\ \bibnamefont
  {Mortensen}}, \bibinfo {author} {\bibfnamefont {J.}~\bibnamefont
  {Blomqvist}}, \bibinfo {author} {\bibfnamefont {I.~E.}\ \bibnamefont
  {Castelli}}, \bibinfo {author} {\bibfnamefont {R.}~\bibnamefont
  {Christensen}}, \bibinfo {author} {\bibfnamefont {M.}~\bibnamefont {Dułak}},
  \bibinfo {author} {\bibfnamefont {J.}~\bibnamefont {Friis}}, \bibinfo
  {author} {\bibfnamefont {M.~N.}\ \bibnamefont {Groves}}, \bibinfo {author}
  {\bibfnamefont {B.}~\bibnamefont {Hammer}}, \bibinfo {author} {\bibfnamefont
  {C.}~\bibnamefont {Hargus}}, \bibinfo {author} {\bibfnamefont {E.~D.}\
  \bibnamefont {Hermes}}, \bibinfo {author} {\bibfnamefont {P.~C.}\
  \bibnamefont {Jennings}}, \bibinfo {author} {\bibfnamefont {P.~B.}\
  \bibnamefont {Jensen}}, \bibinfo {author} {\bibfnamefont {J.}~\bibnamefont
  {Kermode}}, \bibinfo {author} {\bibfnamefont {J.~R.}\ \bibnamefont
  {Kitchin}}, \bibinfo {author} {\bibfnamefont {E.~L.}\ \bibnamefont
  {Kolsbjerg}}, \bibinfo {author} {\bibfnamefont {J.}~\bibnamefont {Kubal}},
  \bibinfo {author} {\bibfnamefont {K.}~\bibnamefont {Kaasbjerg}}, \bibinfo
  {author} {\bibfnamefont {S.}~\bibnamefont {Lysgaard}}, \bibinfo {author}
  {\bibfnamefont {J.~B.}\ \bibnamefont {Maronsson}}, \bibinfo {author}
  {\bibfnamefont {T.}~\bibnamefont {Maxson}}, \bibinfo {author} {\bibfnamefont
  {T.}~\bibnamefont {Olsen}}, \bibinfo {author} {\bibfnamefont
  {L.}~\bibnamefont {Pastewka}}, \bibinfo {author} {\bibfnamefont
  {A.}~\bibnamefont {Peterson}}, \bibinfo {author} {\bibfnamefont
  {C.}~\bibnamefont {Rostgaard}}, \bibinfo {author} {\bibfnamefont
  {J.}~\bibnamefont {Schiøtz}}, \bibinfo {author} {\bibfnamefont
  {O.}~\bibnamefont {Schütt}}, \bibinfo {author} {\bibfnamefont
  {M.}~\bibnamefont {Strange}}, \bibinfo {author} {\bibfnamefont {K.~S.}\
  \bibnamefont {Thygesen}}, \bibinfo {author} {\bibfnamefont {T.}~\bibnamefont
  {Vegge}}, \bibinfo {author} {\bibfnamefont {L.}~\bibnamefont {Vilhelmsen}},
  \bibinfo {author} {\bibfnamefont {M.}~\bibnamefont {Walter}}, \bibinfo
  {author} {\bibfnamefont {Z.}~\bibnamefont {Zeng}},\ and\ \bibinfo {author}
  {\bibfnamefont {K.~W.}\ \bibnamefont {Jacobsen}},\ }\bibfield  {title}
  {\enquote {\bibinfo {title} {The atomic simulation environment—a python
  library for working with atoms},}\ }\href
  {https://doi.org/10.1088/1361-648X/aa680e} {\bibfield  {journal} {\bibinfo
  {journal} {Journal of Physics: Condensed Matter}\ }\textbf {\bibinfo {volume}
  {29}},\ \bibinfo {pages} {273002} (\bibinfo {year} {2017})}\BibitemShut
  {NoStop}%
\end{thebibliography}
%

%
\end{document}


\def\mytitle{Protons accumulate at the graphene-water interface}
\title{Supporting Information for: \mytitle}

\author{Xavier R. Advincula}
\affiliation{Yusuf Hamied Department of Chemistry, University of Cambridge, Lensfield Road, Cambridge, CB2 1EW, UK}
\affiliation{Cavendish Laboratory, Department of Physics, University of Cambridge, Cambridge, CB3 0HE, UK}
\affiliation{Lennard-Jones Centre, University of Cambridge, Trinity Ln, Cambridge, CB2 1TN, UK}
\author{Kara D. Fong}
\email{kdf22@cam.ac.uk}
\affiliation{Yusuf Hamied Department of Chemistry, University of Cambridge, Lensfield Road, Cambridge, CB2 1EW, UK}
\affiliation{Lennard-Jones Centre, University of Cambridge, Trinity Ln, Cambridge, CB2 1TN, UK}
\author{Angelos Michaelides}
\email{am452@cam.ac.uk}
\affiliation{Yusuf Hamied Department of Chemistry, University of Cambridge, Lensfield Road, Cambridge, CB2 1EW, UK}
\affiliation{Lennard-Jones Centre, University of Cambridge, Trinity Ln, Cambridge, CB2 1TN, UK}
\author{Christoph Schran}
\email{cs2121@cam.ac.uk}
\affiliation{Cavendish Laboratory, Department of Physics, University of Cambridge, Cambridge, CB3 0HE, UK}
\affiliation{Lennard-Jones Centre, University of Cambridge, Trinity Ln, Cambridge, CB2 1TN, UK}
\keywords{}

{\maketitle}
\tableofcontents

\onecolumngrid

\FloatBarrier

\section{Molecular dynamics simulations}
\subsection{System setup}
\begin{figure}
    \includegraphics[width=\textwidth]{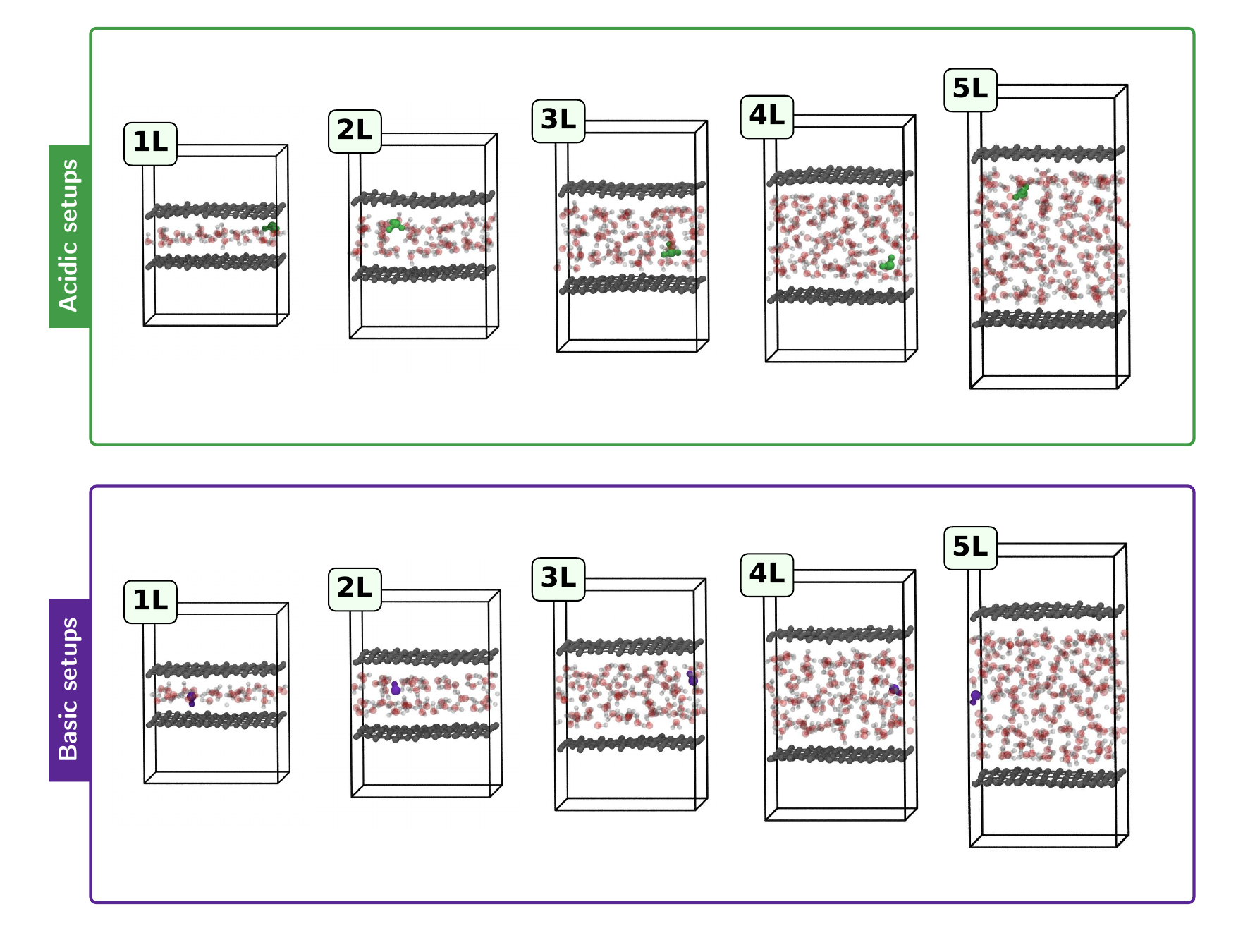}
    \caption{Representative configuration snapshots of the systems studied in this work.
    In the acidic setups, the hydronium ion is represented in green, while in the basic setups, the hydroxide ion is represented in purple.
    The solid black lines indicate the edges of the periodic simulation box.}
    \label{fig:system_setups}
\end{figure}

The systems studied are labeled as 1L, 2L, 3L, 4L, and 5L corresponding to systems with two parallel free-standing graphene sheets (each containing 112 atoms) separated by heights ranging from approximately 6.5 to 20\;\AA\; and intercalated by one (1L), two (2L), three (3L), four (4L), and five of water (5L).
These different slit widths correspond to varying amounts of water molecules corresponding to system sizes between $\approx$ 300 to 700 atoms.
\response{The initial slit widths considered were 6.91\;\AA, 9.41\;\AA, 12.20\;\AA, 14.41\;\AA, and 19.41\;\AA.
In each system, the graphene sheets have dimensions $L_{x} = 17.290$\;\AA\; and $L_{y} = 17.112$\;\AA, derived by repeating the base unit cell dimensions of $a=\sqrt{3}d_{c}$ and $b=3 d_{c}$ multiple times along the $x$ and $y$ directions, respectively.
Specifically, $L_{x}$ corresponds to approximately seven repetitions of the unit cell dimension $a$ along the $x$-axis, while $L_{y}$ is the result of four repetitions of the unit cell dimension $b$ along the $y$-axis.
This tiling of the unit cell ensures that the graphene sheet maintains its characteristic hexagonal lattice structure with a consistent carbon-carbon bond distance of $d_c = 1.42$\;\AA\;\cite{RevModPhys.81.109} throughout the extended sheet.}
All systems were simulated in orthorhombic simulation cells employing periodic boundary conditions in all three directions.
\response{
To prevent interactions between the periodic images, a vacuum space of 15\;\AA\; vacuum was added in the $z$ direction of the initial configurations.
In the MACE architecture, each layer interacts only with neighboring atoms within a specified cutoff distance, which sets the interaction range for that layer. As information is passed through successive layers, the receptive field expands to include neighbors at increasing distances, theoretically extending up to the product of the number of layers and the cutoff. However, when a vacuum is present and no atoms fall within this cutoff, message-passing is restricted, effectively limiting the receptive field to the local cutoff of 6\;\AA. Thus, the 15\;\AA\; vacuum lies beyond the model's effective receptive field and the energy convergence threshold of our electronic structure settings, ensuring decoupling in the $z$-direction.
}
An overview of the systems studied, including a hydronium or hydroxide ion, is provided in Fig \ref{fig:system_setups} and Table \ref{tab:setup_details}. 

\begin{figure}
    \includegraphics[width=\textwidth]{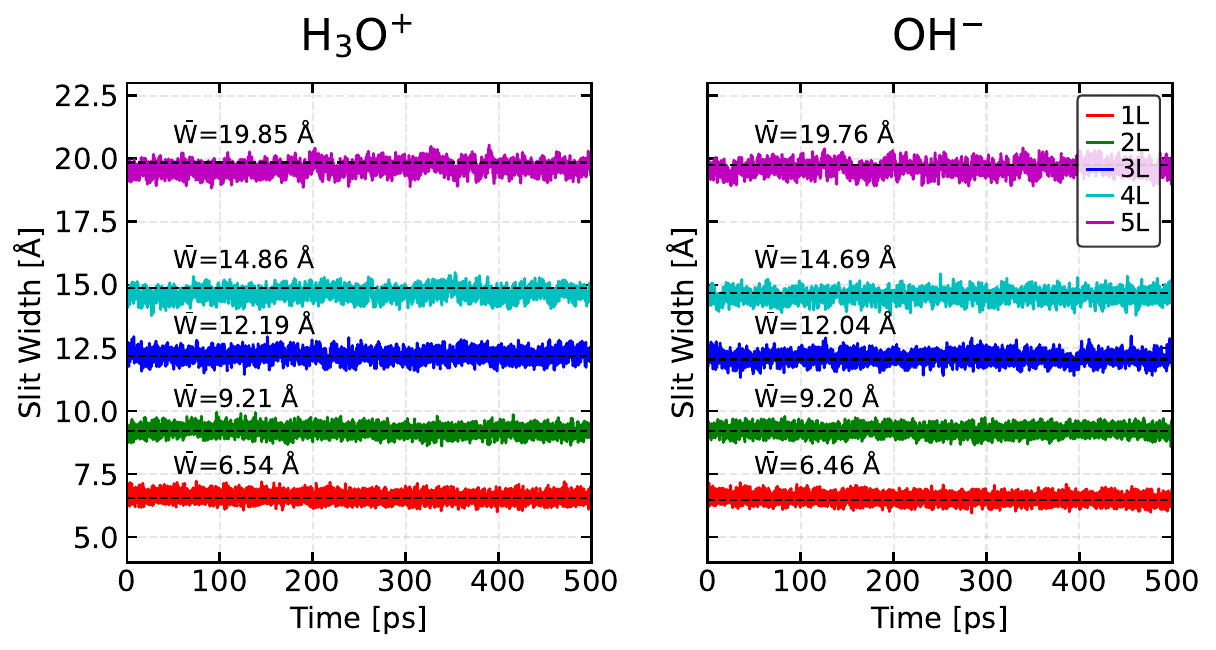}
    \caption{Fluctuations in the slit width of the systems studied here for both the acidic (left) and basic (right) setups.
    For each slit width and each species, the average slit widths $\Bar{\textrm{H}}$ over an entire run are reported to illustrate the size of the systems studied.}
    \label{fig:system_heights}
\end{figure}

The graphene sheets in our simulations were treated in a fully flexible manner to achieve the corresponding equilibrium density.
\response{The simulation cells were initially prepared by randomly packing molecules between the graphene sheets to form one to five distinct layers of water. 
The initial quantity of water molecules was chosen based on prior studies \cite{sergi_nanoconf, kara_pairing_2024} to achieve slit widths comparable to experimentally realizable pore dimensions.}
To provide a sense of the system size, Fig. \ref{fig:system_heights} shows the average slit width and its fluctuations over a 500\;ps period for each run. 

\response{
Our simulation setups are particularly well-suited for experimental comparison, as they avoid artificial pressures that could alter the natural behavior of confined water.
By allowing the graphene sheets to remain fully flexible and reach equilibrium density under ``zero pressure'' conditions, we capture the intrinsic properties of water confined between graphene layers without imposing external constraints that might complicate experimental replication.
This design makes our results directly comparable to experimental setups, where slit pores are often constructed with graphene-graphene distances close to those in our model.
Notably, our selected pore distances align well with experimentally achievable channels, such as those reported in Ref. ~\citenum{radha_b_2016}, and support validation through techniques like sum-frequency generation vibrational spectroscopy, which has been applied to study graphene-water interfaces in Ref.~\citenum{mischa_transp}.
}

\begin{table}
\centering
\caption{Overview of both the acidic and basic setups.
In each system, the graphene sheets have dimensions $L_{x} = 17.290$\;\AA\; and $L_{y} = 17.112$\;\AA\;.
For each system, we report their average slit width, $\Bar{\textrm{W}}$; the total number of atoms, $N_{\textrm{atoms}}$; the corresponding number of water molecules $N_{\textrm{H}_2\textrm{O}}$ and protonic defects (i.e., either a hydronium ion $N_{\textrm{H}_3\textrm{O}^{+}}$ or hydroxide ion $N_{\textrm{O}\textrm{H}^{-}}$); the number of replicate runs, $N_{\textrm{runs}}$; the equilibration time, $t_{\textrm{eq}}$; and the simulation production time, $t_{\textrm{sim}}$.}
\label{tab:setup_details}
\begin{tabular}{@{}ccccccc@{}}
\toprule
System                                                                   &      & 1L & 2L & 3L & 4L & 5L\\ \midrule
\multirow{7}{*}{\begin{tabular}[c]{@{}c@{}}Acidic\\ setups\end{tabular}} & $\Bar{\textrm{W}}$  [\AA{}]          &  6.54      & 9.21   & 12.19   & 14.86    & 19.85      \\
& $N_{\textrm{atoms}}$         & 309              & 378  & 471  & 549  & 699     \\
& $N_{\textrm{H}_2\textrm{O}}$ & 27               & 50 &  81 &  107 &  157     \\
& $N_{\textrm{H}_3\textrm{O}^{+}}$ & 1   &  1 & 1  & 1  & 1      \\
& $N_{\textrm{runs}}$ & 5               & 5 &  5 &  5 &  5     \\
& $t_{\textrm{eq}}$ [ps]            &  90  &  90 & 90  & 90  &  90    \\ 
& $t_{\textrm{sim}}$ [ns]          &  4  &  4 & 4  & 4  &  4  \\ \midrule
\multirow{7}{*}{\begin{tabular}[c]{@{}c@{}}Basic\\ setups\end{tabular}} & $\Bar{\textrm{W}}$   [\AA{}]          &  6.46      & 9.20   & 12.04 & 14.69    & 19.76     \\
& $N_{\textrm{atoms}}$         & 307              & 376  & 469  & 547  & 697     \\
& $N_{\textrm{H}_2\textrm{O}}$ & 27               & 50 &  81 &  107 &  157     \\
& $N_{\textrm{O}\textrm{H}^{-}}$ & 1   &  1 & 1  & 1  & 1      \\
& $N_{\textrm{runs}}$ & 5               & 5 &  5 &  5 &  5     \\
& $t_{\textrm{eq}}$ [ps]          &  90 &  90 & 90  & 90 &  90  \\ 
& $t_{\textrm{sim}}$ [ns]           &  4  &  4 & 4  & 4  &  4 \\ \bottomrule
\end{tabular}
\end{table}

\newpage
\subsection{Identification of hydronium and hydroxide ions}
The hydronium and hydroxide ions do not have a static structure; therefore, their identification requires dynamic identification in each configuration.
Here, we follow the definitions proposed in Ref.~\citenum{chandra_pt_mono}.
To recognize the hydronium ion, we initially assigned the two closest hydrogen atoms to each oxygen atom. Subsequently, the remaining unassigned hydrogen was linked to its nearest oxygen, which we identified as the oxygen of the hydronium ion, denoted as O$^{*}$.

Similarly, the identity of the hydroxide ion also changes throughout the simulation due to proton transfer events.
To determine the oxygen of this anionic defect, we again assigned the two nearest hydrogen atoms to each oxygen.
Next, we identified the hydrogen that was assigned to two oxygens and reassigned it to the oxygen to which it was closest.
The oxygen left with only one hydrogen is then recognized as the hydroxide ion's oxygen, O$^{*}$.

\subsection{Simulation setup}
In this work, we conducted four different types of molecular dynamics (MD) simulations: (i) short \textit{ab initio} MD (AIMD) simulations to generate training data for the development of the MACE machine learning potential (MLP); (ii) additional AIMD simulations to generate reference data for the validation of the MLP; (iii) extensive unbiased MD simulations using the developed MLP, comprising the main results reported in this work;  and (iv) additional MLP-based biased simulations using umbrella sampling to compare the free energy profiles reported.
All simulations involved a flexible treatment of the graphene sheets and employed hydrogen atoms unless explicitly stated otherwise.

\textbf{Short AIMD simulations}

The AIMD simulations were performed using the CP2K/Quickstep code \cite{cp2k_2020} in the NVT ensemble with a time step of 1 fs.
The temperatures were set at 100, 300, and 600\;K and maintained using a combination of a CSVR thermostat \cite{csvr_2007} and an adaptive Langevin thermostat \cite{ad_langevin_2011}.
The revPBE generalized gradient approximation exchange-correlation functional \cite{revpbed3_1} with the zero-damping variant of the Grimme’s D3 dispersion correction \cite{revpbed3_2} was used, in combination with the dual-space Goedecker-Tetter-Hutter pseudopotentials \cite{gth_1996} to represent the atomic cores, a 450 Ry plane wave cutoff, and the TZV2P basis set to expand the Kohn-Sham orbitals of oxygen and hydrogen atoms or the DZVP basis set to expand those of carbon atoms \cite{basis_set_2007}. 
To maintain stable simulations with a computationally manageable time step, deuterium masses were used. 
For the final training of the MLP, all the training structures were reevaluated by performing single-point density functional theory (DFT) calculations with an increased plane wave cutoff of 1200 Ry.

\textbf{Reference AIMD simulations}

The reference AIMD simulations to validate the MLP were performed using the CP2K code \cite{cp2k_2020} in the NVT ensemble with a time step of 1 fs.
The temperature was set to 300\;K and maintained using a CSVR thermostat \cite{csvr_2007} with a 30 fs coupling constant.
To maintain stable simulations with a computationally manageable time step, deuterium masses were used.
A 15\;ps equilibration period was followed by a 150\;ps production period.

\textbf{Unbiased MLP-based MD simulations}

The unbiased MLP-based MD simulations were performed using ASE \cite{ase_2017} in the NVT ensemble where the temperature was maintained at 300 K, unless explicitly stated otherwise.
For this, a Langevin thermostat with a friction coefficient of 2.5 ps$^{-1}$ was used. The timestep was set to 0.5 fs.
For each slit width and protonic defect, we conducted five replicate runs involving a 90\;ps equilibration period followed by a 4\;ns production period, from which statistics of the observables of interest were sampled. 
The equilibration period consisted of an initial 45\;ps phase during which the graphene sheets were fully immobilized to equilibrate the water molecules with the protonic defects. This was followed by a subsequent 45\;ps phase where the graphene sheets were treated as fully flexible, ensuring thorough equilibration.

\textbf{Biased MLP-based MD simulations}

The additional biased MLP-based MD simulations were performed using the LAMMPS simulations package \cite{lammps_2022} with the PLUMED plugin \cite{plumed_2013} in the NVT ensemble where the temperature was maintained at 300 K.
For this, a Nos\'e-Hoover thermostat with a damping constant of 0.05 ps was used. The timestep was set to 0.5 fs.
For each system, a 40\;ps equilibration period was followed by a 75\;ps production period, from which statistics of the observables of interest were sampled.
The equilibration period comprised an initial phase of 20\;ps with the graphene sheets fully immobilized to properly equilibrate the water molecules with the protonic defects, followed by a subsequent 20\;ps phase where the graphene sheets were treated as fully flexible.

\newpage
\section{Machine learning potential} \label{sec:mlp}

\subsection{Model development}
The MLP model was progressively developed over five generations.
The initial generation leveraged a training set from prior work\cite{quasionedimensional} and incorporated an active learning approach \cite{cnnp_2020, complex_2021} to integrate structures designed for water-carbon interactions at slit widths of 5 and 6.5 \;\AA.
This stage accommodated a spectrum of conditions from low- to high-density water at temperatures including 100, 300, and 600 K.
The second iteration introduced structures from path integral MD simulations, capturing the quantum nature of nuclei.
In the third generation, the model was expanded to include various slit widths --6, 10, 15, and 20\;\AA-- adding structures tailored to these dimensions.
The fourth iteration involved another round of active learning to refine further the model based on previously analyzed conditions.
Finally, the fifth generation integrated configurations containing both a hydronium and a hydroxide ion at slit widths of 6, 10, and 13\;\AA\;in bulk water.
This choice deliberately avoided the inclusion of isolated protonic defects to circumvent the need for external charges for maintaining charge neutrality, as suggested in Ref.~\citenum{marsalek_mlp_pt}.
This approach prevents variations in box energies that could arise from different box volumes, thus enhancing the stability and reliability of the simulation results.

\subsection{Model validation}

\textbf{Energy and force errors}

\begin{figure}[htp!]
    \includegraphics[width=\textwidth]{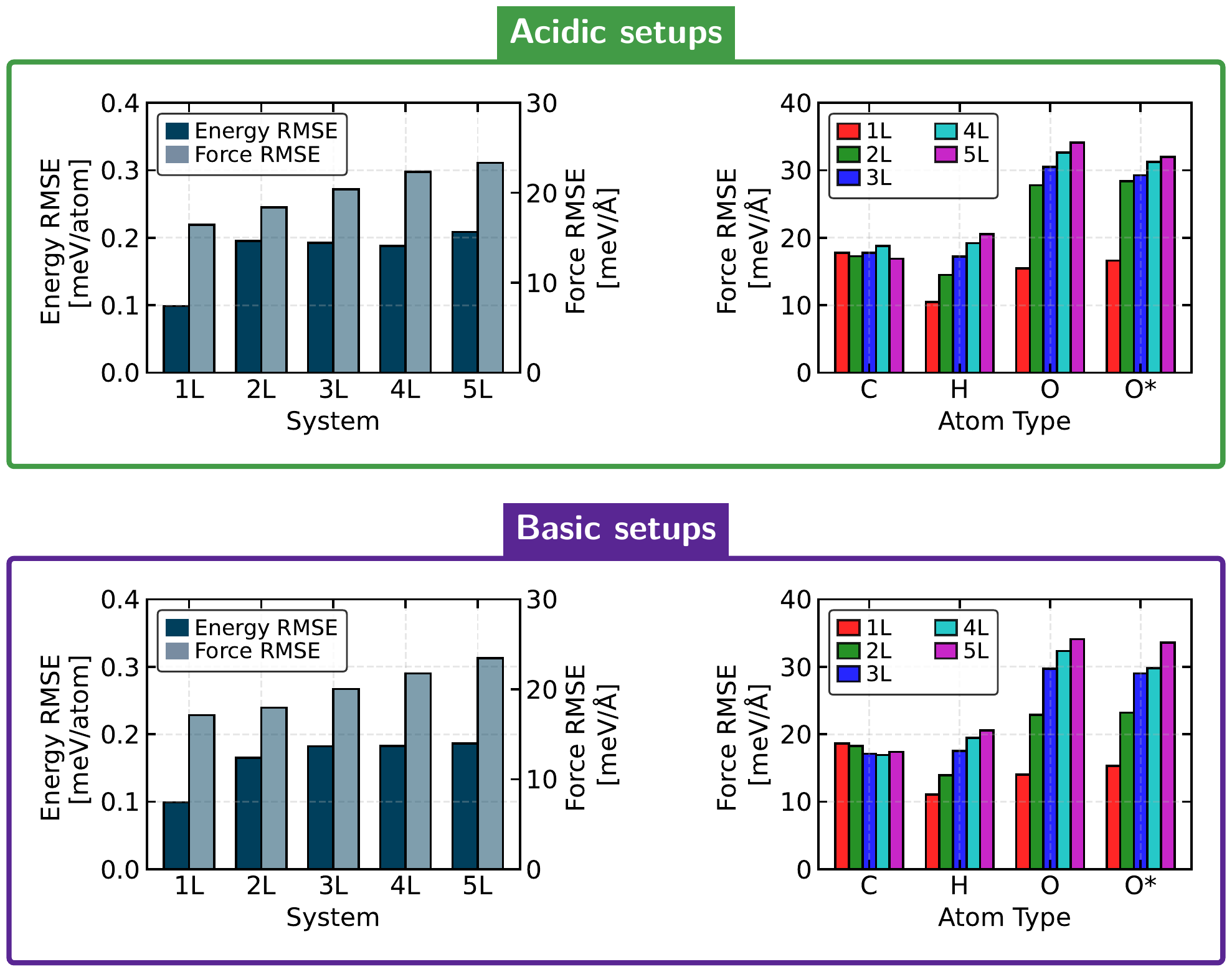}
    \caption{RMSE of the energies and forces obtained using the MLP at 300K, compared to the reference DFT calculations across the five slit widths (left), along with their force RMSE broken down by different atom types, including the specific oxygen of the protonic defect O$^{*}$ (right), for both the acidic and basic setups.}
    \label{fig:energies_forces_benchmark}
\end{figure}

\begin{figure}[htp!]
    \includegraphics[width=\textwidth]{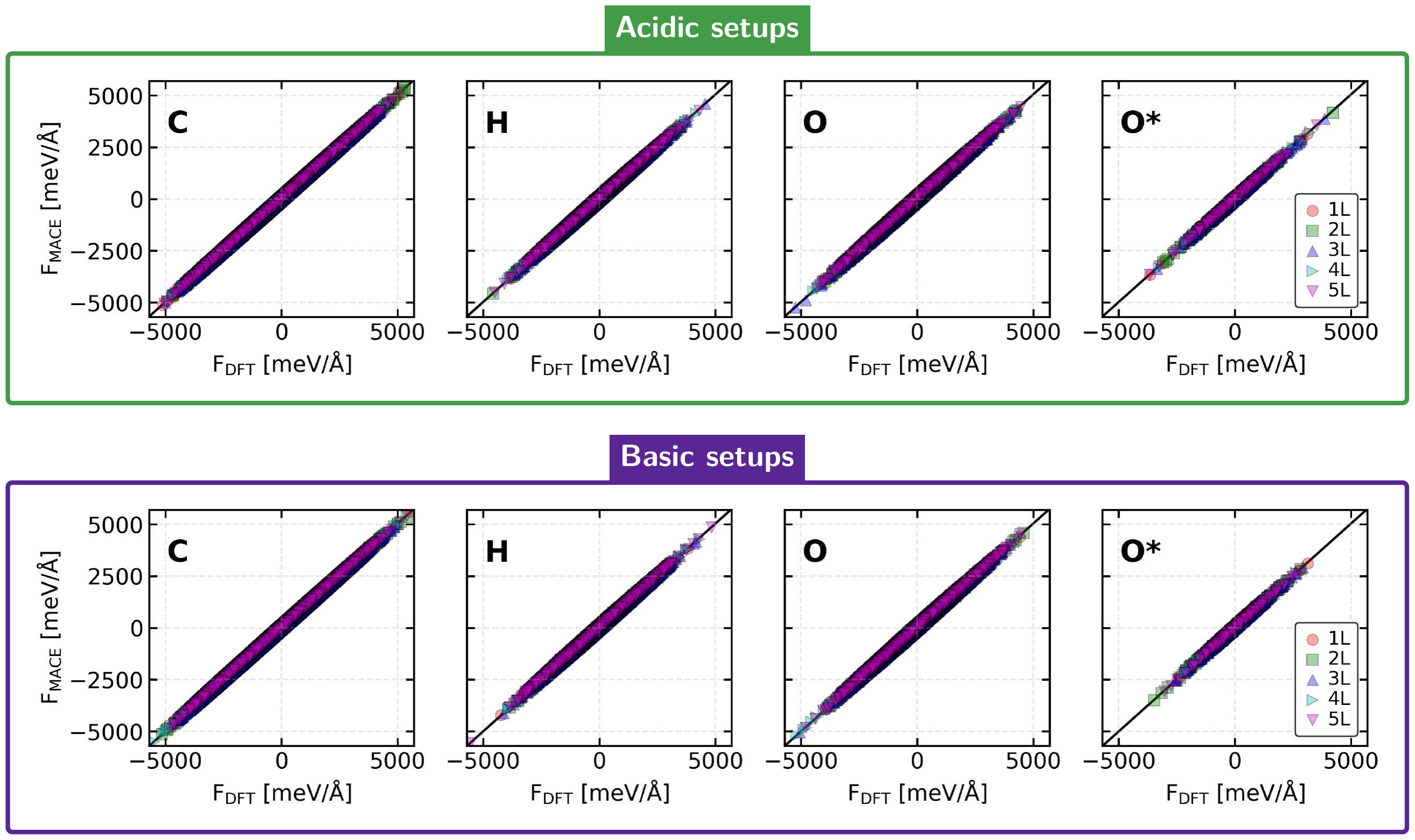}
    \caption{Parity plots for the forces obtained using the MLP, compared to the reference DFT calculations, broken down by different atom types including the specific oxygen of the protonic defect O$^{*}$, across the five slit widths for both the acidic and basic setups.}
    \label{fig:forces_parity}
\end{figure}

To quantify the root-mean-square error (RMSE) of the energies and forces predicted by the MLP, we conducted a detailed analysis using structures generated from 500\;ps MLP-based MD simulations \response{at 300\;K}.
These simulations covered various slit widths and both types of protonic defects.
From these simulations, 100 snapshots were randomly selected for each setup, and their energies and forces were calculated using single-point DFT calculations.
\response{
In cases where the system contained a hydronium or hydroxide ion, a homogeneous background charge was applied to maintain charge neutrality.
The energies predicted by the MLP were then adjusted by subtracting a constant energy offset to account for the shift introduced by the homogeneous background charge.
}
To mitigate the high computational costs typically associated with electronic structure calculations on large systems, we scaled down the size of the systems by setting the dimensions of the graphene sheets to $L_{x}=12.350$\;\AA\; and $L_{y}=12.834$\;\AA.
This validation approach is particularly robust as it assesses structures derived directly from the MLP's potential energy surface, employed in the active learning protocol to develop the model.
As seen in Figs. \ref{fig:energies_forces_benchmark} and \ref{fig:forces_parity}, there is an excellent agreement between the MLP and the results with the underlying level of theory, demonstrating the MLP model's ability to effectively reproduce the energies and forces obtained from the reference DFT calculations.

\begin{figure}[htp!]
    \includegraphics[width=0.9\textwidth]{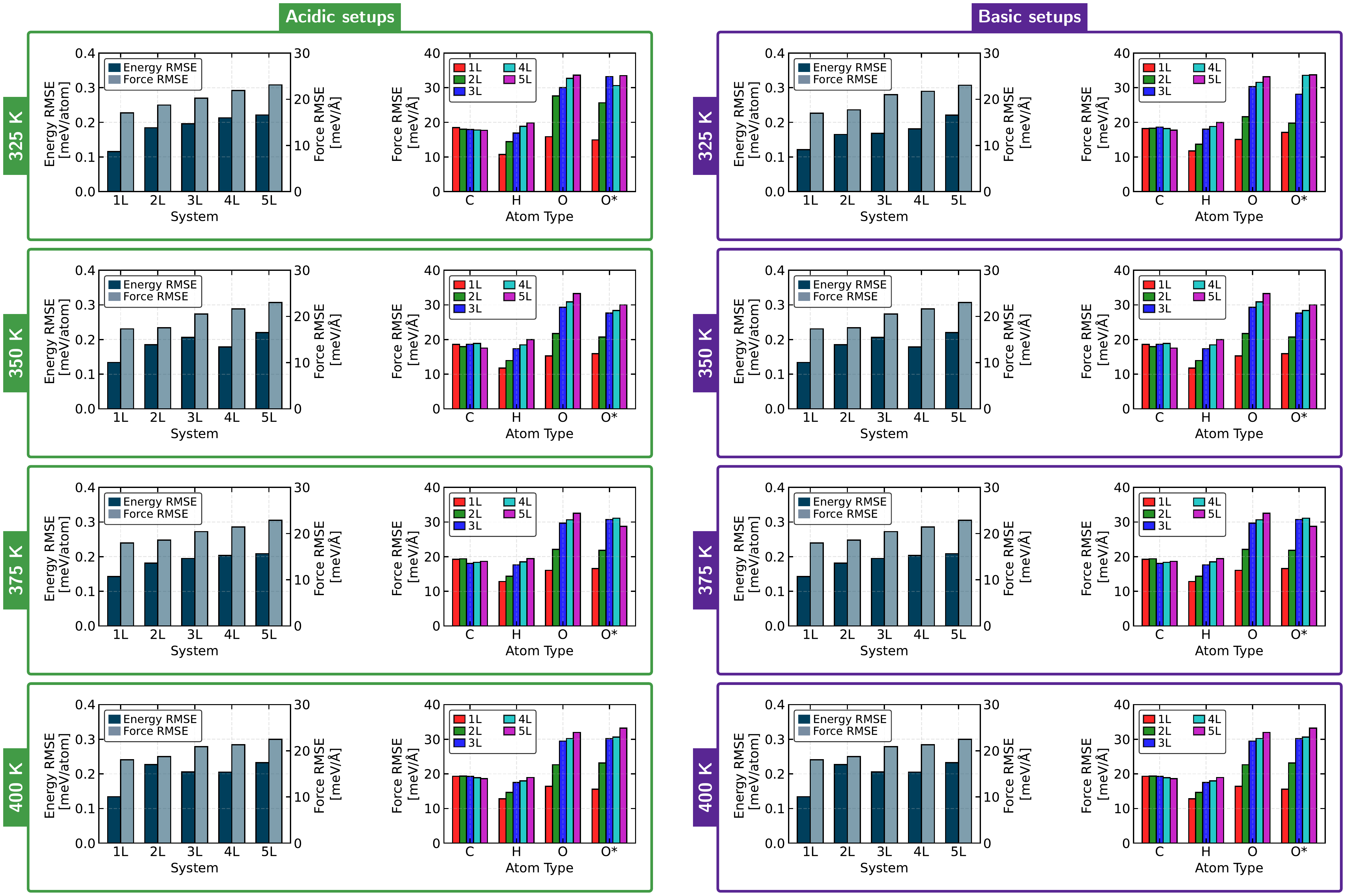}
    \caption{RMSE of the energies and forces obtained using the MLP at 325, 350, 375, and 400 K, compared to the reference DFT calculations across the five slit widths (left), along with their force RMSE broken down by different atom types, including the specific oxygen of the protonic defect O$^{*}$ (right), for both the acidic and basic setups.}
    \label{fig:energies_forces_benchmark_all_k}
\end{figure}

\newpage
\textcolor{black}{
Since the model is also used to simulate conditions across a temperature range from 300 to 400 K, we conducted similar analyses to those presented above for additional temperatures of 325, 350, 375, and 400 K. This is presented in Fig. \ref{fig:energies_forces_benchmark_all_k}.
}

\newpage

\response{For completeness, we also report that our MLP effectively captures temperature-induced density changes, resulting in an expansion of the water region in our setups, as shown in Table \ref{tab:setup_expansion}.}
\vspace{-0.15cm}
\begin{table}[htp!]
\centering
\caption{Overview of the water region expansion of the 3L system as a function of temperature.
For both the acidic and basic conditions, we report the temperature, $T$; the average slit width, $\Bar{\textrm{W}}$; and the standard deviation of the slit width, $\sigma_{\Bar{\textrm{W}}}$.
}
\label{tab:setup_expansion}
\begin{tabular}{@{}cccc@{}}
\toprule
System                                                                   & $\textrm{T}$ [K]     & $\Bar{\textrm{W}}$ & $\sigma_{\Bar{\textrm{W}}}$ \\ \midrule
\multirow{5}{*}{\begin{tabular}[c]{@{}c@{}}Acidic\\ 3L setup\end{tabular}} & 300   & 12.194 & 0.200 \\
& 325 & 12.314  & 0.208        \\
& 350 & 12.453 & 0.222       \\
& 375 & 12.563  & 0.231       \\
& 400 & 12.723  & 0.261     \\ \midrule
\multirow{5}{*}{\begin{tabular}[c]{@{}c@{}}Basic\\ 3L setup\end{tabular}} & 300   & 12.043 & 0.197 \\
& 325 & 12.163  & 0.200        \\
& 350 & 12.295 & 0.209       \\
& 375 & 12.413  & 0.220        \\
& 400 & 12.538  & 0.236     \\ \bottomrule
\end{tabular}
\end{table}

\newpage
\textbf{Comparison to reference AIMD simulations}

To further validate the MLP, we generated AIMD trajectories for water with protonic defects under both bulk and confined conditions. 
This step was crucial for benchmarking structural and dynamical properties, particularly focusing on radial distribution functions (RDFs) and the free energy profiles associated with the transfer of protonic defects.
Ensuring an accurate representation of proton transfer (PT) is key to the reliability of our model.

\begin{figure}[htp!]
    \includegraphics[width=0.8\textwidth]{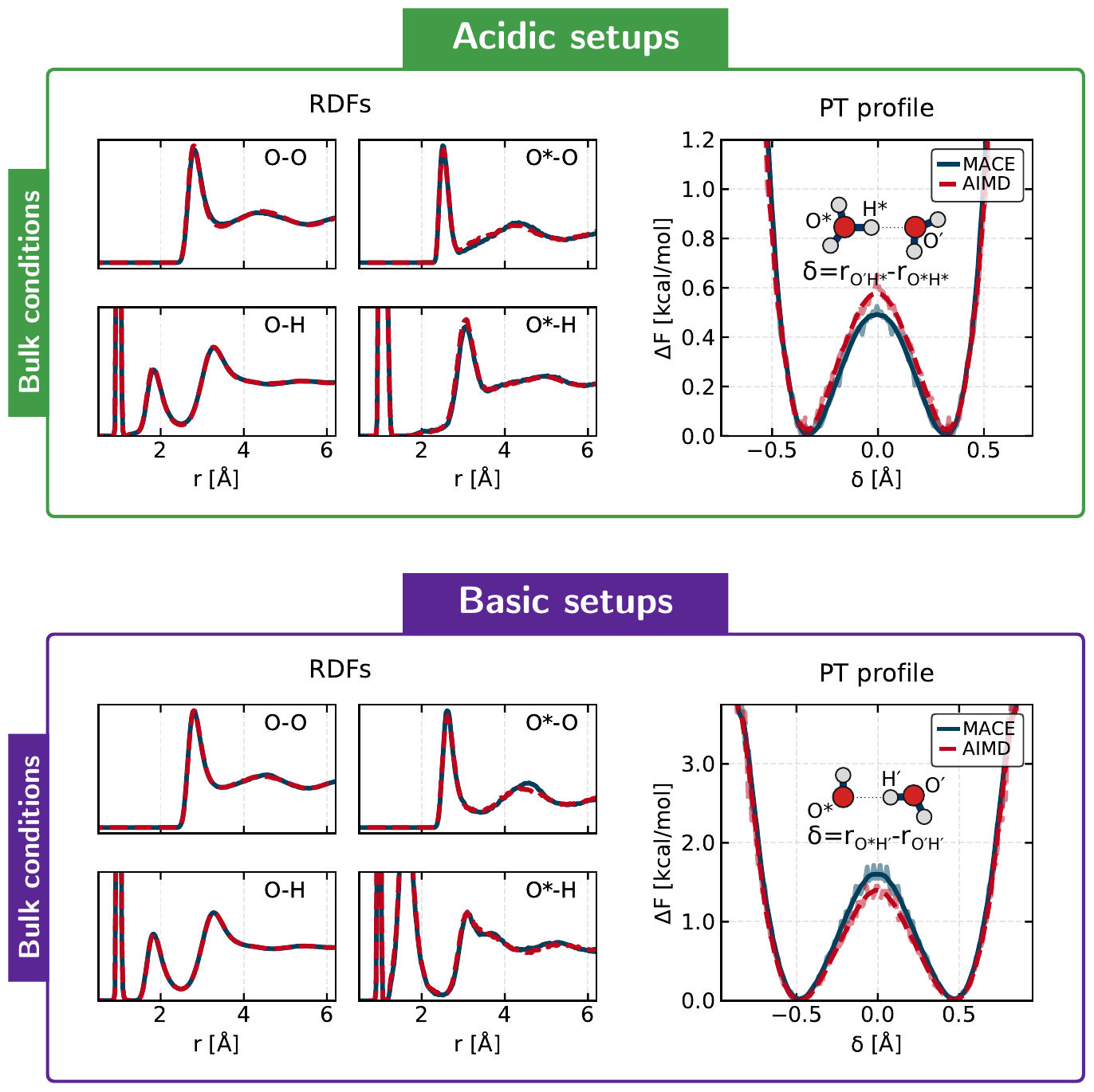}
    \caption{Comparison of RDFs and free energy profiles for protonic defect transfer in bulk conditions, using the PT coordinate defined in the schematic.}
    \label{fig:benchmark_species_bulk}
\end{figure}

\begin{figure}[htp!]
    \includegraphics[width=\textwidth]{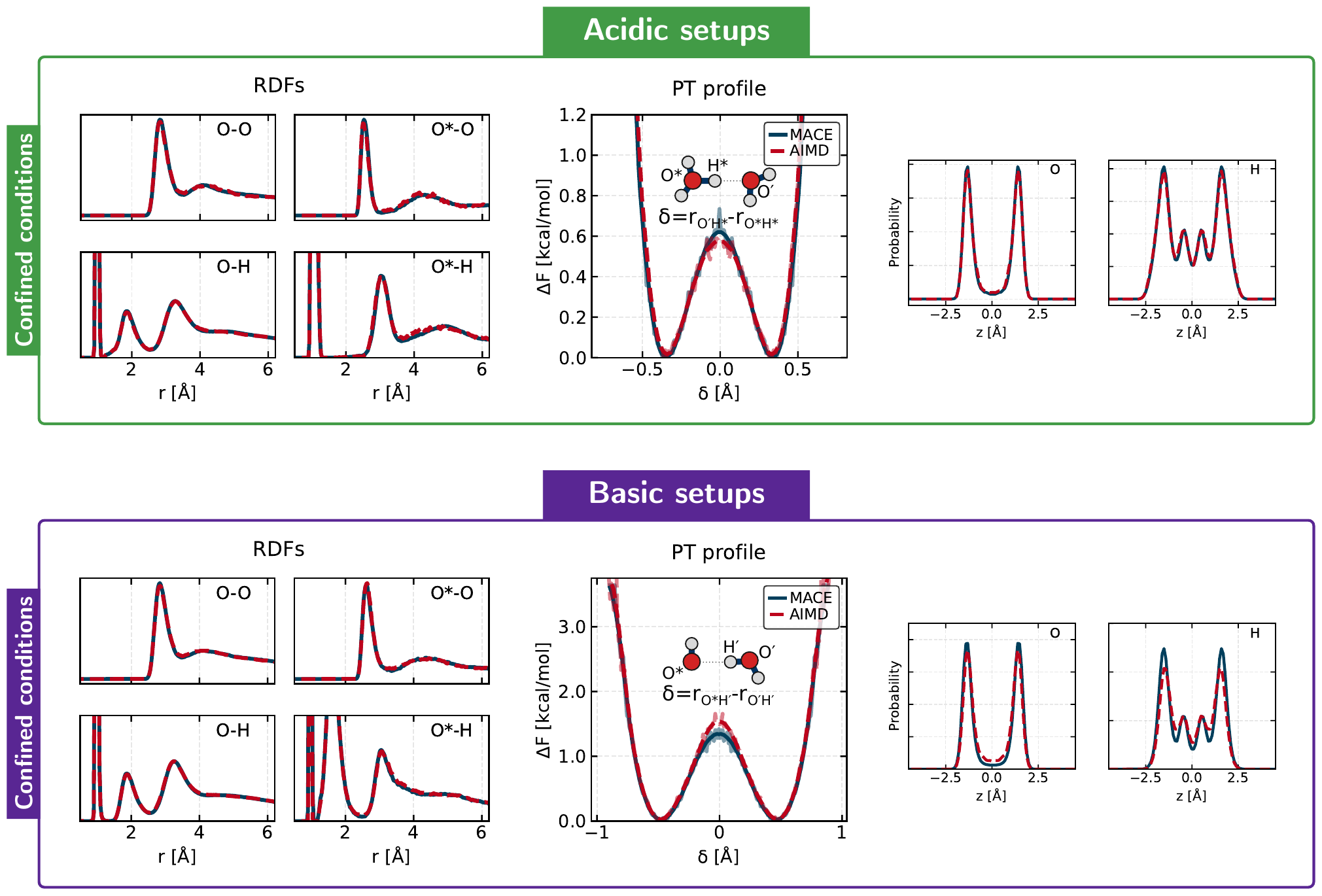}
    \caption{Comparison of RDFs, free energy profiles for protonic defect transfer, and density profiles in confined conditions, using the PT coordinate defined in the schematic.}
    \label{fig:benchmark_species_confined}
\end{figure}

In the bulk simulations, we used 63 water molecules and 1 protonic defect (hydronium or a hydroxide ion).
For the confined simulations, graphene sheets with dimensions of 12.35\;\AA{} and 12.834\;\AA{} were used to create a 3L system.
For each of these conditions, 200\;ps long trajectories were produced to gather validation data.
As seen in Figs. \ref{fig:benchmark_species_bulk} and \ref{fig:benchmark_species_confined}, there is excellent agreement between the MLP and AIMD results,  underscoring the MLP model’s ability to faithfully reproduce the structural and dynamical predictions of the reference DFT calculations and effectively capture the critical physics of protonic defect behavior in both bulk and nanoconfined environments.
\newpage

\newpage

\begin{figure}[htp!]
    \includegraphics[width=\textwidth]{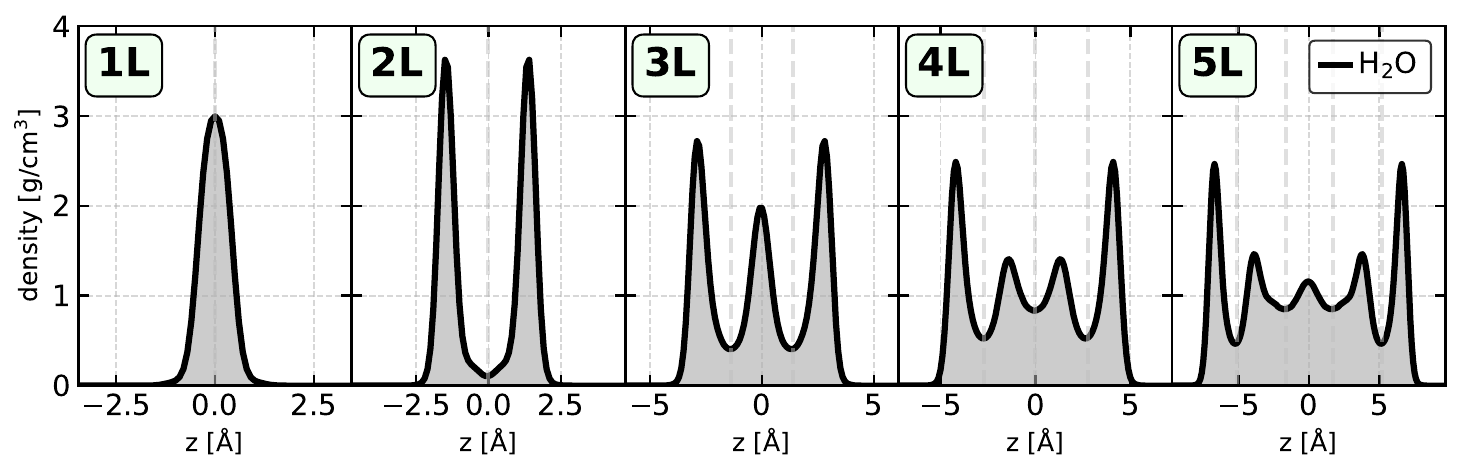}
    \caption{Water density profiles along the $z$ axis perpendicular to the free-standing graphene sheets obtained from the oxygen atoms in a neutral water system. The vertical dashed lines indicate the partitioning among the different water layers.}
    \label{fig:density_recov}
\end{figure}

\newpage
\textbf{\response{Recovering bulk-like density}} \\
\response{
We evidence that bulk-like density is achieved at the center of the thicker slit widths by including a scale for the water density in the simulated slits presented in the main manuscript. As shown in Fig. \ref{fig:density_recov}, the central density closely matches that reported in Ref. S2, and serves as a benchmark for ensuring accurate densities within the slit, with the same central value referenced as `bulk-like'.
}

\textbf{\response{Graphene-hydronium and graphene-hydroxide interactions}} \\
\response{To further validate the graphene-hydronium and graphene-hydroxide interactions modeled with revPBE-D3, we retrained our MLP at the revPBE0-D3 level, addressing potential delocalization errors commonly associated with GGA functionals. The inclusion of exact Fock exchange in this hybrid approach allows for a more rigorous assessment of discrepancies stemming from electronic structure approximations, providing a more reliable framework for accurately capturing the interactions in question. As shown in Fig. \ref{fig:hybrid_check}, the density profiles for hydronium and hydroxide ions show near-complete overlap between revPBE-D3 and revPBE0-D3 levels, confirming that our results are robust across these electronic structure treatments. This consistency underscores the generality of our findings and suggests that the observed interactions reflect the physical behavior intrinsic to these systems, rather than artifacts of functional choice.
}

\begin{figure}[htp!]
    \includegraphics[width=0.9\textwidth]{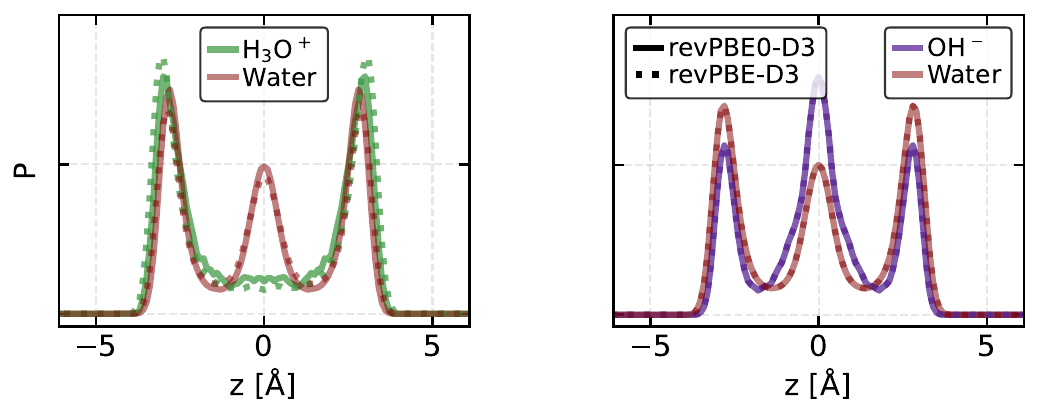}
    \caption{Normalized and symmetrized number density profiles along the $z$ axis perpendicular to the free-standing graphene sheets obtained from the specific oxygen of the protonic defect and the oxygen of the surrounding water molecules of acidic (left) and basic (right) setups for the 3L system.
    The horizontal axis limits in each plot correspond to the average carbon layer positions.}
    \label{fig:hybrid_check}
\end{figure}

\begin{figure}[htp!]
    \includegraphics[width=0.85\textwidth]{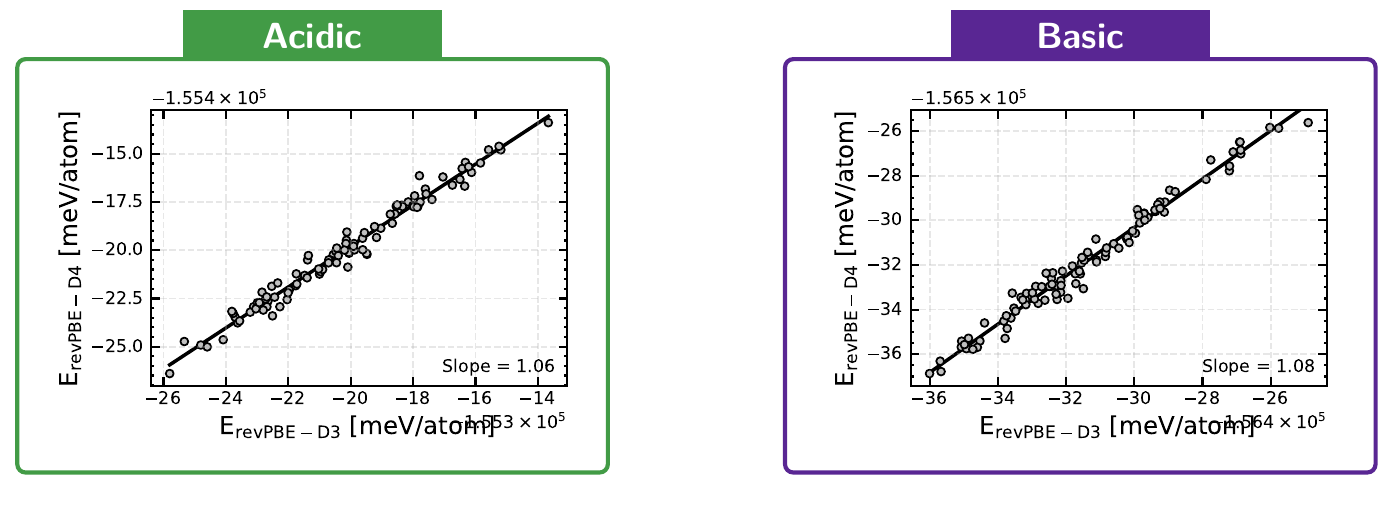}
    \caption{Parity plot for the energies using revPBE-D3 and revPBE-D4 for the acidic (left) and basic (right) 3L system.}
    \label{fig:d3_vs_d4}
\end{figure}

\begin{figure}[htp!]
    \includegraphics[width=0.75\textwidth]{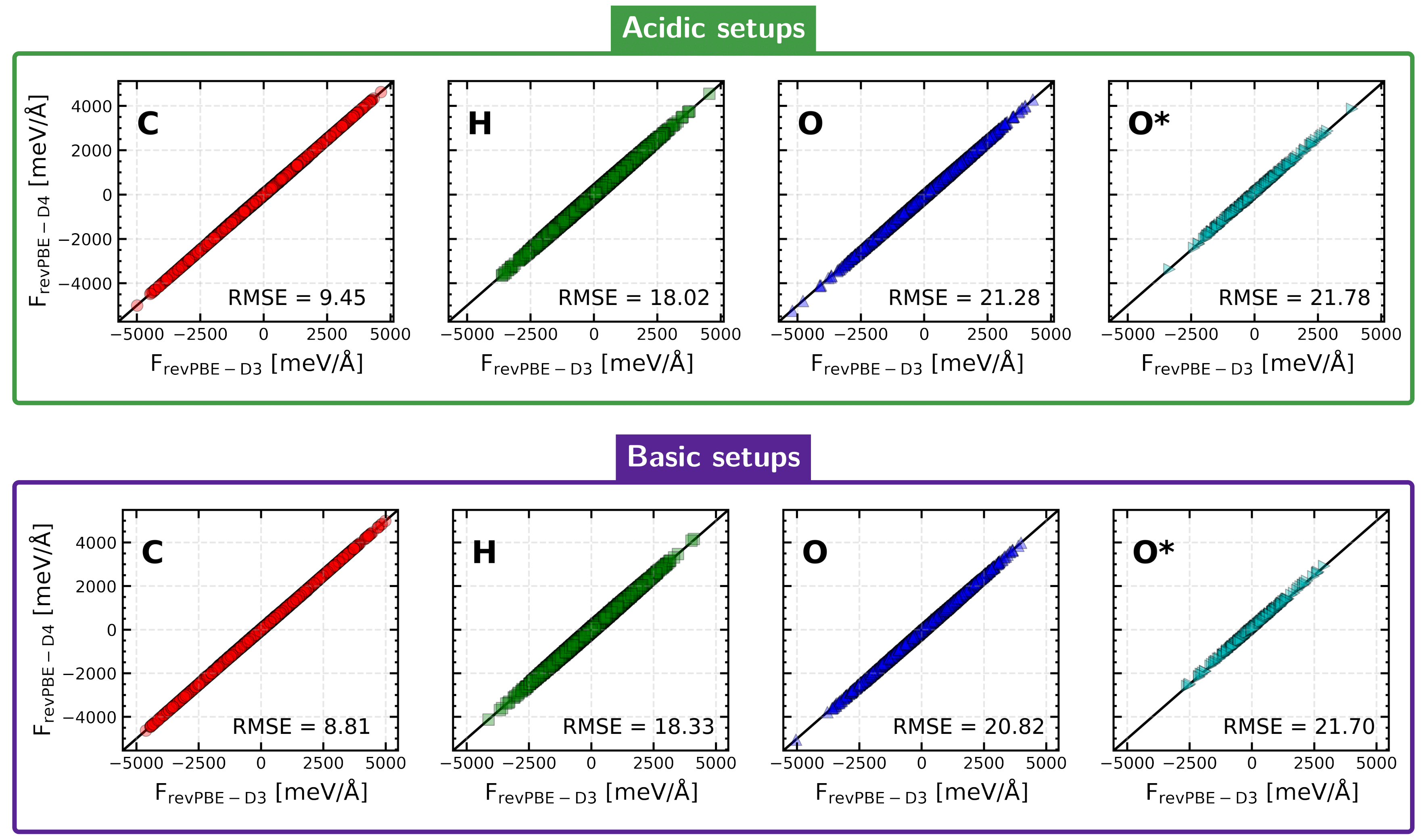}
    \caption{Parity plots for the forces obtained using revPBE-D3 and revPBE-D4, broken down by different atom types including the specific oxygen of the protonic defect O$^{*}$, for the acidic (left) and basic (right) 3L system.}
    \label{fig:forces_parity_d3_d4}
\end{figure}

\response{We further demonstrate that the choice of dispersion correction does not significantly influence the underlying energies or forces. This is evidenced by the nearly perfect correlation observed in the parity plot comparing revPBE-D3 and revPBE-D4 energies (see Fig. \ref{fig:d3_vs_d4}), which shows no substantial differences between the two functionals. Furthermore, a detailed breakdown of the impact of the forces on the different atom types, including protonic defects, confirms the robustness of our approach across these corrections (see Fig. \ref{fig:forces_parity_d3_d4}). This supports the reliability of our results, confirming that our conclusions remain robust regardless of the dispersion correction method used.}

\newpage

\newpage
\section{Effect of graphene flexibility}
To evaluate the impact of graphene flexibility on the reported trends, we conducted additional simulations with the graphene sheets fully immobilized. 
To minimize computational costs, we carried out a 1\;ns simulation on the smallest system, the 3L system, which features an intermediate region. 
This setup allowed us to efficiently observe the trends, providing a clear basis for comparison of the phenomena.
Interestingly, even with the graphene immobilized, the trends in our results remained consistent. 
This indicates that the flexibility of the graphene sheets does not significantly affect the key trends observed in our findings, suggesting that the phenomena are robust across various mechanical constraints imposed on the graphene structure.

\begin{figure}[htp!]
    \includegraphics[width=0.9\textwidth]{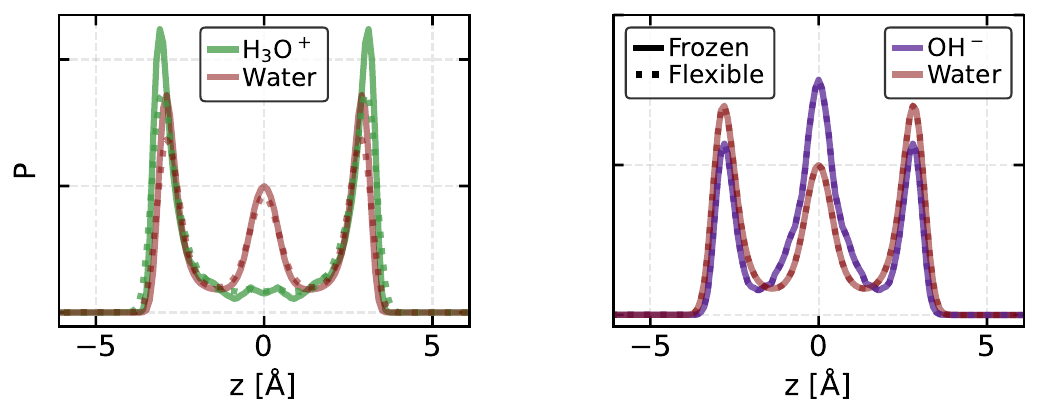}
    \caption{Normalized and symmetrized number density profiles along the $z$ axis perpendicular to the free-standing (dashed) or frozen (solid) graphene sheets obtained from the specific oxygen of the protonic defect and the oxygen of the surrounding water molecules of acidic (left) and basic (right) setups for the 3L system.
    The horizontal axis limits in each plot correspond to the average carbon layer positions.}
    \label{fig:graphene_flexibility}
\end{figure}

\response{We also evaluated the influence of density (and, by extension, pressure) within these immobilized graphene slits. Our results confirm that the accumulation of protons at the graphene-water interface is insensitive to density variations within such conditions.}

\newpage

\begin{figure}[htp!]
    \includegraphics[width=0.9\textwidth]{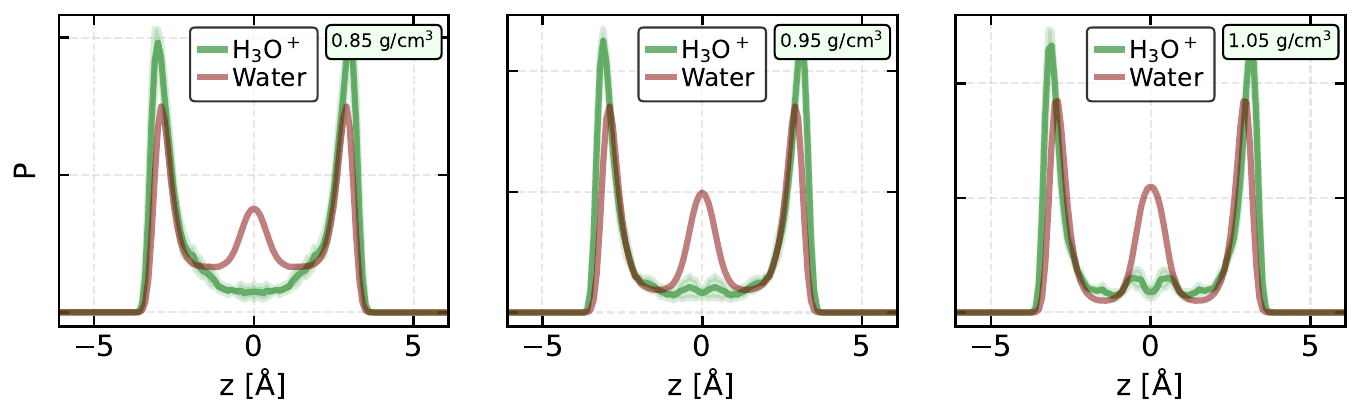}
    \caption{Normalized and symmetrized number density profiles along the $z$ axis perpendicular to the frozen graphene sheets obtained from the specific oxygen of the protonic defect and the oxygen of the surrounding water molecules of acidic setups with varying densities (obtained using the effective volume of water) for the 3L system.}
    \label{fig:hybrid_check}
\end{figure}

\newpage
\section{Potential of mean force calculation with umbrella sampling}

To verify the unbiased simulations reported in the main text, we obtained the potential of mean force (PMF) of the protonic defects relative to the graphene layers using umbrella sampling \cite{umbrella_int_2005, umbrella_int_2}. For this, we conducted MD simulations where we restrained the oxygen atom O$^{*}$ of the protonic defect at different target height values $z_{0}$ above an immobile (i.e., fully frozen) surface carbon atom by using a restraining potential of the form,
\begin{equation}
U_{\textrm{bias},1} (z) =  k_{\textrm{bias},1} (z-z_{0})
\end{equation}
where $z$ is the instantaneous height of the O$^{*}$ above the graphene sheet, and $k_{\textrm{bias},1}=30$\;kcal/mol/\AA{}. To prevent unexpected shifts in the chemical structure of protonic defects due to proton hopping --as suggested in Ref. \citenum{mlb_oh}, where it was applied only to the hydroxide ion--  we restrained the hydrogen coordination value of the protonic defect species $n_{\textrm{O}^{*}-\textrm{H}}$ around a target value $n_{0}$ using a harmonic potential of the form, 
\begin{equation}
U_{\textrm{bias},2} = \frac{k_{\textrm{bias},2}}{2} (n_{\textrm{O}^{*}-\textrm{H}} - n_{0})^{2}
\end{equation}
with $k_{\textrm{bias},2}=400$\;kcal/mol and
\begin{equation}
n_{\textrm{O}^{*}-\textrm{H}} = \sum^{N}_{i=1} \frac{1 - \left( \frac{r_i}{R_0} \right)^{12} }{1 - \left( \frac{r_i}{R_0} \right)^{20}}
\end{equation}
 where $i$ iterates over each hydrogen atom within the simulation box (for a total of $N$), $r_{i}$ is the distance between the hydrogen $i$ and O$^{*}$, and $R_{0}$ is a switch distance set to 1.2\;\AA{}. In systems with a hydronium ion, the coordination number is maintained at 3.0, whereas in systems featuring a hydroxide ion, it is maintained at 1.3, corresponding to their ideal solvated configurations \cite{mlb_oh}. To obtain the PMF profiles, umbrella integration \cite{umbrella_int_2005, umbrella_int_2} is performed using the Python implementation from Ref.~\citenum{atb-code}.

\begin{figure}[htp!]
    \includegraphics[width=\textwidth]{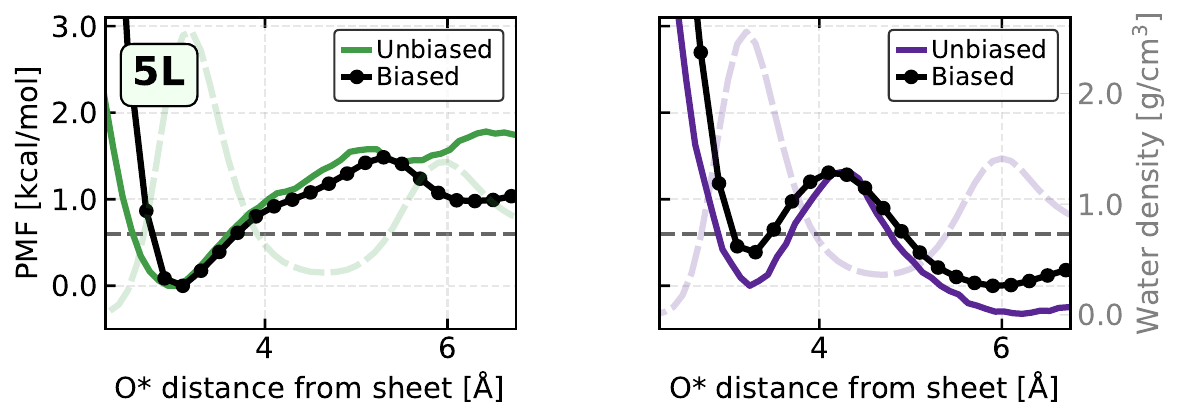}
    \caption{Potential of mean force for the hydronium (left) and hydroxide (right) ions as a function of their oxygen distance O$^{*}$ to the graphene sheet for the 5L system. The structuring of the water layers is represented by the water density profiles, which are indicated with corresponding lighter colors and dashed lines. The horizontal dashed line indicates the thermal energy $k_{\mathrm{B}}T \approx 0.6$\;kcal/mol ($T=300$\;K).}
    \label{fig:biased_sims}
\end{figure}

 \newpage
 The PMF profiles obtained from umbrella integration are compared to the free energy profiles from unbiased simulations, calculated using $\Delta F = -k_{\textrm{B}}T\ln{P(z)}$. This comparison, presented in Fig. \ref{fig:biased_sims}, evidences the appropriate sampling of the phase space and the correct behavior captured in our unbiased simulations. 

\begin{figure}[htp!]
    \includegraphics[width=0.5\textwidth]{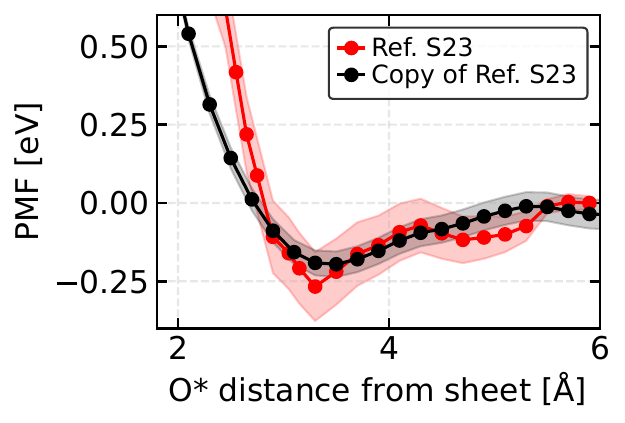}
    \caption{Potential of mean force for the hydroxide ion as a function of its oxygen distance O$^{*}$ to the graphene sheet obtained from Ref.~\citenum{mlb_oh} (red) and that obtained using our PBE-D3 retrained MLP with the same setup (black).}
    \label{fig:mlb_comp_fig_1}
\end{figure}

\begin{figure}[htp!]
    \includegraphics[width=0.5\textwidth]{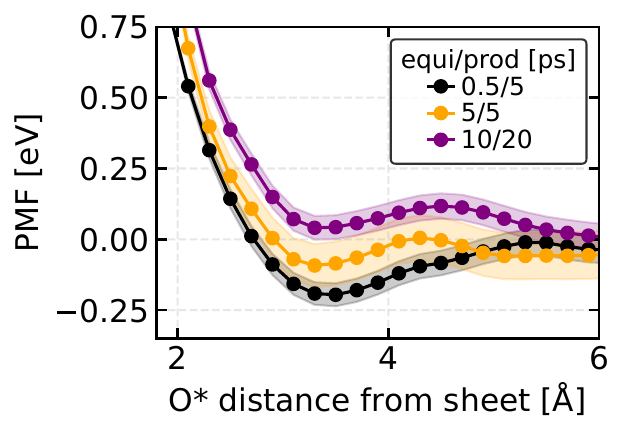}
    \caption{Potential of mean force for the hydroxide ion as a function of its oxygen distance O$^{*}$ to the graphene sheet obtained by copying the setup and simulation protocol in Ref.~\citenum{mlb_oh} (black) and with increasingly larger equilibration/production times (yellow, purple).}
    \label{fig:mlb_comp_fig_2}
\end{figure}

The PMFs reported here indicate a slight destabilization of hydroxide at the interface, which is at odds with the physisorbed state reported in Ref.~\citenum{mlb_oh}, where a free energy barrier of $−0.27\pm0.13$\;eV was observed for the hydroxide ion using umbrella sampling.
We note in passing that no free energy barrier for the hydronium ion was reported, as this has not yet been addressed in the literature.
To provide a fair comparison of these biased PMFs and determine the root of these differences, we retrained the MLP developed herein to the PBE-D3 level to use the same settings as reported in Ref.~\citenum{mlb_oh}.
By copying their initial setup (i.e., a single layer of graphene with water on top), their simulation protocol, and their simulation times (here, we assumed a standardized interval time of 5.0 ps per umbrella window, based on their range of 4–6 ps varied according to parameter adjustments), we obtained quantitatively consistent results to those reported in their study (see Fig. \ref{fig:mlb_comp_fig_1}).
However, using longer simulation times for the umbrella sampling windows (30 ps in total, including both equilibration and production time, compared to the shorter 4.5-6.5 ps), we recovered a similar free energy profile to the one reported in this manuscript, as shown in Fig. \ref{fig:mlb_comp_fig_2}.
This indicates that the short simulation times used in Ref. \citenum{mlb_oh} led to unequilibrated structures, causing gradually increasing errors in the PMF sampling.
These findings reveal the limitations of the brief simulation times employed in Ref.~\citenum{mlb_oh}, which were constrained by the high computational cost of AIMD simulations.
Notably, this also highlights the advantages of using MLPs, which have been increasingly employed in recent years to overcome the challenges associated with expensive AIMD simulations.
This approach is crucial to our study as it enables us to achieve unprecedented accuracy and extensively sample the phase space.

\newpage
\section{Water hydrogen bonding}
\begin{figure}[htp!]
    \includegraphics[width=0.9\textwidth]{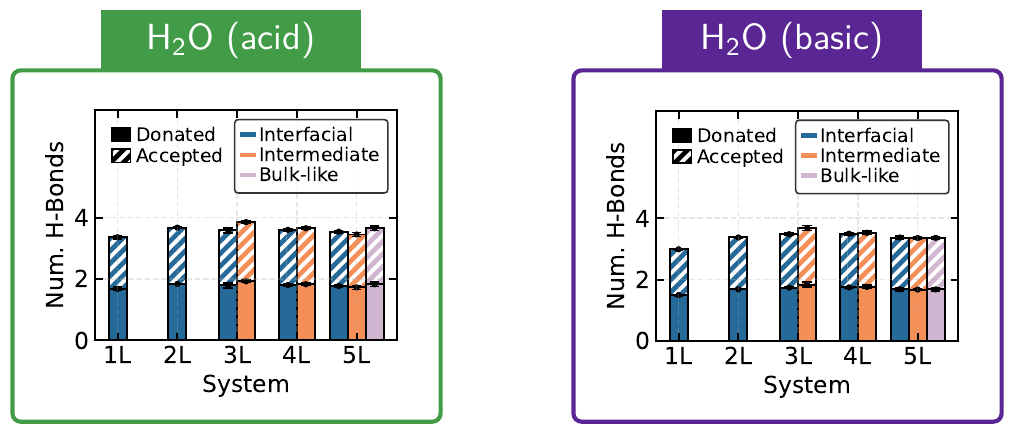}
    \caption{Average number of hydrogen bonds (accepted or donated, as indicated in each legend) for the water molecules in the acidic (left) and basic (right) systems across the different water layers.
    Hydrogen bonds are counted using the geometric definition provided in Ref.~\citenum{luzar_chandler}.
    The error bars are obtained from the standard deviation of the five replicate simulations.}
    \label{fig:hbs_water}
\end{figure}

\response{
To complement the hydrogen bonding analysis of the acidic and basic systems analyzed in this work, we show in Fig. \ref{fig:hbs_water} the average number of hydrogen bonds donated and accepted by surrounding water molecules across different slit systems (1L to 5L). This figure demonstrates how hydrogen bonding varies with position relative to the interface.
}

\newpage
\section{Partial charge analysis}
\response{To determine the extent to which graphene polarizes in response to nearby hydronium or hydroxide ions, we assessed partial charges using Bader charge analysis \cite{bader_1, bader_2, bader_3, bader_4} as a function of the in-plane distances from a hydronium ion or a hydroxide ion.}
Due to the high computational cost associated with electronic structure calculations on large systems, we analyzed the system where ions are always in close contact with the interface to ensure that polarization can be easily represented, namely the 1L system.
Moreover, to further reduce the computational cost associated with this calculation, a  1050 Ry cutoff was used.
While the MLP model does not explicitly include charges on carbon atoms, it does capture graphene polarizability implicitly through the DFT calculations that serve as its training data.
Therefore, we utilized the MLP to generate a total of 5,000\;structures sampled across the 4\;ns trajectories from the MLP-based MD simulations. 

\begin{figure}[htp!]
    \includegraphics[width=0.7\textwidth]{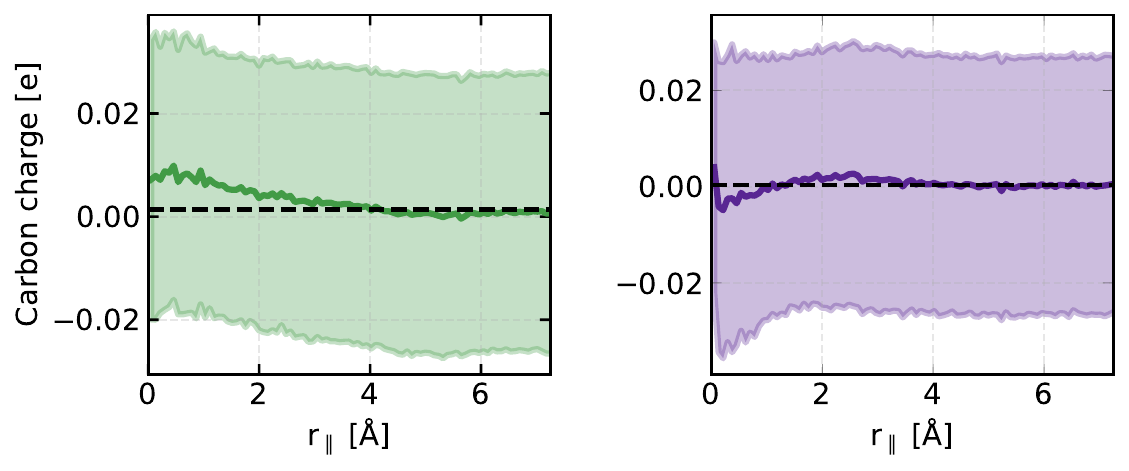}
    \caption{Carbon charges as a function of the in-plane distances from a hydronium ion (left) or a hydroxide ion (right) to the carbon atoms in the 1L system, averaged across 5,000 configurations generated by the MLP model.
    The shaded areas show the range of one standard deviation in the charge distribution, and the black dashed lines indicate the average carbon charge across all configurations.}
    \label{fig:c_charges_bader}
\end{figure}

\begin{figure}[htp!]
    \includegraphics[width=1\textwidth]{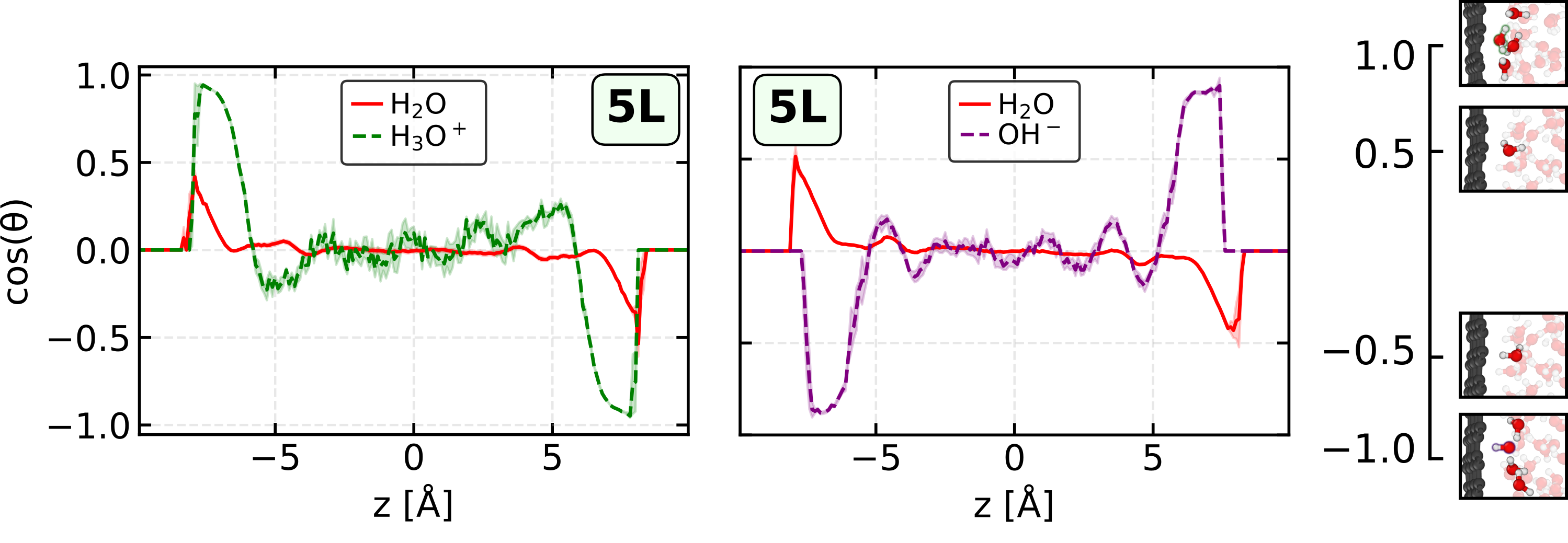}
    \caption{Average angle of the dipole of water molecules and OH$^-$ (left) and H$_3$O$^+$ (right) along the $z$ axis perpendicular to the free-standing graphene sheets for the 5L system with representative snapshots.}
    \label{fig:orients}
\end{figure}

\response{
It is worth noting that water molecules with an OH bond oriented toward the graphene surface can induce polarization effects similar to those of hydroxide ions. However, our analysis reveals a key difference: hydroxide ions consistently orient their OH group toward the graphene interface, creating a more stable and sustained polarization effect compared to the transient orientations observed for regular water molecules. This persistent alignment of hydroxide at the interface underscores its distinctive interaction pattern, as clearly illustrated in Fig. \ref{fig:orients}. %
To further enrich this analysis, we also present the orientational profiles for hydronium ions, offering a comparative perspective on their behavior at the graphene interface.}

\newpage

\newpage
%